\documentclass[letterpaper,12pt]{article}

\usepackage{fullpage,mathrsfs}

\usepackage{amsfonts}
\usepackage{amsmath,amssymb,amsthm}
\usepackage{graphicx}
\usepackage{color}
\usepackage{setspace}
\usepackage{bbm} 

\def\no{\nonumber}
\newcommand{\half}{\frac{1}{2}}

\newcommand{\modz}{u}
\newcommand{\p}{\partial}
\newcommand{\dd}{\delta}
\newcommand{\N}{\mathcal{N}}
\newcommand{\ti}{\widetilde}

\newcommand{\aaa}{\mathbf{a}}
\newcommand{\ccc}{\mathbf{c}}
\newcommand{\OO}{\Lambda}

\newcommand{\bbb}{\mathbbm{b}} 

\newcommand{\be}{\begin{equation}}
\newcommand{\ee}{\end{equation}}
\newcommand{\bea}{\begin{eqnarray}}
\newcommand{\eea}{\end{eqnarray}}

\newcommand{\lp}{\left(}
\newcommand{\rp}{\right)}
\newcommand{\tr}{\textrm{Tr}}

\newcommand{\diff}{\mathrm{d}}
\newcommand{\de}{\partial}
\newcommand{\anglepsi}{\varsigma}
\newcommand{\betaph}{\omega}

\newcommand{\pp}{P}
\newcommand{\qq}{Q}  

\newcommand{\Acks}{A^{\mathrm{cs}}}

\newcommand{\bC}{\ensuremath{\mathbb{C}}}

\newcommand{\bN}{\ensuremath{\mathbb{N}}}

\newcommand{\bR}{\ensuremath{\mathbb{R}}}

\newcommand{\bZ}{\ensuremath{\mathbb{Z}}}

\newcommand{\scA}{\ensuremath{\mathcal{A}}}
\newcommand{\scB}{\ensuremath{\mathcal{B}}}

\newcommand{\scF}{\ensuremath{\mathcal{F}}}

\newcommand{\scI}{\ensuremath{\mathcal{I}}}

\newcommand{\scL}{\ensuremath{\mathcal{L}}}

\newcommand{\scO}{\ensuremath{\mathcal{O}}}

\newcommand{\scR}{\ensuremath{\mathcal{R}}}

\newcommand{\scW}{\ensuremath{\mathcal{W}}}

\numberwithin{equation}{section}       

\begin{document}

\begin{titlepage}

\begin{center}

\begin{flushright}
KCL-MTH-14-09
\end{flushright}

\today

\vskip 3 cm 

{\LARGE \bf Localization on Hopf surfaces}

\vskip 1.5cm

{\large Benjamin Assel, Davide Cassani and Dario Martelli}

\vskip 1cm

{\textit{Department of Mathematics, King's College London, \\ [1mm]
The Strand, London WC2R 2LS,  United Kingdom\\ [5mm]}}

\end{center}

\vskip 3 cm

\begin{abstract}

\noindent  We discuss localization of the path integral for supersymmetric gauge theories with an R-symmetry on Hermitian four-manifolds. After presenting the localization locus equations for the general case, we focus on backgrounds with $S^1 \times S^3$ topology, admitting two supercharges of opposite R-charge. These are Hopf surfaces, with two complex structure moduli $p,q$. We compute the localized partition function on such Hopf surfaces, allowing for a very large class of Hermitian metrics, and prove that this is proportional to the supersymmetric index with fugacities $p,q$. Using zeta function regularisation, we determine the exact proportionality factor, finding that it depends only on $p,q$, and on the anomaly coefficients $a$, $c$ of the field theory. This may be interpreted as a supersymmetric Casimir energy, and provides the leading order contribution to the partition function in a large $N$ expansion.

\end{abstract}

\end{titlepage}

\pagestyle{plain}
\setcounter{page}{1}
\newcounter{bean}
\baselineskip 6 mm

\newpage

\tableofcontents

\section{Introduction}
\label{sec:intro}

The complete information of a quantum field theory is contained in the generating functional of correlation functions; however, 
in an interacting theory this is very hard to compute exactly.  In favourable situations the technique of supersymmetric localization \cite{Pestun:2007rz} allows one to perform exact non-perturbative computations of special types of generating functionals and other observables. In particular, in certain supersymmetric field theories defined on compact Riemannian manifolds, it is possible to evaluate a class of BPS observables by reducing  the functional integrals over all the field configurations to Gaussian integrals around a supersymmetric locus.  In this paper we will present a detailed calculation of the partition function of  ${\cal N}=1$ supersymmetric field theories, defined on a four-dimensional complex manifold.

A systematic procedure for constructing supersymmetric  field theories in a fixed background geometry has been put forward in \cite{Festuccia:2011ws}. In four dimensions, one way to obtain supersymmetric theories is by taking a suitable limit of  new minimal supergravity \cite{Sohnius:1981tp,Sohnius:1982fw,Sohnius:1982xs}, that contains two auxiliary vector fields,  one of which is the gauge field for a local chiral symmetry. In such rigid limit, these, together with the metric, provide background fields coupled to a supersymmetric gauge theory  with an R-symmetry, comprising ordinary vector and chiral multiplets.  
Explicit expressions for supersymmetric Lagrangians and supersymmetry transformations can be obtained from \cite{Sohnius:1981tp,Sohnius:1982fw,Sohnius:1982xs}
and will be presented below.   
 
Supersymmetric theories may be defined only on backgrounds admitting solutions to certain Killing spinor equations (see  (\ref{KeqnZeta}),  (\ref{KeqnTiZeta})   below), 
which in Euclidean signature  are equivalent to the requirement that the four-dimensional manifold is complex and the metric Hermitian  \cite{Klare:2012gn,Dumitrescu:2012ha}. 
In this paper we will construct Lagrangians  that are total supersymmetry variations, and therefore can be utilised to implement the localization technique in \hbox{${\cal N}=1$} field theories defined on arbitrary Hermitian manifolds.  We will then employ these to compute in closed form the partition function of general supersymmetric gauge theories, in the case that the manifold admits at least two supercharges of opposite R-charge, and has the topology of $S^1\times S^3$. 
These manifolds are then {\em Hopf surfaces}, with complex structure characterised by two parameters $p,q$, that we will denote as ${\cal H}_{p,q}\simeq S^1\times S^3$. 

The main result of this paper is the derivation of a formula for the partition function $Z$ of an ${\cal N}=1$ supersymmetric field theory with an R-symmetry, defined on a Hopf surface ${\cal H}_{p,q}$, endowed with a very general Hermitian metric. Namely, we will show that 
\bea
Z  [{\cal H}_{p,q} ] & =  & \mathrm{e}^{- \scF (p,q)} \ \scI(p,q) \, ,
\label{mainre}
\eea
where  $\scI(p,q)$ is the \emph{supersymmetric index}  with $p,q$ fugacities  and $\scF (p,q)$ is a function of the complex structure parameters given by
\be
  \scF (p,q)  \ =\ \frac{4\pi}{3} \lp |b_1| + |b_2| - \frac{|b_1| + |b_2|}{|b_1 || b_2|} \rp (\aaa - \ccc) 
+ \frac{4\pi}{27} \frac{(|b_1|+|b_2|)^3}{|b_1|| b_2|}   (3\, \ccc - 2\, \aaa) \ ,
\label{eps0}
\ee
where $p=\mathrm{e}^{-2\pi |b_1|}, q= \mathrm{e}^{-2\pi |b_2|}$, and $\aaa$, $\ccc$ are the R-symmetry traces, appearing in the Weyl and R-symmetry anomalies of superconformal field theories
\cite{Anselmi:1997ys,Cassani:2013dba}. As we will explain, the real parameters $b_1,b_2$ characterise an almost contact structure in the three-dimensional theory obtained from dimensional 
reduction on $S^1$, allowing us to make contact with the results of \cite{Alday:2013lba}, where the localized partition function of three-dimensional ${\cal N}=2$ supersymmetric gauge theories was computed. 
The supersymmetric index was introduced in \cite{Romelsberger:2005eg,Kinney:2005ej,Romelsberger:2007ec}
in the context of superconformal field theories, and has been used in \cite{Dolan:2008qi,Spiridonov:2009za,Spiridonov:2010hh,Spiridonov:2012ww} (and many others) to test non-perturbative dualities.

The authors of \cite{Closset:2013vra} have shown that very  generally the path integral of a supersymmetric field theory defined on a Hermitian manifold 
can depend only on complex structure deformations of the background. 
Based on this result, they have conjectured that the partition function defined on a Hopf surface ${\cal H}_{p,q}$ is proportional to the supersymmetric index  $\scI(p,q)$, up to possible local counterterms.  
Our explicit computation confirms the validity of  this conjecture,\footnote{For simplicity we will restrict attention to the case where the parameters $p,q$ are real.} although 
we expect  that the ratio $\mathrm{e}^{-\mathcal F}$ between these two quantities generically cannot be expressed in terms of local counterterms.  This provides an interesting quantity characterising a 
four-dimensional supersymmetric field theory, that we will refer to as \emph{supersymmetric Casimir energy}. 

Some progress towards obtaining the partition function (\ref{mainre}) using localization was made in \cite{Closset:2013sxa}, where the 
one-loop determinant of an ${\cal N}=1$ chiral multiplet on a Hopf surface was computed. In particular, in this reference the authors considered a specific Hermitian
metric compatible with  $|p|=|q|$. Localization computations of supersymmetric gauge theories on $S^1\times S^3$ with a conformally flat metric have appeared in \cite{Nawata:2011un,Peelaers:2014ima}.

One of our motivations for computing the partition function from first principles arose from holography \cite{Cassani:2014zwa}. 
 In situations where there exist simple AdS$_5$ gravity duals, the gravity side predicts that the logarithm of the partition function, at leading order in a large $N$ expansion, 
should be proportional to  $N^2$. In one dimension lower, the analogous problem is well understood:  the  $N^{3/2}$ scaling of the on-shell action on the gravity side can be matched to  the large $N$ limit of the localized free energy \cite{Drukker:2010nc}; it has been shown in \cite{Farquet:2014kma} that this agreement can be extended to a broad class of ${\cal N}=2$ gauge theories, whose partition function was computed in \cite{Alday:2013lba}. 
In four dimensions the supersymmetric index scales like $N^0$ at large $N$ \cite{Kinney:2005ej,Gadde:2010en}, implying that 
 the $N^2$ scaling of the logarithm of the partition function must arise as an extra contribution. We find that this contribution is contained in~(\ref{eps0}).

Thus, for superconformal field theories with Sasaki-Einstein gravity duals (so that  $\ccc =\aaa$ at leading order in $N$), 
we obtain   a prediction  for the holographically renormalised action of five-dimensional gauged supergravity, 
evaluated on a solution dual to a supersymmetric field theory defined on a Hopf suface ${\cal H}_{p,q} = \de M_5$. 
In particular, we expect that for a solution $M_5\simeq S^1 \times \mathbb{R}^4$, the renormalised on-shell action will be given by
\bea
S_\mathrm{5d\, sugra} [M_5] & = &   \frac{\pi^2}{54 G_5} \frac{(|b_1|+|b_2|)^3}{|b_1 || b_2|}  \,  ,
\label{5dprediction}
\eea 
up to finite local counterterms.

The rest of this paper is organized as follows.
Section \ref{sec:susy} contains a discussion of the background geometry of four-manifolds allowing for at least one supercharge, and sets the stage for implementing localization in general four-dimensional $\mathcal N=1$ gauge theories with an R-symmetry. In section \ref{Hopf:sec}  we discuss the specific background geometry for Hopf surfaces with $S^1 \times S^3$ topology and $U(1)^3$ isometries. In section \ref{sec:LocalizationHopf}
 we perform the localization computation on the Hopf surfaces. In section \ref{sec:ThePartFct} we compare our result for the exact partition function with the supersymmetric index. We emphasize the presence of the extra pre-factor and define the supersymmetric Casimir energy. We also comment on the implications of our results for gravity duals. We conclude in section \ref{conc:sec} by outlining some perspectives for future work. We also included several appendices. Appendix \ref{app:conventions} contains our conventions. Appendix \ref{app:Weyl} provides a proof that the partition function is independent of the conformal factor of the metric. Appendix \ref{Berger:sec} describes  familiar examples of the background geometries considered in section  \ref{Hopf:sec}. Appendix \ref{nondirect:sec} elaborates on possible generalizations of our results by considering non-direct product metrics, associated to complex values of the complex structure moduli. Appendix \ref{details:sec} includes computations used in section \ref{Hopf:sec}. Appendix \ref{app:4dto3dred} contains details of the reduction of four-dimensional backgrounds to three dimensions. Appendix \ref{app:1LoopDetReg}
 contains the details of the regularization of one-loop determinants.

\section{Supersymmetric backgrounds and Lagrangians}
\label{sec:susy}

We begin our analysis by reviewing and elaborating results about the new minimal formulation of rigid supersymmetry on curved space. Our considerations in this section will be entirely local, while global properties will be discussed in section \ref{Hopf:sec}.

\subsection{Background geometry}
\label{sec:2scharges}

As shown in \cite{Festuccia:2011ws}, in the presence of an R-symmetry the supersymmetry transformations and the Lagrangian of a field theory defined on a 
curved manifold can be derived by coupling the theory to the new minimal formulation of off-shell supergravity \cite{Sohnius:1981tp,Sohnius:1982fw,Sohnius:1982xs}
and freezing the fields in the gravity multiplet to background values, in such a way that the gravitino variation vanishes.
The bosonic fields in the gravity multiplet are the metric and two auxiliary vector fields $A_{\mu}, V_{\mu}$; after the rigid limit, these play the role of background fields. In Euclidean signature, $A_{\mu}$ and $V_{\mu}$ are allowed to take complex values, 
whereas for simplicity the metric will be constrained to be real.

The real part of $A_{\mu}$ is associated to $u(1)_R$ R-symmetry transformations, and transforms (locally) as a gauge field, while the imaginary
part must be a well-defined one-form.
Being  the Hodge dual of a closed  three-form, $V = * \diff B$ is assumed to be a globally defined  one-form,  
constrained by $\nabla^{\mu} V_{\mu} =0$. In Euclidean signature, the condition that the gravitino variation vanishes 
corresponds to two independent first-order differential equations
\bea
\lp \nabla_{\mu} - i A_{\mu} \rp \zeta  + i V_{\mu} \zeta + i V^{\nu} \sigma_{\mu\nu} \zeta \!& = &\! 0  \ , \label{KeqnZeta}
 \\ [2mm]
\lp \nabla_{\mu} + i A_{\mu} \rp \ti\zeta  - i V_{\mu} \ti\zeta - i V^{\nu} \ti\sigma_{\mu\nu} \ti\zeta \!& = &\! 0 \ ,\label{KeqnTiZeta}
\eea
where $\zeta$ and $\ti\zeta$ are two-component complex spinors of opposite chirality, and with opposite charge under the
background gauge field $A$, associated with the R-symmetry. Solutions to these equations are either identically zero or nowhere vanishing. Throughout the paper, spinors with no tilde transform in the 
$(\bf{2,1})$ representation of the $Spin(4) = SU(2)_{+}\times SU(2)_{-}$ Lorentz group, while
spinors with a tilde transform in the $(\bf{1,2})$. 
See appendix~\ref{app:conventions} for further details on our notation and conventions.

It was shown in \cite{Klare:2012gn,Dumitrescu:2012ha} that a necessary and sufficient condition for a Riemannian four-manifold
to have a solution $\zeta$ to~\eqref{KeqnZeta}  is that it admits an integrable complex structure $J^\mu{}_\nu$. Lowering an index with the Hermitian metric, the corresponding fundamental two-form
can be constructed as a  spinor bilinear,
\be
 J_{\mu\nu} \  =\  \frac{2i}{|\zeta|^2} \, \zeta^{\dagger}\sigma_{\mu\nu}\zeta \ .
 \label{CStruct}
\ee
One can also introduce a complex two-form bilinear as   $P_{\mu\nu}  =  \zeta \sigma_{\mu\nu}\zeta$ , 
which is anti-holomorphic with respect to the complex structure $J^\mu{}_\nu\,$. Together these define a $U(2)$ structure on the four-manifold. The solution of (\ref{KeqnZeta}) can be expressed in terms of a nowhere vanishing complex 
function $s$ as $\zeta_{\alpha} = \sqrt{\frac{s}{2}}${\footnotesize{$\left(\!\!\begin{array}{c} 0\\ [-2pt] 1\end{array}\!\!\right)$}}, and the background fields are determined by
\begin{align}
V_\mu \ & =\ -\frac{1}{2}\nabla^\rho J_{\rho\mu} + U_\mu  \ , \label{avav1}\\
A_\mu \ & =\ A^c_\mu - \frac{1}{4}(\delta_\mu^\nu - i J_\mu{}^\nu)\nabla^\rho J_{\rho\nu} + \frac{3}{2} U_\mu \  , 
\label{avav2}
\end{align}
where $A^c_\mu$ is defined as 
\be
A^c_\mu \ = \ \frac{1}{4} J_\mu{}^\nu\partial_\nu\log \sqrt{g} - \frac{i}{2} \partial_\mu \log s \, ,
\ee
with $g$ the determinant of the metric in complex coordinates.
The solution contains an arbitrariness parametrised by the vector field $U^{\mu}$,  which is constrained to be holomorphic, namely 
$J^{\mu}{}_{\nu}U^{\nu} = i U^{\mu}$, and to obey $\nabla_{\mu} U^{\mu}=0$. 
 Note that the combination $\Acks_\mu \equiv A_\mu  - \frac 32 V_\mu$ is independent of the choice of $U_\mu$.\footnote{We denote this as $\Acks$ as it is the background field arising when the theory is coupled to conformal supergravity.} 
Of course a solution $\widetilde \zeta$ to~\eqref{KeqnTiZeta} is also equivalent to the existence of an integrable complex structure defined by
\be
\ti J_{\mu\nu} \ =\  \frac{2i}{|\ti\zeta|^2} \, \ti\zeta^{\,\dagger\,}\ti\sigma_{\mu\nu}\ti\zeta \,,
\ee
and leads to expressions for the background fields $A_\mu$ and $V_\mu$ analogous to the ones above, with a few sign changes;  
see \cite{Dumitrescu:2012ha} for the explicit formulae.

When there exist \emph{both} a non-zero solution $\zeta$ to~\eqref{KeqnZeta} and a non-zero solution $\widetilde \zeta$ to~\eqref{KeqnTiZeta}, namely in the presence of two supercharges of opposite R-charge, the four-dimensional manifold is endowed with a pair of  commuting complex structures $J^\mu{}_\nu$, $\widetilde  J^\mu{}_\nu$, inducing opposite orientations, and 
subject to certain compatibility conditions \cite{Dumitrescu:2012ha}. This means that the manifold admits a specific  
\emph{ambihermitian} structure\footnote{Note that the similar term ``bihermitian''  refers to the different case where
the two commuting complex structures induce the same orientation on the manifold.} \cite{Apostolov:2013oza}. 
Solutions with two supercharges of opposite R-charge may be more efficiently characterised by a complex vector field $K^\mu$, constructed as a spinor bilinear as
\be\label{KilVec}
K^{\mu} \ =\ \zeta \sigma^{\mu}\ti\zeta  \,.
\ee
In particular, one can  show that  $K^\mu$  is holomorphic with respect to both complex structures and satisfies the algebraic property $K_{\mu}K^{\mu}=0$ as well as the differential
condition \hbox{$\nabla_{(\mu} K_{\nu)}=0$}, therefore it comprises  two real Killing vectors. If $K^{\mu}$ commutes with its complex conjugate, 
$K^{\nu} \nabla_{\nu} \overline K{}^{\mu} - \overline K{}^{\nu} \nabla_{\nu} K^{\mu} =0$, 
then the vector field $U^\mu$ above is restricted to take the form $U^{\mu} = \kappa K^{\mu}$, where $\kappa$ is a complex function such that 
$K^{\mu} \p_{\mu}\kappa = 0$, but otherwise arbitrary \cite{Dumitrescu:2012ha}.\footnote{It is shown in \cite{Dumitrescu:2012ha} that if 
$[K,\overline K] \neq 0$,  then the manifold is locally isometric to $\mathbb R \times S^3$, with the standard round metric on $S^3$.} Moreover, introducing adapted complex coordinates $w,z$ (holomorphic with respect to $J^\mu{}_\nu$) such that the complex Killing vector is $K = \partial_w$, the metric takes the form
 \be\label{seibmetric}
 \diff s^2 \ = \ \Omega^2  [(\diff w + h \diff z)(\diff\bar w + \bar h \diff\bar z)+ c^2 \diff z\diff\bar z]\,,
 \ee 
where  $\Omega (z,\bar z)$ and $c(z,\bar z)$ are real, positive functions, while $h(z,\bar z)$ is a complex function. It is useful 
 to introduce the complex frame\footnote{Here $e^1$ and $e^2$ are exchanged with respect to those appearing in~\cite{Dumitrescu:2012ha}. This implies that the $\ti\zeta$ given in \eqref{zetaTizetaSols} below has swapped components with respect to the one in~\cite{Dumitrescu:2012ha}.}
\be\label{complexframe}
e^1 \ =\ \Omega\, c\, \diff z\,, \quad\qquad 
e^2 \ =\ \Omega (\diff w + h \diff z)\,.
\ee
We choose the orientation by fixing the volume form as $ {\rm vol}_4 = - \frac{1}{4} e^1 \wedge \bar e^{\bar 1} \wedge e^2 \wedge \bar e^{\bar 2}$.
Then, as a one-form, $K$ reads
\be
 K \ =\ \frac{1}{2}\Omega^2 (\diff \bar w + \bar h \diff \bar z) \ = \ \frac{1}{2}\Omega\, \bar e^{\bar 2}\,,
 \label{koneform}
\ee
and the real two-forms associated with the commuting complex structures are 
\bea
  J & = &   \frac{2i}{\Omega^2} \, K \wedge \overline{K} - \frac i 2 \Omega^2 c^2 \diff z \wedge \diff\bar z  \ = \  -\frac{i}{2}\left( e^1 \wedge \bar e^{\bar 1} + e^2 \wedge \bar e^{\bar 2} \right),\no\\ [2mm]
  \ti J & = & \; \frac{2i}{\Omega^2} \, K \wedge \overline{K} + \frac i 2 \Omega^2 c^2 \diff z \wedge \diff\bar z\ = \ \frac{i}{2}\left( e^1 \wedge \bar e^{\bar 1} - e^2 \wedge \bar e^{\bar 2} \right)\, .\label{JtildeJ2scharges}
 \eea
With our choice of orientation $J$ is self-dual while $\widetilde J$ is anti-self-dual.\footnote{Our convention for the Hodge star is $*\theta^{a_1 \ldots a_k} = \frac{1}{(4-k)!}\epsilon^{a_1 \ldots a_k}{}_{a_{k+1}\ldots a_4}\theta^{a_{k+1}\ldots a_4}$, where $\theta^a$ denotes a real frame. This is related to the complex frame as $e^1 = \theta^1 + i \theta^2$, $e^2 = \theta^3 + i \theta^4$; so the volume form introduced above is ${\rm vol}_4 = \theta^1\wedge \theta^2\wedge \theta^3\wedge \theta^4\,$.
\label{foot:conventionHodgeStar}}
 Following  \cite{Dumitrescu:2012ha}, we will require also that 
\bea
\overline{ K}{}^\mu \partial_\mu \kappa \ =\ \overline{K}{}^\mu \partial_\mu |s| \ =\  K^\mu \partial_\mu |s| \ =\ 0\, ,
\label{furtherconditions}
\eea  
 so that both $K$ and $\overline{K}$ preserve $A$ and $V$ in addition to the metric.  With these restrictions,  
 the functions  $\kappa$ and $|s|$ do not depend on $w, \bar w$, but can still have an arbitrary dependence on $z$ and $\bar z$. 
In the frame~\eqref{complexframe}, the spinors $\zeta$ and $\widetilde\zeta$ solving \eqref{KeqnZeta} and \eqref{KeqnTiZeta} read
\be\label{zetaTizetaSols}
\zeta_\alpha  \ =\ \sqrt{\frac{s}{2}} \left(\!\!\begin{array}{c} 0\\ 1 \end{array}\!\!\right) \,,\qquad \widetilde \zeta^{\dot \alpha} \ = \ \frac{\Omega}{\sqrt{2s}}\left(\!\!\begin{array}{c} 1\\ 0 \end{array}\!\!\right)\, .
\ee

Let us present more explicit formulae for $A$ and $V$. 
Noting that $\nabla^\rho J_{\rho\mu}\diff x^\mu = *\, \diff * J = * \,\diff J$ and using the expression for $J$ in~\eqref{JtildeJ2scharges}, simple manipulations show that (\ref{avav1})  
and (\ref{avav2}) can be written as\footnote{For any function $f$ we define  $\diff^c f \ =\ J_\mu{}^\nu \partial_\nu f \diff x^\mu \ =\ - i (\partial - \bar \partial)f$.}
\bea
V &=& \diff^c\log \Omega + \frac{2}{\Omega^2c^2}\,{\rm Im}\left( \partial_{\bar z} h\, K \right) + \kappa K\,,\label{reexpressV} \\ [2mm]
A &=& \frac{1}{2}\diff^c\log\left(\Omega^3c\right) - \frac{i}{2}\diff \log\left(\Omega^{-1}s\right) + \left( \frac{3}{2}\kappa - \frac{i}{\Omega^2c^2} \partial_{\bar z}h \right)\! K\,,\label{reexpressA}
\eea
where we  used $\sqrt g = \Omega^4 c^2$.

For later applications it is important to observe that we can use the  freedom in choosing $\kappa$ and $s$ to arrange for  $A$ to be  \emph{real}. 
 Indeed, requiring ${\rm Im}\,A =0$ in~\eqref{reexpressA} and separating the different components, we obtain  
the conditions
\be \label{kapparesult}
|s|\ =\ \Omega\,,\qquad \quad\kappa \ = \ \frac{2i}{3\Omega^2c^2} \partial_{\bar z} h \,,
\ee
where we fixed an irrelevant multiplicative constant in $|s|$. With these choices of $\kappa$ and $|s|$,  the gauge field $A$ takes the simple form
\be
A \ = \  \frac{1}{2} \diff^c  \log (\Omega^3 c) \, + \frac{1}{2} \diff \betaph \,,\label{finalA}
\ee
where  $\betaph$ denotes the phase of $s$, \emph{i.e.} $s \,=\, |s|\, \mathrm{e}^{i\betaph}$. Note that $\betaph$ has not been fixed so far, while  it will be determined 
by our global analysis in section~\ref{Hopf:sec}.  The one-form $V$ in general remains complex 
\be\label{finalV}
V \ =\ \diff^c\log \Omega - \frac{i}{3\Omega^2c^2} \partial_{\bar z} h\, K + \frac{i}{\Omega^2c^2} \partial_{z} \bar h\, \overline{K} \,.
\ee
Recalling that $\Omega$ and $c$ are real and depend only on the $z$, $\bar z$ coordinates, we can also write  more explicitly
\bea
A & = &  {\rm Im}\!\left[\partial_z \log (\Omega^3 c)\, \diff z \right]  + \frac{1}{2} \diff \betaph \,,\no\\ [1mm]
V & = &  2\,{\rm Im}\!\left[\partial_z \log \Omega \, \diff z \right] - \frac{i}{3\Omega^2c^2} \partial_{\bar z} h\, K + \frac{i}{\Omega^2c^2} \partial_{z} \bar h\, \overline{K} \,.\label{alternativeExprAV}
\eea
Finally, the spinors~\eqref{zetaTizetaSols} take the form 
\be\label{zetaTizetaSolsRealA}
\zeta_\alpha \ =\ \sqrt{\frac{\Omega}{2}} \, \mathrm{e}^{i\tfrac{\betaph}{2}} \left(\!\!\begin{array}{c} 0\\ 1 \end{array}\!\!\right) \,,\qquad  \widetilde \zeta^{\dot\alpha} \ =\ \sqrt{\frac{\Omega}{2}} \, \mathrm{e}^{-i\tfrac{\betaph}{2}} \left(\!\!\begin{array}{c} 1\\ 0 \end{array}\!\!\right)  .
\ee

\subsection{Supersymmetry transformations and Lagrangians}
\label{subsec:susyTransfAndLagr}

In this section we present the supersymmetry variations and relevant Lagrangians of the theories that we consider in this paper. 
In Euclidean signature, defining $\N=1$ supersymmetry requires to double the number of degrees of freedom
in each multiplet. This can be realized formally by thinking about a given field and its Hermitian conjugate as transforming 
independently under supersymmetry.  To define the path integral over the fields of a multiplet,
one then has to make a choice of reality conditions, reducing the number of degrees of freedom in a multiplet to the usual one. 
In the following, we will first consider a vector multiplet and then a chiral multiplet.

\subsubsection{Vector multiplet}

The $\N=1$ vector multiplet contains a gauge field $\scA_{\mu}$, 
a pair of two-component complex spinors $\lambda$, $\ti \lambda$ of opposite chirality and an auxiliary field $D$, 
all transforming in the adjoint representation of the gauge group $G$.  As already noted, \emph{a priori} in Euclidean signature
the fermionic fields $\lambda$, $\ti \lambda$ are independent, and the bosonic fields $\scA_{\mu}$, $D$ are not Hermitian.
We define a covariant derivative as
\be\label{DefCovDer}
D_{\mu} \ =\ \nabla_{\mu} - i \scA_{\mu}\!\cdot - i q_R A_{\mu}~,
\ee 
where $\cdot$ denotes the action in the relevant representation, and the R-charges $q_R$ of the fields $(\scA_{\mu}, \lambda, \ti\lambda, D)$ are given respectively by $(0,1,-1,0)$.
The supersymmetry transformations of the fields in the multiplet are
\bea
\dd \scA_{\mu}  &=& i \zeta \sigma_{\mu} \ti\lambda + i \ti\zeta \,\ti\sigma_{\mu} \lambda\;, \no\\ [2mm]
\dd\lambda  &=& \scF_{\mu\nu}\,\sigma^{\mu\nu} \zeta + i D \zeta\;, \no\\ [2mm]
\dd\ti\lambda  &=& \scF_{\mu\nu}\,\ti\sigma^{\mu\nu} \ti\zeta - i D \ti\zeta \,, \no \\ [2mm]
\dd D &=& -\zeta \sigma^{\mu} \big( D_{\mu}\ti\lambda -  \tfrac{3i}{2} V_{\mu} \ti\lambda\, \big) 
 + \ti\zeta \, \ti\sigma^{\mu} \lp D_{\mu}\lambda +  \tfrac{3i}{2} V_{\mu} \lambda \rp, \label{VecTransfo}
\eea
where $\scF_{\mu\nu} \equiv  \p_{\mu} \scA_{\nu}-\p_{\nu} \scA_{\mu}   - i [\scA_{\mu},\scA_{\nu}]$.  Note that the two independent 
spinorial parameters $\zeta$, $\ti\zeta$ need to be solutions 
to the equations \eqref{KeqnZeta}, \eqref{KeqnTiZeta}, and are commuting variables. 
It is understood that when one of the two equations only admits the trivial solution, the corresponding spinor is set to zero in
the supersymmetry transformations. The fermionic fields $\lambda$, $\ti \lambda$ are anti-commuting, and therefore correspondingly the supersymmetry variation
$\delta$ is defined as a Grassmann-odd operator.  Note also that in the above transformations only the conformal invariant and $U^\mu$-independent 
combination of background fields $\Acks_{\mu} \equiv  A_\mu - \frac{3}{2} V_{\mu}$ appears, in the covariant derivative
$D^\mathrm{cs}_\mu\equiv \nabla_{\mu} - i \scA_{\mu}\!\cdot  - i q_R \Acks_{\mu}$. 

The supersymmetry algebra is given by
\bea
&& \{\dd_{\zeta},\dd_{\zeta}\} = \{\dd_{\ti\zeta},\dd_{\ti\zeta}\} = 0  \ , \no \\ [2mm]
&& [\dd_{\zeta}, \dd_K ]\, =\, [\dd_{\ti\zeta}, \dd_K ] = 0 \ , \no \\ [2mm]
&& \{\dd_{\zeta},\dd_{\ti\zeta}\} =\, 2i \, \dd_K \ ,
\eea
where $\dd_{\zeta}$ (respectively, $\dd_{\ti\zeta}$) means that $\ti\zeta$ (respectively, $\zeta$) is set to zero in the supersymmetry transformations (\ref{VecTransfo}), and on a field of R-charge $q_R$ we have $\dd_K = \scL_K  - i K^\mu \scA_\mu \cdot - i q_R \, K^{\mu}A_{\mu}$, where $\mathcal L_K$ is the Lie derivative along $K$.
If there is only one Killing spinor $\zeta$, then one just has $\dd_{\zeta}^2 = 0\,$.

A tedious calculation shows that the Lagrangian 
\bea
\scL_{\rm vector}  & = &  \tr \bigg[\;  \frac 1 4 \scF^{\mu\nu}\scF_{\mu\nu} - \half D^2 + \frac{i}{2} \lambda\, \sigma^{\mu} D_{\mu}^\mathrm{cs} \ti\lambda +  \frac{i}{2} \ti\lambda\, \ti\sigma^{\mu} D^\mathrm{cs}_{\mu}\lambda  \; \bigg]
\label{Lagrangians:vec}
\eea
is invariant under the supersymmetry transformations (\ref{VecTransfo}). Here $\tr$ is the trace in the adjoint representation of the gauge group. 
We will show momentarily that if both spinors $\zeta$, $\ti\zeta$ exist, then this Lagrangian is the sum of two  supersymmetry variations; this will
be important for applying the localization argument.

Given that in Euclidean signature the degrees of freedom are doubled, it is conceptually clearer to impose reality conditions on the fields only after computing the supersymmetry variations. 
Therefore, to define various supersymmetry-exact terms, we introduce an involution $^\ddagger$ acting as
\begin{align}
& (\scA_{\mu},  D)^\ddagger \ =  \ (\scA_\mu,-D),          \qquad\qquad   \zeta^{\ddagger} \ =\ \zeta^{\dagger} \ ,
\end{align}
and as complex conjugation on numbers.\footnote{We will not need to define the action of $^\ddagger$ on $\lambda$ and $\ti \lambda$.} 
Then we define
\begin{align}
\scL_{\rm vector}^{(+)} & = \ \dd_{\zeta} V^{(+)} \ =  \ -\dd_{\zeta} \lp \frac{1}{4 |\zeta|^2}   \tr \  (\dd_{\zeta}\lambda)^{\ddagger} \lambda   \rp    \no\\
& = \  \frac{1}{4 |\zeta|^2}  \tr \  (\dd_{\zeta} \lambda)^{\ddagger} \dd_{\zeta}\lambda  -  \frac{1}{4 |\zeta|^2} \tr \  \dd_{\zeta} \lp (\dd_{\zeta}\lambda)^{\ddagger} \rp \lambda \no \\[2mm]
& \equiv  \ \dd V_{\rm bos}^{(+)}  + \dd V_{\rm fer}^{(+)} \,. \label{constructdeltaVgauge} 
\end{align}
The bosonic term is straightforward to evaluate and reads 
\begin{align}
\dd V_{\rm bos}^{(+)} &\ =\  \frac 1 4 \,\tr \lp  \scF^{(+)}_{\mu\nu}\scF^{(+) \, \mu\nu} -  D^2 \rp  ,
\label{vpbos}
\end{align}
where $\scF^{(\pm)}_{\mu\nu} = \half( \scF \pm\ast \scF)_{\mu\nu}\,$.
The fermionic term reads
\begin{align}
\dd V_{\rm fer}^{(+)}  &\ =\ \frac{1}{4 |\zeta|^2} \,\tr   \left[ \ - (\zeta^{\dagger} \sigma^{\mu\nu} \lambda) \dd_{\zeta}\scF_{\mu\nu} + i  (\zeta^{\dagger}\lambda)\dd_{\zeta}D  \ \right] \, 
\label{eq1}
\end{align}
and with some manipulations can be rewritten as
\begin{align}
\dd V_{\rm fer}^{(+)}\ &=\ \tr  \Big[ \ \frac{i}{2} \lambda\, \sigma^{\mu} \Big(D_{\mu}\ti\lambda -\frac{3i}{2} V_{\mu} \ti\lambda \Big) \Big] \ .
\end{align}
To obtain this we used the following expression for the supersymmetry variation of the gauge field strength ${\scF}_{\mu\nu}\,$ 
\bea
\dd \scF_{\mu\nu} &=& \ 2i \,\zeta\sigma_{[\nu} D_{\mu]}\ti\lambda +  V_{[\mu} \zeta\sigma_{\nu]} \ti\lambda 
   +  \epsilon_{\mu\nu\kappa\lambda} V^{\kappa}(\zeta \sigma^{\lambda} \ti\lambda) \no \\ [2mm]
   &&\!\!\!\!+\, 2i\, \ti\zeta \ti\sigma_{[\nu} D_{\mu]}\lambda  -  V_{[\mu} \ti\zeta \ti\sigma_{\nu]} \lambda 
   +  \epsilon_{\mu\nu\kappa\lambda} V^{\kappa}(\ti\zeta \ti\sigma^{\lambda} \lambda)  ~.
\eea
We have thus shown that 
\be
\scL_{\rm vector}^{(+)} \ = \ \tr \lp  \frac 1 4 \scF^{(+)}_{\mu\nu}\scF^{(+) \, \mu\nu} - \frac 1 4 D^2  + \frac{i}{2} \lambda\, \sigma^{\mu} D_{\mu}^\mathrm{cs} \ti\lambda \rp  \,.
\ee

If there exists a second Killing spinor $\ti\zeta$, then the previous computations can be repeated with trivial modifications. 
Namely, we can define
\begin{align}
\scL_{\rm vector}^{(-)} & = \ \dd_{\ti\zeta\,} V^{(-)} \ =  \ -\dd_{\ti\zeta} \bigg( \frac{1}{4 |\ti\zeta|^2}   \tr \  (\dd_{\ti\zeta}\ti\lambda)^{\ddagger} \ti\lambda   \bigg) \no\\
& = \  \frac{1}{4 |\ti\zeta|^2}  \tr \  (\dd_{\ti\zeta} \ti\lambda)^{\ddagger} \dd_{\ti\zeta}\ti \lambda  -  \frac{1}{4 |\ti\zeta|^2} \tr \  \dd_{\ti\zeta} \lp (\dd_{\ti\zeta}\ti\lambda)^{\ddagger} \rp \ti\lambda \no \\[2mm]
& \equiv  \ \dd V_{\rm bos}^{(-)}  + \dd V_{\rm fer}^{(-)}~, \label{constructdeltaVgaugetilde}
\end{align}
with 
\be
\scL_{\rm vector}^{(-)} \ = \ \tr \lp  \frac 1 4 \scF^{(-)}_{\mu\nu}\scF^{(-) \, \mu\nu} - \frac 1 4 D^2  +  \frac{i}{2} \ti\lambda\, \ti\sigma^{\mu} D^\mathrm{cs}_{\mu}\lambda \rp  \,.
\ee

The sum of the two terms is
\be
  \scL_{\rm vector}^{(+)} + \scL_{\rm vector}^{(-)} \ =\  \tr \Big[ \ \frac 1 4  \scF_{\mu\nu}\scF^{\mu\nu} - \half  D^2 +  
  \frac{i}{2} \lambda\, \sigma^{\mu} D_{\mu}^\mathrm{cs} \ti\lambda +  \frac{i}{2} \ti\lambda\, \ti\sigma^{\mu} D^\mathrm{cs}_{\mu}\lambda   \Big] 
  \ =\  \scL_{\rm vector}\,.\label{Vvector}
\ee
Therefore, we have shown that the vector multiplet Lagrangian $\scL_{\rm vector}$ in (\ref{Lagrangians:vec}) 
is the sum of a $\dd_{\zeta\,}$-exact term and a $\dd_{\ti\zeta}\,$-exact term.
Note  that to derive this result we have \emph{not} imposed any reality condition, and correspondingly at this stage the bosonic part of the Lagrangian is not positive semi-definite.

In order to apply the localization arguments, it will be important that  $\scL_{\rm vector}^{(+)}$ and  
$\scL_{\rm vector}^{(-)}$  are separately invariant under {\it both} supersymmetries associated with $\zeta$ and $\ti\zeta$, so that  
\bea
\dd_\zeta  \scL_{\rm vector}^{(-)} & = & \dd_\zeta \dd_{\ti\zeta}  V^{(-)} \ = \    {\rm tot\,der} \, ,\no\\ [1mm]
  \dd_{\ti\zeta}\scL_{\rm vector}^{(+)}  & =  &  \dd_{\ti\zeta} \dd_{\zeta}  V^{(+)} \ = \   {\rm tot\,der}\,,
\eea
where ``tot$\,$der'' denotes a total derivative.
Recalling that $\delta_\zeta^2 = \delta_{\widetilde \zeta}^2 = 0$, 
these are equivalent to the fact that the vector multiplet Lagrangian is invariant under both supersymmetry variations, namely 
$\dd_\zeta  \scL_{\rm vector}    =  \dd_{\ti\zeta\,} \scL_{\rm vector} = {\rm tot\,der}\,$.

\subsubsection{Chiral multiplet}
\label{chiralmultiplet:sec}

The $\N=1$ chiral multiplet contains  two complex scalars $\phi, \ti\phi$, a pair of two-component complex spinors $\psi$, $\ti \psi$ of opposite chirality, 
and two complex auxiliary fields $F, \ti F$. As for the fields of the vector multiplet, in Euclidean signature
the fermionic fields $\psi$, $\ti \psi$  and the complex scalars $\phi, \ti\phi$, and  $F, \ti F$ are all independent. The fields $(\phi,\psi,F)$ transform in a representation $\scR$, 
while $(\ti\phi,\ti\psi,\ti F)$ transform in the conjugate representation $\scR^{\ast}$. The R-charges $q_R$ entering in \eqref{DefCovDer} for the fields $(\phi,\psi,F,\ti\phi,\ti\psi,\ti F)$ are given by $(r,r-1,r-2,-r,-r+1,-r+2)$ respectively, with $r$ arbitrary.  The supersymmetry transformations of the fields in the multiplet  can be read off from \cite{Sohnius:1982fw,Festuccia:2011ws,Closset:2012ru} and are 
\bea
 \dd \phi &=&  \sqrt{2}\, \zeta \psi \;,\no\\ [2mm]
\dd \psi  &=&  \sqrt{2}\, F \zeta + i\sqrt{2} (\sigma^{\mu}\ti\zeta)  D_{\mu} \phi \;, \no\\ [2mm]
\dd F  &=& i\sqrt{2}\, \ti\zeta \ti\sigma^{\mu} \Big( D_{\mu}\psi - \frac{i}{2} V_{\mu} \psi \Big) - 2i (\ti\zeta \ti\lambda) \phi \;, \no\\ [2mm]
\dd \ti\phi  &=&  \sqrt{2}\, \ti\zeta \ti\psi \;, \no\\ [2mm]
\dd \ti\psi  &=&  \sqrt{2}\, \ti F \ti\zeta + i\sqrt{2} (\ti\sigma^{\mu}\zeta) D_{\mu} \ti\phi \;, \no \\ [2mm]
\dd \ti F  &=& i\sqrt{2}\, \zeta \sigma^{\mu} \Big( D_{\mu}\ti\psi + \frac{i}{2} V_{\mu} \ti\psi \Big) + 2i \ti\phi \,(\zeta \lambda) \;. \label{ChiTransfo}
\eea
These preserve the Lagrangian 
\begin{align}
\scL_{\rm chiral}\ &=\ D_{\mu}\ti \phi D^{\mu} \phi   + V^{\mu} \big( i  D_{\mu}\ti \phi\,\phi - i \ti \phi D_{\mu}\phi \big)+ \frac{r}{4} \lp R + 6 V_{\mu}V^{\mu} \rp \ti\phi \phi + \ti \phi D \phi             - \ti F F
   \no\\
 & \quad\;  +      i \ti \psi \,\ti\sigma^{\mu}D_{\mu}\psi  + \half V^\mu \ti\psi\, \ti \sigma_{\mu} \psi  + i \sqrt 2 \big( \ti\phi \lambda \psi - \ti\psi \, \ti\lambda \phi \big) \,.
\label{Lagrangians:chi}
\end{align}
This depends on both background fields $A$ and $V$, except when the R-charge takes the value $r = 2/3$, in which case these only appear in the combination $\Acks = A - \frac{3}{2}V$ and the Lagrangian is conformal invariant. Below we will show that the existence of a single supersymmetry parameter $\zeta$
is enough to express $\scL_{\rm chiral}$ as a total supersymmetry variation, up to an irrelevant boundary term. 

In general, one can consider several chiral multiplets with different R-charges $r_I$, with Lagrangian given by the sum of the \eqref{Lagrangians:chi} for each multiplet, and 
also add to this a superpotential term ${\cal L}_W$, as in flat space. 
The explicit expression in component notation is given in \cite{Festuccia:2011ws}.
The superpotential $W$ can be an arbitrary holomorphic function\footnote{In this paper we assume that $W$ is a polynomial in the fields $\phi_I$.} of the fields $\phi_I$, and in order not to break the 
R-symmetry of the theory it must be homogeneous of degree two in the R-charges. 
This follows from the fact that  the fermions $\psi_I$ have R-charges $r_I-1$ and in components the superpotential 
contains a fermionic piece
\bea
\frac{\de^2 W}{\de \phi_I\de \phi_J}\psi_I \psi_J \ \in\ {\cal L}_W~,
\eea
whose R-charge is $r[W]-r_I-r_J + (r_I-1)+ (r_J-1)$.
On integrating out the auxiliary fields $F_I$ one obtains\footnote{{\em A priori} $\widetilde W$ is an arbitrary function of $\widetilde \phi_I$, but reality conditions will relate this to $W$.}
\bea
\ti F_I & = & \frac{\partial W}{\partial\phi_I}~, \qquad F_I \ = \ \frac{\partial \widetilde W}{\partial\ti\phi_I}~.
\label{intoutF}
\eea
 
In order to write the supersymmetry-exact
 terms we extend the action of the involution $^\ddagger$ used for the vector multiplet to the bosonic fields of the chiral multiplet as 
\begin{align}
& (\phi,F,\ti\phi,\ti F)^{\ddagger}\ =\  (\ti\phi,-\ti F,\phi,-F) \,.
\end{align}
While we will not need to define how $^\ddagger$ acts on $\psi$, $\ti \psi$ and on $V_\mu$, we will need its action on $A_\mu$. There are two natural definitions we can take, which in general are not equivalent. If we define $A_{\mu}^{\ddagger}  = A_{\mu}$, then the computation below shows that the Lagrangian $\mathcal L_{\rm chiral}$ is $\delta_\zeta$-exact (up to a boundary term) without any restriction on $A_\mu$. However, notice that this Lagrangian is \emph{not} invariant under changes of $U_\mu$, and its bosonic part is not positive semi-definite even after imposing reality conditions on the dynamical fields. 
If instead we define $A_{\mu}^{\ddagger}  = A_{\mu}^\dagger$,  then the localizing term that we will choose in the next section does not depend on $U_\mu$ and its bosonic part is positive semi-definite after choosing suitable reality conditions. However, for complex $A_\mu$, this does not reconstruct the Lagrangian (\ref{Lagrangians:chi}). In the following we will assume that $A_\mu$ is real, so that the two definitions are equivalent; as showed at the end of section~\ref{sec:2scharges}, this is certainly possible in the presence of two supercharges of opposite R-charge. Later we will make some comments about relaxing this choice.

We consider
\begin{align}\label{deltaVchiral}
\dd V_{\rm chiral} &= \dd_{\zeta} V_1 + \dd_{\zeta} V_2  +  \dd_{\zeta} V_3  + \dd_\zeta V_U \no\\
 & = \dd_{\zeta} \lp  \frac{1}{2|\zeta|^2} \Big[ (\dd_{\zeta}\psi)^{\ddagger} \psi - \ti\psi (\dd_{\zeta} \ti\psi)^{\ddagger}  \Big]  \rp  + \dd_{\zeta} V_3 + \dd_\zeta V_U \no\\
 & = \frac{1}{2|\zeta|^2}  \Big[ (\dd_{\zeta} \psi)^{\ddagger} \dd_{\zeta}\psi   +   \dd_{\zeta} \big( (\dd_{\zeta}\psi)^{\ddagger} \big) \psi   + (\dd_{\zeta} \ti\psi)^{\ddagger} \dd_{\zeta} \ti\psi  +  
  \ti\psi \,\dd_{\zeta} \big((\dd_{\zeta} \ti\psi)^{\ddagger} \big)   +    2i \delta_\zeta \big( \ti\phi \,\zeta^{\dagger}\lambda \,\phi \big)       \no\\[2mm] 
&\qquad\qquad   - \sqrt 2\,\dd_{\zeta} \big( \, U^{\mu}\zeta^{\dagger}\sigma_{\mu} \ti\psi \,\phi  \big) \Big]\no \\[2mm]
&\equiv  \dd V_{\rm bos \, 1}  + \dd V_{\rm fer \, 1} + \dd V_{\rm bos \, 2}  + \dd V_{\rm fer \, 2} + \dd V_{\rm bos \, 3} +  \dd V_{\rm fer \, 3}  + \dd V_{{\rm bos} \, U}  + \dd V_{{\rm fer} \, U}  \, .
\end{align}
For the bosonic part, the supersymmetry transformations (\ref{ChiTransfo}) lead to 
\bea
\dd V_{\rm bos \, 1} \!\!\!&=&\!\!\! - \ti F F~, \no\\ [2mm]
\dd V_{\rm bos \, 2}\!\!\! &=&\!\!\!  \left( g^{\mu\nu} - i J^{\mu\nu}  \right)D_{\mu}\ti\phi  D_{\nu}\phi   \no\\ [2mm]
\!\!\!&=&\!\!\! D_{\mu}\ti\phi D^{\mu}\phi  - 2 i ( V^{\mu} -  U^{\mu} ) \ti\phi D_{\mu}\phi  + \tfrac{r}{4}\! \lp R + 6 V_{\mu}V^{\mu} \rp \ti\phi \phi + \tfrac{1}{2}J^{\mu\nu} \ti\phi \scF_{\mu\nu}\phi - i \nabla_\mu\big( J^{\mu\nu}  \ti \phi D_\nu \phi \big), \no\\ [2mm]
\dd V_{\rm bos \, 3} \!\!\! &=&\!\!\!  - \half J^{\mu\nu} \ti\phi \scF_{\mu\nu}\phi + \ti\phi D\phi\, ,\no \\ [2mm]
 \dd V_{{\rm bos} \, U}  \!\!\! &=&\!\!\! 2i\, U^\mu D_\mu \ti \phi\,\phi\,,
\eea
where to go from the first to the second line in the second term we have used the identity (\ref{Prop1}), and in the last line we used the holomorphicity of $U^\mu$, namely $J^{\mu}{}_{\nu} U^{\nu} = i U^{\mu}$.
As for the fermionic terms, after some computations involving the Fierz identities in (\ref{fierce}) we find
\bea
\dd V_{\rm fer \, 1} &=& -\frac{i}{2} D_{\mu}\ti\psi \,\ti\sigma^{\mu}\psi  - \frac{1}{2} J^{\mu\nu}D_{\mu}\ti\psi\, \ti\sigma_{\nu}\psi  + \frac{1}{2|\zeta|^2} V^{\mu} (\zeta\sigma_{\mu}\ti\psi)(\zeta^{\dagger}\psi)
- i \frac{\sqrt 2}{|\zeta|^{2}} \ti\phi (\zeta\lambda)(\zeta^{\dagger}\psi) \, ,\no \\ [2mm]
\dd V_{\rm fer \, 2} &=& \frac{i}{2} \ti\psi\, \ti\sigma^{\mu}D_{\mu}\psi  - \frac{1}{2} J^{\mu\nu}\ti\psi \,\ti\sigma_{\nu}D_{\mu}\psi  + V^{\mu}\ti\psi\,\ti\sigma_{\mu}\psi - \frac{1}{2|\zeta|^{2}} V^{\mu} (\zeta^{\dagger} \sigma_{\mu}\ti\psi)(\zeta\psi) - i \sqrt 2 \, \ti\psi\,\ti\lambda\,\phi\,,\no \\
\dd  V_{\rm fer \, 3} &=&    i\, \frac{\sqrt 2}{|\zeta|^{2}}  \ti\phi (\zeta^{\dagger}\lambda)(\zeta\psi)\no \\ [2mm]
 \dd V_{{\rm fer} \, U}  \!\!\! &=&\!\!\! -  U^{\mu} \ti\psi \,\ti \sigma_{\mu} \psi \,,
\eea
where in the last equality we used holomorphicity of $U^\mu$, in the form $U^\mu \,\ti\sigma_\mu \zeta = 0\,$.
The total fermionic part can be written as
\bea
\dd V_{\rm fer\,1} \!+ \dd V_{\rm fer\, 2} \!+ \dd V_{\rm fer\,3} + \dd V_{{\rm fer} \, U} \!&=&\!   i \ti\psi \,\ti\sigma^{\mu}D_{\mu}\psi + \frac{1}{2} V^{\mu}\ti\psi\,\ti\sigma_{\mu}\psi
 + i \sqrt 2 \, (\ti\phi \lambda\psi - \ti\psi \,\ti\lambda\phi) \no \\ [2mm]
\!&&\!    -  \frac i2 D_\mu \left( (\delta^\mu{}_\nu -i  J^{\mu}{}_\nu)\ti\psi \,\ti\sigma^{\nu}\psi \right)\, .
\eea
Adding everything up, we obtain
\be
\dd V_{\rm chiral} \ = \ \scL_{\rm chiral} + \nabla_\mu Y^\mu\, ,
\ee
where $\mathcal L_{\rm chiral}$ is the Lagrangian \eqref{Lagrangians:chi} and the total derivative term is
\be
Y^\mu \ =\   - i  J^{\mu\nu} \ti \phi D_\nu \phi - i (V^\mu - 2 U^\mu ) \ti\phi\phi   -  \frac i2  (\delta^\mu{}_\nu -i  J^{\mu}{}_\nu)\ti\psi \,\ti\sigma^{\nu}\psi \,.
\ee
In a similar way, one can see that $\mathcal L_{\rm chiral}$ is also exact under the variation generated by $\ti \zeta$.


\subsection{Supersymmetric locus equations}
\label{subsec:locuseq}

Let us now discuss how to use the results above to compute the path integral of supersymmetric field theories, using the localization method. 
The standard localization arguments require to deform the path integral defined by a supersymmetric action by adding a term
that is a supersymmetry variation, and whose bosonic part is positive semi-definite.  In this way the complete path integral is given by the 
one-loop determinant around the locus where this bosonic part vanishes.  We will address the vector multiplet and chiral multiplet separately.

\subsubsection{Vector multiplet}
\label{sssec:VectorLocus}

If the manifold admits one Killing spinor $\zeta$, then we can deform the vector multiplet Lagrangian (\ref{Lagrangians:vec}) by adding to it the $\delta_\zeta$-exact term 
(\ref{constructdeltaVgauge}) with an arbitrary parameter $t$, namely 
\bea
S & =   & \int \diff^4 x \sqrt{g}  \left(     \scL_{\rm vector}   + t  \,  \delta_\zeta  V^{(+)} \right)\,.
\eea
We see that imposing the reality conditions $\scA_{\mu}^{\dagger} = \scA_{\mu}$, $D^{\dagger}= -D$ implies that the bosonic part~\eqref{vpbos} of the deformation term is positive semi-definite.\footnote{We note that actually the  weaker reality condition $\scF_{\mu\nu}^{(+)\dagger}= \scF_{\mu\nu}^{(+)}$ 
is sufficient to guarantee positivity of the deformation term. The condition $\scA_{\mu}^{\dagger} = \scA_{\mu}$ implies that also the 
original Lagrangian \eqref{Lagrangians:vec} has positive bosonic part, but this is not necessary for the localization argument.}
The localization locus is given by $\dd V_{\rm bos}^{(+)} = 0$, yielding the conditions
\begin{align}
\scF^{(+)}_{\mu\nu} \,=\, 0 \, , \qquad D \,=\, 0 \ .
\label{locus:1susy}
\end{align}
Of course this is also equivalent to $\delta \lambda=0$,  whose independent components 
give $J_{\mu\nu}{\cal F}^{\mu\nu}= P_{\mu\nu} {\cal F}^{\mu\nu} = 0 = D\,$.
The conclusion is that when there exists only one supercharge associated with $\zeta$, the localization locus is given by anti-instanton configurations.  In the case of a supercharge associated with $\ti\zeta$, the same argument works by considering the term $\dd_{\ti\zeta\,} V^{(-)}$ in \eqref{constructdeltaVgaugetilde}, with the conclusion being that the localization locus is given by instanton configurations.

If the manifold admits both $\zeta$  and $\ti\zeta$, then we can deform the  vector multiplet Lagrangian (\ref{Lagrangians:vec}) by adding 
both the $\delta_{\zeta\,}$-exact and $\delta_{\ti\zeta}\,$-exact terms,  namely 
\bea
S & =  & \int \diff^4 x \sqrt{g}  \left(     \scL_{\rm vector}   + t_+    \delta_\zeta  V^{(+)}    + t_-   \delta_{\ti\zeta}\,  V^{(-)}\right)\,.
\eea
To see that the path integral is independent of the parameter $t_+$ one notes that  
$\delta_\zeta \scL_{\rm vector} =   \delta_\zeta  \delta_{\ti\zeta\,}  V^{(-)}= {\rm tot\,der}$. Similarly, the path integral is also independent of the parameter $t_-$.
In the end one can take $t_+=t_-=t$ and omit the first term, without affecting  the conclusions. 
 The localization locus then is given by $\dd V_{\rm bos}^{(+)} = \dd V_{\rm bos}^{(-)}  = 0$, which is equivalent to the conditions 
\be
\scF_{\mu\nu}\, =\, 0 \, , \qquad D\, =\, 0 \ ,
\label{locus:vec}
\ee
so that both the self-dual and the anti-self-dual parts of the gauge field strength vanish.
We will discuss the solutions to these equations in section \ref{sec:susylocus}, after specializing the topology of the four-dimensional manifold. 

Notice that the conclusions above are manifestly independent of the choice of the holomorphic vector field $U^\mu$, as well as of the reality properties of the background fields $A_\mu, \;V_\mu\,$.

\subsubsection{Chiral multiplet}
\label{sssec:ChiralLocus}

If the manifold admits one Killing spinor $\zeta$, then we can deform the  chiral multiplet Lagrangian (\ref{Lagrangians:chi}), possibly supplemented by a superpotential,  
by adding to it the $\delta_\zeta$-exact term $\delta_\zeta (V_1 + V_2)$ defined in section \ref{chiralmultiplet:sec}. 
Namely, we consider
\bea
S & = &  \int \diff^4 x \sqrt{g}  \left[    \scL_{\rm chiral} + {\cal L}_W  + t \,   \delta_\zeta  (V_1+V_2)\right]\,,
\label{deformchi}
\eea
where $t$ is an arbitrary parameter. 
We must then choose reality conditions such that $\dd V_{\rm bos \, 1}$ and $\dd V_{\rm bos \, 2}$  are positive semi-definite.\footnote{The reason why we are not using simply $t\,\scL_{\rm chiral}$, which is also $\delta_\zeta$-exact, is that its bosonic part contains the terms $\delta V_{\rm bos \, 3}$ and $\delta V_U$, which 
are not positive after imposing the reality conditions.} The former requirement 
is satisfied imposing $\ti F = - F^{\dagger}$. In order to ensure that 
$2 |\zeta|^2 \dd V_{\rm bos \, 2} = (\dd_{\zeta} \ti\psi)^{\ddagger} \dd_{\zeta} \ti\psi$ is positive we require $\ti\phi \,=\, \phi^{\dagger}$ (hence the involution $^\ddagger$ acts as the Hermitian conjugation $^\dagger$).
 Note that $\dd V_{\rm bos \, 2}$ does not depend on the background field $V_\mu$, thefore there are no reality constraints to impose on the latter. On the other hand, 
it does depend on the background field $A_\mu$, hence its choice may a priori affect positivity. When $A$ is real, the localization locus is defined by the conditions $\dd V_{\rm bos \,1}= \dd V_{\rm  bos \, 2}=0$, so that in particular 
$\dd_{\zeta} \psi = \dd_{\zeta} \ti\psi = 0$. These are equivalent to
\begin{align}
F \,=\, 0 \, , \qquad 
J^{\mu}{}_{\nu} D^{\nu} \ti\phi \,=\, i D^{\mu} \ti\phi  ~.
\label{locus:chi}
\end{align}
The second equation means that $D^{\mu}\ti\phi$ is a holomorphic vector, or equivalently that $\ti\phi$ is a 
holomorphic section on a suitable line bundle. These configurations are still very complicated and in this paper we will not analyse them further. Before moving to the case
of two supercharges, let us briefly comment on the role of $U^\mu$. 
Since this is a holomorphic vector,  it drops out from the supersymmetry transformations \eqref{ChiTransfo}, and therefore, if we define $A_{\mu}^{\ddagger}  = A_{\mu}^\dagger$, 
it also drops out from the localizing term and hence from the locus equations \eqref{locus:chi}.
In this case the positivity property of $\dd V_{\rm bos \, 2\,}$  is not affected by the choice of $U^{\mu}$.

Let us now discuss the case when the manifold admits both $\zeta$  and $\ti\zeta$. 
In this case, the same deformation term in \eqref{deformchi} can be written \emph{also}
as $\delta_{\ti\zeta\,}$-exact term $\delta_{\ti\zeta} (\ti V_1+ \ti V_2)$, with tilded and untilded objects appropriately swapped.
Assuming the same reality conditions, and in particular choosing $A_\mu$ real\footnote{For example, on K\"ahler manifolds, the canonical choice is to take $A$ real and $V=0$.} (with again no reality condition on $V_\mu$),
 the localization locus  becomes $\dd_{\zeta} \psi = \dd_{\zeta} \ti\psi = \dd_{\ti\zeta} \,\psi = \dd_{\ti\zeta} \,\ti\psi = 0\,$.
 Contracting with appropriate spinors this can be recast into the equations
\be
F \,=\, 0\;,  \qquad  J^{\mu}{}_{\nu} D^{\nu} \ti\phi \,=\, i D^{\mu} \ti\phi \; , \qquad \ti J^{\mu}{}_{\nu} D^{\nu} \phi \,=\, i D^{\mu} \phi \  .
\label{locus:chi2}
\ee
The last two equations imply $K^{\mu}D_{\mu}\phi = K^{\mu} D_{\mu}\ti\phi = 0$.
Notice that the locus equations $J^{\mu}{}_{\nu} D^{\nu} \ti\phi \,=\, i D^{\mu} \ti\phi$ and $\ti J^{\mu}{}_{\nu} D^{\nu} \phi \,=\, i D^{\mu} \phi$ are derived from two deformation terms that are equal up to a total derivative (exactly equal when integrated over the compact four-manifold). This means that although the two equations may be different locally, they admit the same global solutions.

As in the case of the vector multiplet, the solutions to the locus equations \eqref{locus:chi2} depend 
on the global structure of the four-manifold considered.  
In section \ref{sec:susylocus} we will solve \eqref{locus:chi2} in 
the case of $M_4=S^1 \times M_3$, where $M_3$ is topologically a three-sphere, 
allowing for a very general class of metrics.

Before moving to the analysis of the localization on Hopf surfaces, it is interesting to note that, for manifolds amitting two Killing spinors of opposite R-charge, 
one can prove that the localization locus and one-loop determinants do not depend on the conformal factor $\Omega$ of the metric. 
This argument is presented in appendix \ref{app:Weyl}. It is in agreement with \cite{Closset:2013sxa}, that showed that the partition function is independent of small metric deformations that do not affect the complex structures. We will see in section \ref{subsec:1LoopDet} how indeed the dependence on $\Omega$ drops from the computation.

\section{Hopf surfaces}
\label{Hopf:sec}

In this section we focus on a particular class of geometries admitting two spinors of opposite R-charge, 
requiring that the four-dimensional manifold has the \emph{topology} of $S^1 \times S^3$. 
This will play an important role in the calculation of the localized partition function in section  \ref{sec:LocalizationHopf}. 
Furthermore, in order to make contact with the results of \cite{Alday:2013lba}, 
we will assume that there exists a third Killing vector commuting with $K$, and  that  the metric is a direct product.

\subsection{Generalities}


A \emph{Hopf surface} is essentially a four-dimensional complex manifold with the topology of $S^1\times S^3$, and it may be defined as a compact complex surface whose universal covering is $\bC^2 - (0,0)$.  Any such surface  arises as the quotient by a finite group $\Gamma$ of a \emph{primary} Hopf surface, which is defined as having fundamental group isomorphic to
$\bZ$ \cite{kodaira2,kodaira1}. 
In the following we will restrict our attention to primary Hopf surfaces,  referring  to them simply as Hopf surfaces. 
These  are described as a quotient of $\bC^2-(0,0)$, with coordinates $z_1,z_2$ by a cyclic group 
\begin{align}
(z_1,z_2) \sim (p z_1 + \lambda z_2^m, q z_2) \ , 
\label{defhopf}
\end{align}
where $\sim$ denotes identification of coordinates, $m \in \bN$,  and $p,q, \lambda$ are  complex parameters, such that
$0 < |p| \le |q| < 1$ and $(p-q^m)\lambda=0$. See \emph{e.g.} \cite{GO}.  It was shown in \cite{kodaira2,kodaira1}
that all primary Hopf surfaces are diffeomorphic to $S^1\times S^3$. Moreover, it is shown in  \cite{Closset:2013vra}
  that Hopf surfaces with\footnote{These are referred to as of ``class 0'' in 
\cite{GO}, while those with $\lambda =0$ are referred to as of ``class 1''.} 
$\lambda \neq 0$ admit only one Killing  spinor $\zeta$, and we will not  consider them  further. We will only 
consider Hopf surfaces with $\lambda=0$, showing that these admit a very general class of metrics, compatible with both 
complex structures $J$ and $\widetilde J$, and hence both solutions $\zeta$ and $\widetilde \zeta$.

From the geometric point of view, the question that usually arises is whether on  a  manifold
there exists a particular type of metric. In the case of Hopf surfaces, a class of metric that appears to be of interest is that 
of \emph{locally conformally K\"ahler} (LCK) metrics \cite{GO}. This means that there exists, at least locally, a conformal rescaling of the metric, to a K\"ahler one. 
A simple way to state this property is that the Lee form associated to the complex structure is closed: $\diff \theta =0$.  Indeed Ref.~\cite{GO} constructed a large class
of LCK metrics on a Hopf surface. However, from the point of view of rigid supersymmetry, there is no natural condition on the curvature of a metric, and indeed the
LCK property is too restrictive. From the expressions \eqref{avav1}, \eqref{avav2} we see that this property 
is equivalent to the requirement that the curvature of the conformally 
invariant background field $\Acks$ is purely real:
\bea
 \mathrm{Im} [ \diff \Acks ] & = & 0  \quad \Leftrightarrow \quad \mathrm{LCK}~.
\eea
Although the Hermitian metric discussed in \cite{Cassani:2014zwa} (see \emph{e.g.} equation (5.38) of this reference) is indeed LCK, as can be seen from the expression of $\Acks$ in
(5.10), in general this property is not satisfied by Hermitian metrics admitting two Killing spinors of opposite R-charge.

Notice also that the  metrics  written in equation (4.7) of  \cite{Closset:2013vra} arise from the particular 
choice of complex coordinates on $\bC^2 - (0,0)$ made in this reference. 
Below we will present a different construction, where we will start with a 
smooth metric on $S^1\times S^3$, containing arbitrary functional degrees of freedom. This will make transparent the fact 
that the constants $p,q$ parameterise the complex structure of the Hopf surface, while the metric is largely independent of these.

\subsection{Global properties}
\label{toric:sec}

We will discuss the geometries of interest starting from a four-dimensional metric that is \emph{by construction} a non-singular complete metric 
on $S^1\times S^3$. Requiring that this is compatible with an integrable complex structure ensures
that it is a metric on a Hopf surface \cite{kodaira2,kodaira1}.  The existence of two Killing spinors 
$\zeta$, $\widetilde\zeta$ is guaranteed 
imposing that the metric admits a complex Killing vector $K$ commuting with its complex conjugate and satisfying $K_\mu K^\mu = 0$.

The global analysis of the geometry is facilitated if we assume that there exists an additional real
Killing vector commuting with $K$, so that 
generically the isometry group of the metric is $U(1)^3$, with a $U(1)$  acting on $S^1$ and a $U(1)\times U(1)$ acting on a transverse metric 
on $S^3$. The three-dimensional part is therefore \emph{toric}, and in particular admits an almost contact structure and a dual Reeb vector
field whose orbits in general do not close.\footnote{It would be straightforward to analyse the case where the isometry group of the four-dimensional metric is
$U(1)^2$. Since a $U(1)$ factor acts on $S^1$, the other $U(1)$ is generated by a Reeb vector field on $M_3\simeq S^3$ of regular type. 
This case is however less interesting.}
In appendix \ref{nondirect:sec} we analyse the most general metric with $U(1)^3$ isometry, while 
in the rest of the present section we will consider the following metric of direct product form\footnote{Note that this Riemannian metric 
is related to a supersymmetric Lorentzian metric with time coordinate $t= i\tau$ \cite{Cassani:2012ri}. 
This implies that the partition function we will 
compute in section~\ref{sec:LocalizationHopf} can also be thought of  as arising from the Euclidean (and compactified) time path integral of a theory 
defined on $\mathbb R_t \times M_3$. This partially motivates our choice of restricting to a direct product metric. Other  motivations are discussed 
in appendix \ref{nondirect:sec}.}
\bea
\label{4dmetrictoric}
\diff  s^2 & = & \Omega^2 \diff\tau^2  + \diff s^2 (M_3)  \ \equiv \ \Omega^2 \diff\tau^2 + f^2 \diff\rho^2 +  m_{IJ}   \diff\varphi_I \diff\varphi_J  \qquad \quad I,J=1,2 \, .
\eea
Here   $\tau \sim \tau + 2\pi$ is a coordinate on $S^1$, while for $M_3 \simeq S^3$ we take coordinates 
$\rho,\varphi_1,\varphi_2$ adapted to the description of $S^3$ as a $T^2\simeq U(1)^2$ fibration over an interval. In these coordinates the Killing vectors generating the  
$U(1)\times U(1)$ isometry  are $\partial/\partial\varphi_1$ and $\partial/\partial\varphi_2$.
Without loss of generality we  take canonical $2\pi$ periodicities for  $\varphi_1,\varphi_2$, and  assume  $0 \le \rho \leq 1$, 
with the  extrema of the interval corresponding to the north and south poles of the three-sphere. For $\rho \in [0,1]$, we require that
$\Omega=\Omega(\rho)>0$, $f=f(\rho)>0$ and that the torus metric $m_{IJ} = m_{IJ}(\rho)$ is positive-definite.
Moreover,  in order for the metric to be non-singular, some conditions need to be satisfied at the poles of $S^3$, which we will spell out below.

Near to an end-point, one of the one-cycles of the torus remains finite, while the other one-cycle must  shrink, in a way such 
that the associated angular coordinate locally describes, together with $\rho$, a copy of $\mathbb R^2$. Let us assume that
$\partial/\partial\varphi_1$ (respectively, $\partial/\partial\varphi_2$) generates the one-cycle that shrinks at $\rho \to 1$ (respectively, $\rho \to 0$).
Then, as $\rho \to 0$ we require that
\be\label{conditionsrhoto0}
f \to f_2 \, , \quad m_{11} \to m_{11}(0) \,,\quad m_{22}= (f_2\rho)^2 + \mathcal O( \rho^3)\,,\quad m_{12} = \mathcal O(\rho^2)~,
\ee
where $f_2>0$ and $m_{11}(0)>0$ are constants. Similarly, as $\rho \to 1$ we require 
\be
f \to f_1 \, ,\;\; m_{11}= f_1^2(1-\rho)^2 + \mathcal O( (1-\rho)^3),\;\; m_{22}\to m_{22}(1) \, , \;\; m_{12} = \mathcal O[(1-\rho)^2]~,
\ee 
where $f_1>0$ and $m_{22}(1)>0$ are constants. 
Note that $m_{IJ}$ must degenerate at the poles, since either one of the vectors $\partial/\partial\varphi_I$ has vanishing norm there. 
Indeed, as $\rho \to 0$ we see that $\det (m_{IJ})$ goes to zero precisely as $m_{11}(0)(f_2 \rho)^2$, while when $\rho \to 1$ it goes to zero 
as $m_{22}(1)f_1^2(1-\rho)^2$.

It is now simple to construct supersymmetric backgrounds preserving two supercharges of 
opposite R-charge, with  metric given by \eqref{4dmetrictoric}. As reviewed in section~\ref{sec:2scharges}, a solution $\zeta$ and a solution 
$\ti\zeta$ to equations \eqref{KeqnZeta}, \eqref{KeqnTiZeta}  exist if  the metric  
admits a complex Killing vector $K$ commuting with its complex conjugate, $[K,\overline K] =0$, and squaring to zero, $K_\mu K^\mu = 0$.
We choose
\be
K \ = \ \frac{1}{2} \left[b_1 \frac{\partial}{\partial \varphi_1} + b_2 \frac{\partial}{\partial \varphi_2} - i \frac{\partial}{\partial\tau}\right]\,,
\label{torickv}
\ee
where $b_1$ and $b_2$ are two real parameters, so that  the orbits of ${\rm Re}K$ generically do not close. Notice that ${\rm Re}K$ is a Reeb vector on $M_3$, whose 
dual one-form defines an almost contact structure. 
This clearly satisfies $[K,\overline K] = 0$, while the condition $K_\mu K^\mu = 0$ is equivalent to
\be\label{OmegaToric}
\Omega^2 \ = \ b^I m_{IJ} b^J  \,  \qquad \mathrm{for} \quad \rho \in [0,1]~.
\ee
Note that this can be regarded 
as a constraint on the $g_{\tau\tau}$ component of the metric \eqref{4dmetrictoric}, hence the three-dimensional 
part of~\eqref{4dmetrictoric} is a non-singular metric on $M_3\simeq S^3$,  \emph{independent} of 
 the two parameters $b_1,b_2$  \cite{Alday:2013lba,Farquet:2014kma}.  In appendix \ref{nondirect:sec} we discuss how this condition is generalised 
 in the case of a non-direct product metric, showing that this is related to \emph{complexifying} the parameters $b_1,b_2\,$.

The background fields $A$ and $V$ can be determined using the formulae in section 
\ref{sec:2scharges}, which require first casting the metric in the canonical complex coordinates $w,z$.
We will do this in two steps. Firstly, we will show that the metric can be written as
\be
\diff s^2 \ =\  \Omega^2 \left[ \diff\tau^2 + (\diff\psi + a)^2 + c^2 \diff z\diff\bar z  \right] ,
\label{prodmetric}
\ee
where  $\psi$ is an angular  coordinate  such that
\be
\frac{\partial}{\partial\psi} \ = \  b_1 \frac{\partial}{\partial \varphi_1} + b_2 \frac{\partial}{\partial \varphi_2}\,,
\ee
and $z$ is a complex coordinate defined in terms of $\rho,\varphi_1, \varphi_2$. 
Moreover, $c=c(z,\bar z)$ is a real, positive function of $z$, while  
$a = a_z(z,\bar z)\diff z +\bar a_{\bar z}(z,\bar z)\diff \bar z$ is a real one-form.
Notice that the three-dimensional part of the metric \eqref{prodmetric} is precisely of the 
form implied by new minimal supersymmetry in three dimensions \cite{Closset:2012ru}, 
and used in the analysis of \cite{Alday:2013lba}. Secondly, we will introduce another complex coordinate, $w$, 
thus arriving at the form~\eqref{seibmetric}.

A convenient\footnote{The only requirement is that the change of coordinates should be invertible.}
choice of Killing vector on $M_3$ independent of (\ref{torickv}) is
\be
\frac{\partial}{\partial\chi} \ = \ b_1 \frac{\partial}{\partial \varphi_1} - b_2 \frac{\partial}{\partial \varphi_2}\,,
\ee
with the  corresponding change of coordinates given by
\be\label{Relphi1phi2Withpsichi}
\varphi_1 \,=\, b_1(\psi+\chi)\,,\qquad \varphi_2 \,=\, b_2(\psi-\chi)\ .
\ee
In terms of the $\psi,\chi$ coordinates, the $M_3$ part of the metric~\eqref{4dmetrictoric} becomes
\be
\diff s^2
(M_3) \ =\ \Omega^2  \left[ (\diff \psi + a)^2 + \Omega^{-2}f^2 \diff \rho^2 + c^2 \diff\chi^2 \right] \,,
\label{gpsichi}
\ee
where $\Omega^2$ is given in \eqref{OmegaToric}, the function $c$ reads
\be\label{cfunction}
c  \ = \  \frac{ 2| b_1b_2|}{\Omega^2} \sqrt{\det( m_{IJ})}\ ,
\ee
and  the one-form $a = a_\chi \diff \chi$ is given by
\be\label{aToric}
a_\chi  \ =\      \frac{1}{\Omega^2}\left(b_1^2 \, m_{11} - b_2^2 \,m_{22} \right) \ .
\ee
Next, we define the complex coordinate $z$ as $z  =  \modz (\rho) + i\, \chi$, where the real function $\modz(\rho)$ is a solution to
\be\label{conditions_defz}
u' \ =\ \frac{f}{\Omega\, c}\;,
\ee
with prime denoting derivative with respect to $\rho$. 
This differential equation can be solved for $\rho \in (0, 1)$, so the complex coordinate $z$, together with $\psi$, covers $S^3$ everywhere except 
 at the poles, which are found at ${\rm Re}\,z \to \pm \infty$ (\emph{cf.} the expansions in \eqref{cexpansions} below).
We then see that the metric takes the desired form~\eqref{prodmetric}. In these coordinates, the vector $K$ becomes
\be 
K \ =\ \frac{1}{2}\left(\frac{\partial}{\partial\psi} - i \frac{\partial}{\partial\tau}\right)\, ,
\ee
while as a one-forms it reads
\be
K \ = \ \frac{1}{2}\,\Omega^2 \left( \diff\psi + a - i\, \diff \tau \right)\, .
\ee
Note that although the metric components in \eqref{gpsichi} 
depend explicitly on $b_1,b_2$,  this is just an artefact of the choice of coordinates. 
In particular, global properties of the metric may be analysed  only in the coordinates $\rho,\varphi_1,\varphi_2$, 
and not in the coordinates $\psi,z$, as neither $\psi$ nor $\chi = {\rm Im}\,z$ are period coordinates in general.

Let us now cast  the metric~\eqref{prodmetric} in the form~\eqref{seibmetric},  introducing a complex coordinate $w$ in addition to $z$. 
We take 
\be 
w \ =\ \psi + i\, \tau + \pp(z,\bar z)\,,
\ee 
where $\pp(z,\bar z)$ is a {\rm complex} function. With this definition, we have $K = \partial/\partial w$, and the two metrics  match if we impose 
\be\label{matchingconditions}
\partial_z \overline \pp \,=\, a_z\qquad {\rm and}\qquad h \,=\,  \partial_z (\overline \pp - \pp)\,,
\ee
where the first equation can be solved for $\pp$, while the second equation determines $h$.
We can now discuss the background fields $V$ and $A$ given, for example, in \eqref{alternativeExprAV}, 
with the latter chosen real for convenience.  Noting that \eqref{matchingconditions} implies
\be
\partial_{\bar z} h \, =\,  \partial_{\bar z} a_z - \partial_z \bar a_{\bar z} \,=\, -\frac{i}{2}*_2\!(\diff a)\,,
\ee 
where $*_2$ denotes the Hodge star of the 2d metric $\diff z \diff\bar z$, with volume form ${\rm vol}_2 = \frac{i}{2}\diff z\wedge\diff \bar z $, 
we see that the choice of $\kappa$ in~\eqref{kapparesult}, ensuring that $A$ is real, reads
\be\label{fixkappa}
\kappa \ = \ \frac{*_2(\diff a)}{3\Omega^2c^2} \,,
\ee 
so that $\kappa$ is real and completely determined  by the metric on $M_3$. 
Then the formula for $V$ in \eqref{alternativeExprAV} 
can be written as
\be\label{Vinpsicoord}
V \ =\ 2\,{\rm Im}\!\left[\partial_z \log \Omega \, \diff z \right]  - \frac{1}{3 c^2} *_2\!(\diff a) (\diff\psi + a) - \frac{i}{6 c^2} *_2\!(\diff a)\, \diff\tau \,.
\ee
In the coordinates $\rho, \varphi_1, \varphi_2$, this becomes
\be\label{VforRealAtoric}
V \ =\ \frac{1}{2f}\left[c\,\Omega'  - \frac{\Omega}{6c} \left(a_\chi^2\right)' \right]\left( \frac{\diff \varphi_1}{b_1} - \frac{\diff \varphi_2}{b_2} \right) - \frac{\Omega}{6cf} (a_\chi)' \left( \frac{\diff \varphi_1}{b_1} + \frac{\diff \varphi_2}{b_2}  + i\,\diff \tau\right),
\ee
where the functions $\Omega(\rho)$, $c(\rho)$ and $a_\chi(\rho)$ are those in~\eqref{OmegaToric}, \eqref{cfunction}, \eqref{aToric}.
Similarly, the expression for the real gauge field $A$ in~\eqref{alternativeExprAV} becomes
\be\label{realAforS1M3}
A \ =\  \frac{1}{4\Omega^2f}\,(\Omega^3c)' \left( \frac{\diff \varphi_1}{b_1} - \frac{\diff \varphi_2}{b_2} \right) + \frac{1}{2} \diff \betaph\,.
\ee

Having obtained $V$ and $A$ in the $\rho,\varphi_1,\varphi_2$ coordinates, we can now discuss their global properties, in particular their regularity at the poles of $S^3$. Recalling our assumptions on $f$ and $m_{IJ}$, it is easy to see that for $\rho$ close to zero the functions $\Omega$, $c$ and $a_\chi$ behave as
\be
\Omega^2 = b_1^2\left[ m_{11}(0) + m_{11}'(0) \,\rho\right] + \mathcal O(\rho^2) \,,\quad\;   c  =  \frac{2f_2}{\sqrt{m_{11}(0)}}\frac{|b_2|}{|b_1|}\,\rho + \mathcal{O}(\rho^2) \,,\quad\;  a_\chi = 1+ \mathcal{O}(\rho^2)\,,
\label{cexpansions}
\ee
with analogous expressions holding for $\rho \to 1$. Hence, at leading order in $\rho \to 0$, we see that $V$ behaves as 
\be
V \ = \ 
k \left(\frac{1}{b_1} \diff\varphi_1 + \frac{i}{2} \diff\tau \right) + {\cal O}(\rho) \,,
\ee
where $k$ is a constant.\footnote{This reads $k = - \frac{b_1^2}{3 |b_2| f_2^2} \left[ \frac 12m_{11}''(0) - (m_{11}'(0))^2 - (f_2 b_2/b_1)^2  \right] $.} 
This is regular, as neither the one-cycle dual to $\diff\varphi_1$ nor the one dual to $\diff\tau$ shrink to zero size at $\rho = 0$.
Regularity of $V$ at $\rho =1$ is seen in a similar way.

On the other hand, regularity of $A$ is not automatic; by imposing this we determine $\betaph$, namely the phase of $s$. 
At leading order in $\rho \to 0$ we have
\be
A \ =\ 
\frac{|b_2|}{2}\left(\frac{\diff\varphi_1}{b_1} - \frac{\diff\varphi_2}{b_2}\right) + \frac{1}{2}\diff\betaph + {\cal O}(\rho) \,,
\ee
while at leading order in $(1-\rho)  \to 0$ we have
\be
A \ =\ -\frac{|b_1|}{2}\left(\frac{\diff\varphi_1}{b_1} - \frac{\diff\varphi_2}{b_2}\right) + \frac{1}{2}\diff\betaph + {\cal O}(1-\rho)\,.
\ee
In order to ensure that $A$ does not have a component along the $S^1$ that shrinks at either poles, we must take
\be\label{choicebeta}
\betaph \ =\ {\rm sgn}(b_1)\,\varphi_1+ {\rm sgn}(b_2)\,\varphi_2\,.
\ee

To summarise, starting with an arbitrary non-singular metric $\diff s^2 (M_3)$ on $S^3$, 
we have constructed a non-singular (direct-product) metric on $S^1\times S^3$, compatible with two commuting complex structures, 
and thus admitting two supercharges with opposite R-charge $\zeta$, $\widetilde \zeta$. The choice  \eqref{choicebeta} guarantees that the background fields $A$, $V$ are non-singular. In appendix~\ref{Berger:sec} we illustrate the formulae above in an explicit example based on the Berger three-sphere.

\subsection{Complex structure}
\label{subsec:Hopf}

The pair $(\diff s^2, J)$ determines a Hopf surface, which must arise as a quotient of  $\bC^2- (0,0)$ as in 
\eqref{defhopf}. We now show this explicitly, by relating the complex coordinates $w,z$ to complex coordinates 
$z_1,z_2$ on  $\bC^2- (0,0)$, and determining the complex structure parameters $p,q$ in terms of the parameters 
$b_1,b_2$ introduced above. This will provide a relation between the complex structure in four dimensions, and 
the almost contact structure in the three-dimensional geometry obtained by reduction along the $S^1$.

Using \eqref{conditions_defz}, and taking $P (z,\bar z)= i \qq  (\rho)$ with $Q(\rho)$ a real function,  the first equation in \eqref{matchingconditions} becomes 
\be
Q' \  =\ \frac{f a_\chi}{\Omega\, c}\, ,
\label{weqn}
\ee
and we claim that an appropriate  choice of complex coordinates on $\bC^2- (0,0)$ is given by 
\bea
z_1  & =  & \mathrm{e}^{-|b_1| (iw+z)} ~,\nonumber\\
z_2 & =  & \mathrm{e}^{-|b_2| (iw-z)}~.
\eea
Since these are related to $w,z$ by a holomorphic change of coordinates, they are automatically compatible with the complex structure induced by supersymmetry. 
In terms of the globally defined coordinates on $S^1 \times S^3$ we have
\bea
z_1  & =  & \mathrm{e}^{|b_1|\tau}  \mathrm{e}^{|b_1| (\qq-\modz)}    \mathrm{e}^{-i \, {\rm sgn}(b_1) \varphi_1}~,\nonumber\\
z_2 & =  &   \mathrm{e}^{|b_2|\tau}  \mathrm{e}^{|b_2| (\qq+\modz)}  \mathrm{e}^{-i \, {\rm sgn}(b_2)\varphi_2}  ~.
\label{z1z2glob}
\eea

If $(z_1,z_2)$  are indeed coordinates on  $\bC^2-(0,0)$, it is  immediate to see that the identification $\tau \sim \tau + 2\pi$ leads to
\begin{align}
(z_1,z_2) \sim (\mathrm{e}^{2\pi |b_1|} z_1, \mathrm{e}^{2\pi |b_2|} z_2) \ ,
\end{align}
corresponding to a Hopf surface with parameters $p= \mathrm{e}^{-2\pi |b_1|}$ and $q =\mathrm{e}^{-2\pi |b_2|}$.\footnote{Strictly speaking, it 
is $p= \mathrm{e}^{-2\pi |b_1|}$, $q =\mathrm{e}^{-2\pi |b_2|}$ if $|b_2| \le |b_1|$ and $p= \mathrm{e}^{-2\pi |b_2|}$, $q =\mathrm{e}^{-2\pi |b_1|}$ if $|b_1| \le |b_2|$. } 
Note that the choice of $p,q$ is \emph{independent} of the metric on $M_3$, and only affects the four-dimensional metric through $\Omega^2$. 

It remains to show that  $z_1,z_2$ are complex coordinates on $\bC^2-(0,0)$ when $\tau$ is decompactified, so that $\tau \in \bR$. From (\ref{z1z2glob})
it  is clear that the phases $-{\rm sgn}(b_j) \varphi_j$ correspond to the angular directions in polar coordinates for the two copies of $\bC$ in  $\bC^2 = \bC \oplus \bC$. 
Therefore we have to show that $|z_1|$, $|z_2|$ are appropriate radial directions, and that the point $(0,0)$ is excluded. The proof is given in appendix \ref{details:sec}, while below we present
a simple example where the function $Q$ derived from \eqref{weqn} can be obtained explicitly.

Consider the Berger sphere $M_3=S^3_v$ with metric 
\be
\diff s^2(S^3_v) \ = \ \diff \theta^2 + \sin^2\theta\, \diff \varphi^2 + v^2 ( \diff \anglepsi+ \cos\theta \,\diff \varphi)^2~,
\label{biax}
\ee
discussed in detail in appendix \ref{Berger:sec}. In the special case $b_1 = -b_2 = \frac{1}{2 v} >0$ 
we have  $\theta=\pi \rho$, $\Omega=1$, $f=\pi$, $c = \frac 1 v \sin\theta$, $a_{\chi} =\cos \theta$. 
The equations \eqref{conditions_defz} and \eqref{weqn} 
become $\p_{\theta} u = v ( \sin \theta)^{-1}$ and $\p_{\theta} \qq = v \, {\rm cotan}\, \theta$ and  are solved by
\begin{align}
u (\theta) \ = \ v \log \tan \frac{\theta}{2} \ , \quad  \quad \qq (\theta) \ = \  v \log \sin \theta \ ,
\end{align}
yielding the coordinates
\bea
z_1 & = &\sqrt 2 \, \mathrm{e}^{\frac{\tau}{2v}} \cos \frac{\theta}{2} \mathrm{e}^{-i \varphi_1} \, , \nonumber\\
z_2 & = &  \sqrt 2 \, \mathrm{e}^{\frac{\tau}{2v}}  \sin  \frac{\theta}{2} \mathrm{e}^{-i \varphi_2} \ ,
\eea
in agreement with \cite{Cassani:2014zwa}.  It is straightforward to see that these indeed cover $\bC^2-(0,0)$ when $\tau \in \mathbb R$.


\section{Localization}
\label{sec:LocalizationHopf}

In this section we will compute the partition function of a four-dimensional ${\cal N}=1$ supersymmetric gauge theory defined on a background geometry admitting two supercharges
of opposite R-charge, comprising a Hopf surface with arbitrary (real) parameters $p,q$, and a very general Hermitian  metric with $U(1)^3$ isometry.
We will consider gauge theories  with  a vector multiplet transforming in the adjoint representation of a  gauge group $G$, and chiral multiplets transforming in arbitrary representations of $G$.

\subsection{Localization locus}
\label{sec:susylocus}

The vector multiplet supersymmetric locus  given by (\ref{locus:vec}) implies that $\scA_\mu$ is a \emph{flat connection}. 
After having specified an $S^1\times S^3$ topology, 
the flat connections are characterized by the holonomy of constant gauge fields around $S^1$. 
In particular, up to gauge transformations, 
the localized fields of the vector multiplet are
\begin{align}
 \scA_{\mu}  =  (\scA_i,\scA_\tau) =  (0,\scA_0 )  \ , \qquad  D = 0 \ ,\label{locus:VecS1M3}
\end{align}
where $\scA_0$ is constant.  Notice that this result holds without any further assumption on the metric, therefore it is true
also if the metric is not a direct product  or/and it has only a $U(1)^2$ isometry.

Let us fix  the vector multiplet fields at their locus values~\eqref{locus:VecS1M3} and  proceed to analyse  
the supersymmetric locus of a  chiral multiplet with R-charge $r$, determined by the  equations (\ref{locus:chi2}). 
Following the discussion of section \ref{sssec:ChiralLocus}, we will choose $A_{\mu}$ real and impose the reality conditions $\ti \phi = \phi^\dagger$ and $\ti F = - F^{\dagger}$ on the bosonic fields. Then
the locus equations read
\bea
F  & =  & 0  \, , \no\\ [2mm]
  (J_{\mu}{}^{\nu}+ \ti J_{\mu}{}^{\nu}) D_{\nu} \phi &  =  & 0    \, , \no \\ [2mm]
 (J_{\mu}{}^{\nu} - \ti J_{\mu}{}^{\nu}) D_{\nu} \phi & = &  -2i D_{\mu} \phi ~.
\eea
Contracting the second equation with $K^{\mu}$ and $\overline{K}{}^{\mu}$ leads to $K^{\mu}D_{\mu}\phi = \overline{ K}{}^{\mu} D_{\mu}\phi = 0$.
Using the expressions for $J$, $\widetilde J$ and $K$ given in section~\ref{sec:2scharges},
the equations for $\phi$ become
\bea
 D_{\tau} \phi &=& \p_{\tau}\phi - i \scA_0 \phi  \ =\ 0 \, ,\no\\ [2mm]
 D_{\psi} \phi &=& \p_{\psi} \phi - i r A_{\psi} \phi \ =\ 0 \, , \\ [2mm]
 D_{\bar z} \phi &=& \p_{\bar z} \phi - i r A_{\bar z} \phi \ =\ 0 \ , \no
 \label{ChiLocusEqn}
\eea
where we have used the fact that $A_{\tau}=0$.  The first equation implies that $\phi$ is proportional to $\mathrm{e}^{i \scA_0 \tau}$, which is not globally defined on $S^1$, except 
when $\scA_0 = 0$ modulo large gauge  transformations.\footnote{We discuss these large gauge transformations below.}  Therefore in this case we immediately conclude that $\phi=0$.
When $\scA_0=0$ the analysis is slightly more subtle.  
The first equation implies that $\phi$ is independent of $\tau$, and using (\ref{finalA}) the two remaining equations 
are solved by
\be\label{solchirallocus}
\phi \ = \  
\ C  (z) \left(\Omega^3c\right)^{-\frac r 2}  \mathrm{e}^{\frac{ir}{2} (\mathrm{sgn}(b_1)\varphi_1 + \mathrm{sgn}(b_2)\varphi_2)}\,,
\ee
with $C (z)$ a (locally) holomorphic function of $z$. In order to obtain a globally defined solution, we must impose periodicity around the two $S^1$
 parametrized by $\varphi_1$ and $\varphi_2$. Recalling that $z = u(\rho) + \frac{i}{2} \lp \frac{\varphi_1}{b_1} -  \frac{\varphi_2}{b_2} \rp$, 
 periodicity under the shift $\varphi_1 \rightarrow \varphi_1 + 2\pi$ sgn$(b_1)$ yields
\be
C \big( z + \tfrac{\pi i}{|b_1|} \big)\, \mathrm{e}^{\pi i r} \ = \ C (z) \ , 
\label{Period1}
\ee
and similarly periodicity under 
$\varphi_2 \rightarrow \varphi_2 + 2\pi$ sgn$(b_2)$ gives
\be
C \big( z - \tfrac{\pi i}{|b_2|} \big)\, \mathrm{e}^{\pi i r} \ = \ C (z) \ ,
\label{Period2}
\ee
so that  in particular $C(z)$ is a periodic function in the imaginary direction\footnote{This is true, independently of whether $\chi$ is a periodic or a non-compact coordinate.} 
$C \big( z  +  i\pi \frac{ |b_1|+|b_2|}{|b_1 b_2|} \big) = C(z)$.   
Since  $|\phi|=|C (z)| (\Omega^3 c)^{-\frac r 2}$,  with $\Omega^3 c$ vanishing only at the poles $\rho=0, \rho=1$ (see appendix \ref{details:sec}),  
we see that in order to have a non-singular solution $\phi$ for $r>0$, $C(z)$ must vanish at $\rho=0,\rho=1$,
that is $\lim_{{\rm Re} \, z \rightarrow \pm \infty} C (z) = 0$.  Extending $C(z)$ to the complex $(u,\chi)$ plane, we see  that it 
  is a bounded entire function,  and therefore Liouville's theorem implies it  is a constant.  The limits at the poles imply $C =0$, thus showing that  for $r>0$, the localization locus is $\phi=0$.

If $r\leq 0$ we get the following restriction.  
The general solution of \eqref{Period1}  is $C (z) = \sum_{n \in  \bZ} C_n \, \mathrm{e}^{-|b_1|(r+2n)z}$, where $C_n$ are constants. 
Inserting this  into \eqref{Period2},  we see that  for each $n \in \bZ$, either $\mathrm{e}^{\pi i |b_1|(r+2n) + \pi i r} = 1$ or $C_n =0$. 
So there can be non-trivial solutions if and only if the R-charge $r$ takes the very special form
\be
r \ = \ - \frac{2|b_1|n + 2|b_2|m}{|b_1| + |b_2|} \le 0 \quad , \quad n,m \in \bZ \ .
\label{specialR}
\ee
Thus simply assuming  that $r$ is not one of the special values \eqref{specialR}, the chiral multiplet 
localization locus is given by 
\be\label{locus:ChiS1M3}
  F \ =\ \phi\ =\ 0 \  \, .
\ee


The full supersymmetric locus  is  thus completely characterized by the constant Lie algebra element $\scA_0$.
Correspondingly, the path integral splits into a matrix integral over $\scA_0$, and a Gaussian integral over all the fluctuations about the saddle point locus \eqref{locus:VecS1M3},  \eqref{locus:ChiS1M3}.  Following a similar discussion in 
\cite{Aharony:2003sx},  we will now explain how to use the residual gauge freedom to extract the correct integration measure of the 
\emph{matrix model}.

\subsection{The matrix model}
\label{sec:thematrix}

First of all,  one can use constant  gauge transformations
to diagonalize $\scA_0$ and  reduce the integration to the Cartan subalgebra of the gauge group $G$, introducing a Vandermonde determinant
\be
 \Delta_0 [\scA_0]\  = \  \prod_{\alpha \in \Delta_+}  (\alpha_{\scA_0})^2  ~,
 \ee 
 where  $\Delta_{+}$ denotes  the set of positive roots and $\alpha_{\scA_0} \equiv \alpha(\scA_0)$.
 In a Cartan basis $\{ H_k \}$ we have  $\scA_0=\sum_{k=1}^{r_G} a_k H_k$, where $r_G$ is the rank of the gauge group $G$. Then for a root $\alpha =\{ \alpha_k \}$, we have $\alpha_{\scA_0} = \sum_k a_k \alpha_k $.
One also has to divide by the order of the Weyl group $|\scW|$ in order to take care of gauge transformations that permute the elements of the Cartan basis.

Furthermore, the path integral must be invariant under large gauge transformations along the $S^1$, that shift $\scA_0\to \scA_0+ \sum_k d_k H_k$, where $d_k \in \bZ$.\footnote{We assume that the gauge field is normalized so that all the matter fields have integer charges.}
Thus we  can restrict  the  range of integration of the constants $\{ a_k \}$ to be  over the maximal  torus $T^{\, r_G}$ of $G$, parametrised by
\be 
z \ =\ \{ z_k \}\ =\ \{ \mathrm{e}^{2\pi i a_k } \} \, \in\,  T^{\, r_G}~.
\ee
The localization argument then reduces the partition function to  the form
\begin{align}
Z \ =\ \frac{1}{|\scW|} \int_{T^{r_G}}  \! \frac{\diff z}{2\pi i z} \ \Delta_0 [\scA_0]  \ Z_{\rm classic}[\scA_0] \  \widetilde Z^{\rm vector}_{\rm 1{\textrm-}loop}[\scA_0] \, \prod_J  Z^{{\rm chiral} \, (J)}_{\rm 1{\textrm-}loop}[\scA_0]   \ ,
\label{ZMModel}
\end{align}
where the integration measure $\diff \scA_0$ has been replaced by 
\be
\diff\scA_0\ \equiv \ \prod_{k=1}^{r_G} \diff a_k \ \rightarrow\ \frac{\diff z}{2\pi i z} \ \equiv\ \prod_{k=1}^{r_G} \frac{\diff z_k}{2\pi i z_k}\,.
\ee  
Here $Z_{\rm classic} [\scA_0]$ is the classical contribution from the vector and chiral multiplets.
However,  for the theories that we consider, with Lagrangians (\ref{Lagrangians:vec}), 
 (\ref{Lagrangians:chi}) (plus superpotential couplings), we have $Z_{\rm classic}\,=\, \mathrm{e}^{-S_{\rm classic}}=1\,$.
The remaining factors  $\widetilde Z^{\rm vector}_{\rm 1{\textrm-}loop}[\scA_0]$ and $Z^{{\rm chiral} \, (J)}_{\rm 1{\textrm-}loop}[\scA_0]$ are 
the one-loop determinants of the vector multiplet and chiral multiplets fluctuations around the
configurations \eqref{locus:VecS1M3} and \eqref{locus:ChiS1M3}.

Denoting by  $\scA_{\tau}$ and  $\scA_{i}$ the components of the gauge field $\mathcal A_\mu$
 along $S^1$ and $M_3$, respectively, we will impose the following gauge-fixing conditions
\begin{align}
 \nabla_\tau \, a \, =\, 0~,      \qquad     \nabla^i \scA_i \,=\,0   \,  ,
\label{gaugefixing}
\end{align} 
where $a \equiv  \frac{1}{{\rm vol} (M_3)}\int_{M_3} \scA_{\tau}$. Let us discuss the first condition, while we will deal with  the second  condition later   \cite{Adams:1996hi,Marino:2011nm}.
The Faddeev--Popov determinant  $\mathrm{det}' \big( \nabla^\tau D_\tau^{(0)} \big)$  associated to  
$\nabla_\tau a  = 0$ can be written in terms of ghost fields $\gamma, \bar \gamma$, yielding an integral over 
the following  gauge-fixing term  
\bea
S^\mathrm{gauge-fixing}_a & = & \int \diff \tau \, \mathrm{Tr} \, \left[  \bar \gamma  \big(  \nabla^\tau D_\tau^{(0)} \big) \gamma  + \xi \nabla_\tau a \right] \ ,
\label{gfix}
\eea
where  $D_\tau^{(0)}  = \nabla_{\tau} -  i [\mathcal A_0, \cdot ]$ and a prime on the determinant means that it does not contain the zero mode along $S^1$. 
The second term is simply a rewriting of the delta function $\delta (\nabla_\tau a )$ enforcing the gauge-fixing condition, with $\xi$ a Lagrange multiplier.  
The gauge fixing action (\ref{gfix}) can be included in the deformation term by replacing $\dd V \rightarrow \dd' V'$,  with $\dd' = \dd + \dd_B$, where $\dd_B$ is the BRST transformation, and $V' = V + \tr \, \bar \gamma \nabla_{\tau} \, a$  \cite{Kapustin:2009kz}. We refer to \cite{Pestun:2007rz} for a more rigorous treatment of the ghosts.


Writing $a  =  \scA_0 +  \nabla_\tau \varphi $ and doing the path integral over $\varphi$ introduces a Jacobian factor $(\det' \nabla^2_\tau)^{-1/2}$, 
which combined with the Faddeev--Popov determinant  yields 
\bea
 \widetilde Z^{\rm vector}_{\rm 1{\textrm-}loop}[\scA_0] & = & \Delta_2 [\scA_0] \,    Z^{\rm vector}_{\rm 1{\textrm-}loop}[\scA_0]~,
 \eea
where 
\bea
\Delta_2 [\scA_0]  & \equiv  & \mathrm{det'}  D_\tau^{(0)}  \ =  \ \prod_{\alpha  \in \mathfrak{g}}  \prod_{n \neq 0}  \left(in - i \alpha_{\scA_0} \right) ~,
\eea
and $\alpha \in \mathfrak{g}$ labels both non-zero roots and Cartan generators. 
A  straightforward computation  yields 
\bea
\Delta_2 [\scA_0]  & = &  (2\pi)^{r_G} \, \prod_{\alpha \in \Delta_+ } \frac{4 \sin^2 (\pi \alpha_{\scA_0} )}{\alpha_{\scA_0} ^2}~,
\eea
 where we used the formula $\sin (\pi z) = \pi z \prod_{n=1}^\infty \big(1 - \tfrac{z^2}{n^2} \big)$,
  and employed zeta function regularisation to regularise the infinite products.
Finally, the matrix model becomes
\begin{align}
Z \ =\  \frac{1}{|\scW|} \int_{T^{r_G}}  \! \frac{\diff z}{2\pi i z} \ \Delta_1 [\scA_0]  \ Z^{\rm vector}_{\rm 1{\textrm-}loop}[\scA_0] \, \prod_J  Z^{{\rm chiral} \, (J)}_{\rm 1{\textrm-}loop}[\scA_0]   \ ,
\label{ZMModel2}
\end{align}
with 
\bea
\Delta_1 [\scA_0] \ = \ \Delta_0 [\scA_0] \, \Delta_2 [\scA_0] & = &  (2\pi)^{r_G}  \prod_{\alpha \in \Delta_{+}} 4\sin^2(\pi \alpha_{\scA_0}) ~.
\eea

\subsection{One-loop determinants}
\label{subsec:1LoopDet}

Our strategy to compute the one-loop determinants on $S^1 \times M_3$ for the vector and chiral multiplets is to take advantage of the three-dimensional 
results\footnote{Previous studies of relations between the index  of four-dimensional gauge theories and the partition function in three dimensions include 
\cite{Dolan:2011rp,Gadde:2011ia,Imamura:2011uw,Agarwal:2012hs,Aharony:2013dha}.}
 of~\cite{Alday:2013lba}. First we expand the fields into Kaluza--Klein (KK) modes along the $S^1$ parametrized by $\tau$. Denoting by $\Phi$ a generic field (bosonic or fermionic), we take
\begin{align}
\Phi(x^i,\tau) \ =\ \sum_{n \in \bZ} \Phi_n(x^i) \, \mathrm{e}^{-i n \tau} \ .
\end{align}
The four-dimensional one-loop determinant may be replaced by the product over one-loop determinants for the KK modes on $M_3$
\be
Z_{\rm 1{\textrm-}loop}^{\rm 4d}[\Phi]\ =\ \prod_{n \in \bZ} Z_{\rm 1{\textrm-}loop}^{\rm 3d}[\Phi_n] \ .
\ee
The one-loop determinants on $M_3$ were computed in \cite{Alday:2013lba} and our aim is to use the results therein for $Z_{\rm 1{\textrm-}loop}^{\rm 3d}[\Phi_n]$. For this to be possible we need to show that the Gaussian action for fluctuations around the localization locus, resulting from the deformation terms $\dd V$, matches the Gaussian action for the three-dimensional fluctuations of \cite{Alday:2013lba}, with an appropriate mapping between fields.
Instead of proving this directly, we will take an alternative route, which is to show that the four-dimensional supersymmetry transformations given by \eqref{VecTransfo}, \eqref{ChiTransfo} reduce under KK decomposition to the three-dimensional supersymmetry transformations of \cite{Alday:2013lba}. Then it will follow that the three-dimensional Gaussian actions
for the KK multiplets are identical to the Gaussian actions of \cite{Alday:2013lba} by construction.

In order to proceed with the reduction to three dimensions, we need to relate the four-dimensional background fields to the three-dimensional ones. This analysis is presented in appendix \ref{app:4dto3dred}; the explicit relations between the four-dimensional background fields $(A_{\mu},V_{\mu})$ and the three-dimensional background fields $(\check{A}_i, \check{V}_i, \check h)$ are given in~\eqref{our4dto3dbackground} (we use a $\check\,$ symbol to denote three-dimensional quantities). With our choice of real $A_{\mu}$, the three-dimensional fields $\check{A}_i, \check{V}_i, \check h$ are also real, as it is assumed in \cite{Alday:2013lba}.

\subsubsection{Vector multiplet}
\label{subsubsec:VecS1M3}

We denote  as $\scB_{\tau}$ and  $\scB_{i}$ the fluctuations of the gauge field $\mathcal A_\mu$ along $S^1$ and $M_3$, respectively, 
 $\sigma = \Omega^{-1} \scB_{\tau}$ and consider the KK fields fluctuations $(\scB_{n\, j},\sigma_n,\lambda_n, \ti\lambda_n,D_n)$ around the localization locus (\ref{locus:VecS1M3}), where it is understood that $(\ti\lambda_n)_{\alpha} = i (\sigma^4)_{\alpha \dot\alpha} \ti\lambda_n^{\dot\alpha}\,$.
The supersymmetry transformations (\ref{VecTransfo}) (with $\ti\zeta=0$) read for these KK fields
\bea
\dd\scB_{n \, j} \!&=&\!  i \zeta \gamma_j \ti\lambda_n \, ,\quad \qquad \dd \sigma_n = \zeta  \ti\lambda_n \,, \no\\ [2mm]
\dd\lambda_n \!&=&\! - \frac{i}{2} \varepsilon^{ijk} \scF_{n \, ij} \, \gamma_k \zeta 
- i \big( \p_j \sigma_n -i \check V_j \sigma_n + \frac{i}{\Omega} [\scA_0,\scB_{n\, j}]  + \frac{i}{\Omega} n \scB_{n\, j}  \big) \, \gamma^j \zeta
+  \lp \check D_n - \check h\sigma_n \rp \zeta \,,\no\\ [2mm]
\dd \ti\lambda_n \!&=&\! 0\,, \label{KKsusyVec}\no \\ [2mm]
\dd \check D_n \!&=&\! -i \zeta \gamma^j \big( \check\nabla_j - i \check A_j + \frac i 2 \check V_j  \big) \ti\lambda_n + \frac{1}{2} \check V_j \, \zeta\gamma^j \ti\lambda_n + \frac{i}{\Omega} \zeta [ \scA_0, \ti\lambda_n] + \frac{i}{\Omega} n\, \zeta \ti\lambda_n  + \frac{\check h}{2} \zeta \ti\lambda_n \,,
\eea
where we defined $\check D_n = i D_n + (\check h- \check V_{\psi}) \sigma_n$ and used the convention $\gamma^j = - i \sigma^4 \ti\sigma^j$ for the three-dimensional gamma matrices (see appendix~\ref{app:4dto3dred} for more details about the 3d conventions).\footnote{In deriving the KK supersymmetry transformations, we have made use of the relation~\eqref{3dprojection}. We also point out the fact that the three-dimensional free parameter $\check\kappa$ of~\eqref{our4dto3dbackground} drops from the supersymmetry transformations and does not affect the whole computation.}

These transformations correspond to the supersymmetry transformations of the three-dimensional $\N=2$ vector multiplet fluctuations $\lp \scA_j,\sigma, \lambda, \lambda^{\dagger},D \rp_{\rm 3d}$ of \cite{Alday:2013lba}  with respect to the three-dimensional spinor $\eta = \sqrt 2 \zeta$,\footnote{The authors of \cite{Alday:2013lba} performed localization using a spinor $\epsilon$ of positive charge under $\check A_{\mu}$ and wrote explicitly the supersymmetry transformations for $\epsilon$. In our derivation, the relations between four-dimensional and three-dimensional background fields imply that the four-dimensional supersymmetry parameter $\zeta$ is mapped to a three-dimensional supersymmetry parameter $\eta$ of negative charge under $\check A_{\mu}$, see appendix~\ref{app:4dto3dred} for details. Thus the supersymmetry transformations~\eqref{KKsusyVec} are mapped to the three-dimensional supersymmetry transformations with respect to a negative charge spinor. These are not detailed  in \cite{Alday:2013lba}, but they can be derived from the $\epsilon$ transformations by changing (in our notations) $\epsilon_{\alpha} \rightarrow \eta_{\alpha}$, $(\check A_j,\check V_j, \scA_j) \rightarrow -(\check A_j,\check V_j, \scA_j)$ and $\ti\lambda \leftrightarrow \lambda$ (also $\ti\Phi \leftrightarrow \Phi$ for all fields for the chiral multiplet). They are also given in \cite{Closset:2012ru}.  The fact that we have a negative charge spinor $\eta$ in three dimensions does not prevent us from using the results of \cite{Alday:2013lba}, since the localization computation is unchanged if $\eta$ is used instead of $\epsilon$.}
with the map
\begin{align}
\lp \scB_{n\, j},\sigma_n, \sqrt 2\,  \lambda_n, \sqrt 2 \, \ti\lambda_n, \check D_n \rp\ &=\ \lp \scA_j, -\sigma, \lambda^{\dagger}, \lambda,-D \rp_{\rm 3d}\ , \no\\
 n + [\scA_0, \, \cdot\,]\ &=\ [\sigma_0, \, \cdot\,] \ .
 \label{4d3dVectorMap}
\end{align}
The evaluation of the one-loop determinant is done by decomposing all KK fields, denoted generically $\Phi_n$, into the Cartan basis of the gauge algebra
\begin{align}
\Phi_n \ =\ \sum_{k=1}^{r_G} \Phi_{n \, k} H_k + \sum_{\alpha \in {\rm roots}} \Phi_{n \, \alpha} E_{\alpha} \ ,
\end{align}
where $H_j$ generate the Cartan subalgebra and $E_{\alpha}$ are the ladder operators. The map~\eqref{4d3dVectorMap} descends to the $\alpha$-component multiplets, with
\be
n + \alpha_{\scA_0} \ =\ \alpha(\sigma_0) \ .
\ee
The multiplets along the Cartan directions can be associated with ``vanishing roots'' $\alpha=0$.

To be able to map the four-dimensional deformation terms to the  three-dimensional ones, we note  that on $S^1 \times  M_3$ the deformation terms~\eqref{constructdeltaVgauge} 
and~\eqref{constructdeltaVgaugetilde}, expanded at quadratic order around the localization locus,   are equal: $\dd_{\zeta} V^{(+)}=\dd_{\ti\zeta} \,V^{(-)}$. For the fermionic part this is obvious, while for the bosonic
part this follows from  the identity 
\bea
\tr \int_{S^1 \times M_3} \! \!\scF \wedge  \scF \ =\  \tr \int_{S^1 \times M_3} \! \!\diff \left(   {\cal B} \wedge \diff {\cal B} - 2i  \scA_0 \wedge {\cal B} \wedge  \cal B  \right)  \ = \ 0~.
\eea
%
  Hence we have $\dd V_{\rm 4d} = - \frac{1}{2 |\zeta|^2} \dd_{\zeta} \!\lp \tr\,  (\dd_{\zeta}\lambda)^{\ddagger} \lambda   \rp $.  
%
In section \ref{sssec:VectorLocus}, we saw that the reality conditions which, along with a real $A_{\mu}$, ensure positivity of the bosonic deformation terms are $\scA_{\mu}^{\dagger} = \scA_{\mu}$, $D^{\dagger}= -D$. For the fermions we choose $i\sigma_{4}\ti\lambda = \lambda^{\dagger}$. For the KK modes these translate into  $\scA_{n\,\mu} = \scA_{-n\,\mu}^{\dagger} $, $D_{n} = - D^{\dagger}_{-n}$ and $\ti\lambda_n = \lambda^{\dagger}_{-n}$. 

Then, using the map~\eqref{4d3dVectorMap} to three-dimensional fields, the Gaussian action for the $n$-th KK mode and $\alpha$ component fluctuations can be expressed as 
\be
\dd V_{\rm (4d)}^{(n,\alpha)} = -\frac{1}{2 |\zeta|^2} \tr \, \dd_{\zeta}\! \lp (\dd_{\zeta}\lambda_{(n,\alpha)})^{\ddagger} \lambda_{(n,\alpha)} \rp   
\, = \, - \frac{1}{4 |\eta|^2} \tr \, \dd_{\eta}\! \lp (\dd_{\eta}\lambda_{(\alpha)})^{\ddagger} \lambda_{(\alpha)} \rp_{\rm (3d)} \, = \,  \dd V_{\rm (3d)}[\sigma_0^{(n,\alpha)}] \, ,
\label{dVvectorMap}
\ee
where the action of ${}^{\ddagger}$ on the KK modes is $\Phi^{\ddagger}_{(n,\alpha)} = \Phi_{(-n,-\alpha)}$, and the constant scalar for the resulting three-dimensional deformation term is $\sigma_0^{(n,\alpha)} = n + \alpha_{\scA_0}$. This three-dimensional deformation term is the same as the one considered in \cite{Alday:2013lba}. 
The reality conditions on the three-dimensional fields in $\alpha$ components obtained from this map are $\Phi^{(3)}_{(\alpha)} = \Phi^{(3)}_{(-\alpha)}{}^{\dagger}$ for bosons
 and $\ti{\lambda}^{(3)}_{(\alpha)} = \lambda^{(3)}_{(-\alpha)}{}^{\dagger}$ for fermions, and match the reality conditions of \cite{Alday:2013lba}. Moreover, the three-dimensional gauge fixing condition $\nabla^j \scB_j=0$ chosen above becomes $\nabla^j \scA^{(3)}_j=0$,  reproducing the gauge fixing condition of \cite{Alday:2013lba}. We can then use the result of \cite{Alday:2013lba} for the three-dimensional one-loop determinant for each $(n,\alpha)$-component multiplet. Note that the contribution from the Faddeev--Popov determinant of the three-dimensional gauge fixing (namely the second in~\eqref{gaugefixing})
 is included in the result of~\cite{Alday:2013lba}.
We obtain the expected relation
\be
Z_{\rm 1{\textrm-}loop}^{\rm vector}[\scA_0] \ =\ \prod_{\alpha \in \mathfrak{g}}  \ \prod_{n \in \bZ}  Z_{\rm 1{\textrm-}loop \, (3d)}^{\rm vector} \big[ \sigma_0^{(n , \alpha)} \big]\,,
\label{VecDet}
\ee
with $\sigma_0^{(n , \alpha)} = n + \alpha_{\scA_0}$ and here $\alpha \in \mathfrak{g}$ labels both roots and Cartan components. 

From \cite{Alday:2013lba}, we extract
\begin{align}
Z_{\rm 1{\textrm-}loop \, (3d)}^{\rm vector} [ \sigma_0^{(\alpha)} ] 
\ &=\  \frac{1}{i \alpha(\sigma_0)} \ \prod_{n_1,n_2 \ge 0} \frac{ n_1 b_1 + n_2 b_2 + i \alpha(\sigma_0) }{-(n_1+1)b_1 -(n_2+1)b_2  + i \alpha(\sigma_0)} \ ,
\end{align}
holding for $b_1,b_2>0$.
A careful re-examination\footnote{We thank J.~Sparks for discussions about this point.} of the three-dimensional one-loop computation in \cite{Alday:2013lba}
shows that for arbitrary real $b_1,b_2$, the one-loop determinant is given by the formula above with $b_1, b_2$ replaced by $|b_1|, |b_2|\,$.

Renaming $n \rightarrow n_0$, our one-loop determinant is expressed by the infinite product:
\begin{align}
Z^{\rm vector}_{\rm 1{\textrm-}loop} &= Z_{\rm Cartan} \prod_{\alpha \in \textrm{roots}}  \ \prod_{n_0 \in \bZ} \lp  \frac{1}{i(n_0 + \alpha_{\scA_0})} \ \prod_{n_1,n_2 \ge 0} \frac{n_1 b_1 + n_2 b_2 + i (n_0 + \alpha_{\scA_0}) }{ -(n_1+1)b_1 -(n_2+1)b_2 + i (n_0 + \alpha_{\scA_0})}  \rp \no\\
&= Z_{\rm Cartan} \ \Delta_1^{-1} \prod_{\alpha \in \textrm{roots}} \lp  \prod_{n_0 \in \bZ} \prod_{n_1,n_2 \ge 0} \frac{n_1 b_1 + n_2 b_2 + i (n_0 + \alpha_{\scA_0}) }{ -(n_1+1)b_1 -(n_2+1)b_2 + i (n_0 + \alpha_{\scA_0})}  \rp .
\label{DetVec00}
\end{align}
We see that the first factor cancels  with the matrix model measure $\Delta_1 [\scA_0]$, while the second factor needs to be regularized. We perform this regularization in appendix \ref{app:1LoopDetReg}, using multiple Gamma functions. These manipulations yield the Jacobi theta function $\theta(z,p)$ and the Pochhammer symbol $(z;p)$, defined for $z,p \in \bC$ and $|p|<1$ respectively by
\be
\theta(z,p) \ =\ \prod_{n \ge 0} \lp 1 - z p^n \rp \lp 1 - z^{-1} p^{n+1} \rp \; , \qquad 
(z;p) \ =\ \prod_{n \ge 0} (1- z p^n)  \;.
\ee
The result is the following expression for the one-loop determinant
\be
Z^{\rm vector}_{\rm 1{\textrm-}loop}\ =\  \mathrm{e}^{i\pi \Psi^{(0)}_{\rm vec}} \, \mathrm{e}^{i\pi \Psi^{(1)}_{\rm vec}} \ (p;p)^{r_G} (q;q)^{r_G} \ \Delta_1^{-1} \ 
 \prod_{\alpha \in \Delta_{+}}  \theta \lp \mathrm{e}^{2\pi i \alpha_{\scA_0}}, p \rp \,  \theta \lp  \mathrm{e}^{-2\pi i \alpha_{\scA_0}}, q \rp ,
 \ee
 with
\be
\Psi^{(0)}_{\rm vec} \ =\  \frac{i}{12} \lp b_1 + b_2 - \frac{b_1+b_2}{b_1 b_2} \rp |G| \,,\qquad\; \Psi^{(1)}_{\rm vec} \ =\ -i \, \frac{b_1+b_2}{b_1 b_2} \sum_{\alpha \in \Delta_{+}} \alpha_{\scA_0}^2 \ ,  \label{Zvec}
\ee
where $p = \mathrm{e}^{-2\pi b_1}$, $q= \mathrm{e}^{-2\pi b_2}$,  $|G|$ is the dimension of $G$, and
we have split the prefactor into a part $\Psi^{(0)}_{\rm vec}$ independent of $\alpha_{\scA_0}$ and a part $\Psi^{(1)}_{\rm vec}$ depending on $\alpha_{\scA_0}$. 
This result looks puzzling, because the factor $\mathrm{e}^{i \pi \Psi^{(1)}_{\rm vec}}$ spoils the invariance under the shifts $\alpha_{\scA_0} \rightarrow \alpha_{\scA_0} + d$ for $d \in\bZ$, associated to large gauge transformations $\scA_0 \rightarrow \scA_0 + \sum_k  d_k H_k$, $d_k \in \bZ$. In other words, $\mathrm{e}^{i \pi \Psi^{(1)}_{\rm vec}}$ is not a function of $z_{\alpha} = \mathrm{e}^{2\pi i \alpha_{\scA_0}}$ as it must be. 
For the final matrix model to be consistent, all such ``anomalous'' terms breaking the symmetry under large gauge transformations must cancel. We will see in section \ref{ssec:IndexAnom} that this is indeed what happens if the theory satisfies relevant physical constraints.

\subsubsection{Chiral multiplet}
\label{subsubsec:ChiS1M3}

The evaluation of the one-loop determinant for the chiral multiplet proceeds in a similar fashion. 
The KK fields $(\phi_{n},\psi_{n},F_{n},\ti\phi_n,\ti\psi_n,\ti F_n)$
 all vanish on the localization locus (\ref{locus:ChiS1M3}), hence we can keep the same notations for their fluctuations around zero.
The supersymmetry transformations (\ref{ChiTransfo}), with respect to the spinor $\zeta = \frac{1}{\sqrt 2}\eta$, and with the vector multiplet localized to (\ref{locus:VecS1M3}), read for these KK fields:
\bea
\dd\phi_{n} \!&=&\! \eta \psi_{n}  \; , \qquad \dd\ti\phi_n \ =\ 0\,, \qquad \dd\psi_{n} \ =\  F_{n} \eta  \; , \qquad \dd F_{n} \ =\ 0\,, \\ [2mm] 
 \dd \ti\psi_{n} & =&  - i \lp \check D_j \ti\phi_n + \frac{r}{2}\partial_j \log\Omega\rp\gamma^j \eta - \frac{i}{\Omega} \lp n
 + \scA_0 \rp \ti\phi_n \, \eta - r \check h \ti\phi_n \, \eta\,, \no \\ [2mm]
\dd \ti F_n &=& i \eta \gamma^j \lp \check D_j - \tfrac{i}{2} \check V_j + \frac{r}{2}\partial_j \log\Omega\rp \ti\psi_n - \frac{i}{\Omega} \lp n+\scA_0 \rp \eta \ti\psi_n - \Big( r - \frac 12 \Big) \check h \eta \ti\psi_n \,,\qquad
\eea
with $\check D_j = \check\nabla_j + i q_R \lp \check A_j - \half \check V_j \rp$ acting on a field fluctuation of R-charge $q_R$, and where $(\ti\psi_n)_{\alpha} \equiv i (\sigma^4)_{\alpha \dot\alpha} \ti\psi^{\dot\alpha}_n\,$. 
The match with the three-dimensional multiplet of \cite{Alday:2013lba} is given by
\begin{align}
(\phi_{n},-\psi_{n},-i F_{n},\ti\phi_n,\ti\psi_n,-i \ti F_n) \ &=\ \Omega^{-r/2}(\phi,\psi,F,\phi^{\dagger},\psi^{\dagger},F)_{\rm 3d}\ ,  \no\\
 n + \scA_0 \ &=\ \sigma_0 \ .
\end{align}
The reality conditions ensuring the positivity of the four-dimensional deformation term are $\phi_n^{\dagger}= \ti\phi_{-n}$ and $F_n^{\dagger}= -\ti F_{-n}$ for bosons, while for the fermions we choose $\psi^{\dagger}_n=-\ti\psi_{-n}$. 

It follows that the Gaussian action around the locus solution for the $n$-th KK mode is
\begin{align}
\dd V_{\rm 4d}^{(n)} &= \dd_{\zeta} \lp  \Big[ (\dd_{\zeta}\psi_n)^{\ddagger} \psi_n - \ti\psi_n (\dd_{\zeta} \ti\psi_n)^{\ddagger}  \Big] \rp 
\ = \ \dd_{\eta} \lp  \Big[ (\dd_{\eta}\psi)^{\dagger} \psi + \psi^{\dagger} (\dd_{\eta} \psi^{\dagger})^{\dagger}  \Big] \rp_{\rm 3d}
= \ \dd V_{\rm 3d}[\sigma_0^{(n)}] \ ,
\end{align}
with $\sigma_0^{(n)} =  n + \scA_0$ and where we have dropped overall factors of $\Omega$ that can be cancelled by irrelevant redefinition of the deformation terms.\footnote{See also appendix~\ref{app:Weyl}, where an alternate   way to see that $\Omega$ does not affect the result is given.} Again we recover the three-dimensional deformation term used in \cite{Alday:2013lba}. The reality conditions on three-dimensional fields following from our map are $\lp \Phi^{(3)}{}^{\dagger} \rp^{\dagger} = \Phi^{(3)}$ for bosons and $\lp \psi^{(3)}{}^{\dagger} \rp^{\dagger} = \psi^{(3)}$ for fermions, matching \cite{Alday:2013lba}, so that we are able to use their three-dimensional one-loop determinant for each KK multiplet.

Decomposing the fields along the weight basis of their representation $\scR$,\footnote{Note that the fields with a tilde transform in the complex conjugate representation ${\cal R}^{\ast}$, whose weights are opposite to the weights of ${\cal R}$.}
\begin{align}
\Phi_n \ = \ \sum_{\rho \ {\rm weight}} \Phi_{n, \,\rho} \,  , 
\end{align}
the 4d-3d map holds for the fields $\Phi_{(n, \rho)}$ with $\sigma_0^{(n,\rho)} =  n + \rho_{\scA_0}$, where \hbox{$\rho_{\scA_0} \equiv \rho(\scA_0) = \sum_{k=1}^{r_G} \rho_k a_k$.}
We obtain the expected result
\begin{align}
Z_{\rm 1{\textrm-}loop}^{\rm chiral}[\scA_0] \ &=  \prod_{\rho \in {\rm weights}} \ \prod_{n \in \bZ}  Z_{\rm 1{\textrm-}loop \, (3d)}^{\rm chiral} \big[ \sigma_0^{(n , \rho)} \big],
\label{ChiDet}
\end{align} 
where $\rho \in {\rm weights}$ denotes a sum over the weights of the chiral multiplet representation $\scR$.
From \cite{Alday:2013lba}, we extract the result (for $b_1,b_2 > 0$)
\begin{align}
Z_{\rm 1{\textrm-}loop \, (3d)}^{\rm chiral} \big[ \sigma_0^{(\rho)} \big] 
\ &=\  \prod_{n_1,n_2 \ge 0} \frac{ n_1 b_1 + n_2 b_2 + i \rho(\sigma_0) - \frac{r-2}{2}(b_1 + b_2) }{n_1 b_1 +n_2 b_2  - i \rho(\sigma_0) + \frac{r}{2}(b_1 + b_2)} \ .
\end{align}
For arbitrary real $b_1,b_2$, the one-loop determinant is given by the formula above with $|b_1|, |b_2|$ instead of $b_1, b_2$.

Renaming $n \rightarrow n_0$, the one-loop determinant is
\begin{align}
Z^{\rm chiral}_{\rm 1{\textrm-}loop} &= \prod_{\rho \in \textrm{weights}} \  \prod_{n_0 \in \bZ}  \ \prod_{n_1,n_2 \ge 0} 
\frac{ \rho_{\scA_0} + i \frac{r-2}{2} (b_1 +b_2) +  n_0  - i n_1 b_1 - i n_2 b_2 }
{ -  \rho_{\scA_0} - \frac{i r}{2} (b_1 +b_2) -  n_0  - i n_1 b_1 - i n_2 b_2}  \ , 
\end{align}
Again the regularization of the infinite product is detailed in appendix \ref{app:1LoopDetReg}.
This involves the elliptic gamma function, defined for $z, p,q \in \bC$ and $|p| < 1$, $|q| <1$ by
\begin{align}
\Gamma_e(z,p,q)= \prod_{n_1,n_2 \ge 0} \frac{1- z^{-1}p^{n_1+1} q^{n_2+1} }{1- z p^{n_1} q^{n_2} } \ .
\end{align}
The result is
\be
Z^{\rm chiral}_{\rm 1{\textrm-}loop} \ =\  \mathrm{e}^{i\pi \Psi^{(0)}_{\rm chi}}\, \mathrm{e}^{i\pi \Psi^{(1)}_{\rm chi}} \ \prod_{\rho \in \Delta_{\scR}}  \, \Gamma_e \lp \mathrm{e}^{2\pi i \rho_{\scA_0}} \, (pq)^{\frac{r}{2}},p ,q \rp,
\ee
with
\bea
\Psi^{(0)}_{\rm chi} &=& \frac{i}{24}\frac{b_1 + b_2}{b_1 b_2} \left[ (r-1)^3 \, \lp b_1 + b_2 \rp^2 - (r-1) \,  \lp b_1^2 + b_2^2 + 2 \rp   \right] |\scR|\,, \label{Zchi}\\ [2mm]
\Psi^{(1)}_{\rm chi} &=& \sum_{\rho \in \Delta_R} - \frac{\rho_{\scA_0}^3}{3 \, b_1 b_2}  -i (r-1)\frac{b_1 + b_2}{2\, b_1 b_2}  \rho_{\scA_0}^2   + [3(r-1)^2 (b_1 + b_2)^2 - 2 - b_1^2 - b_2^2] \frac{\rho_{\scA_0}}{12 \, b_1 b_2} \,, \no
\eea
where $p = \mathrm{e}^{-2\pi b_1}$, $q= \mathrm{e}^{-2\pi b_2}$, $\Delta_{\scR}$ is the set of weights of the representation $\scR$, and $|\scR|$ is its dimension.
As in the case of the vector multiplet, we have split the prefactor into a part $\Psi^{(0)}_{\rm chi}$ independent of $\scA_0$, and an ``anomalous'' part $\Psi^{(1)}_{\rm chi}$ carrying 
the inconsistent $\scA_0$ dependence. To obtain a consistent result, we will require in the final matrix model that these ``anomalous'' terms vanish.


\section{The partition function}
\label{sec:ThePartFct}

In this section we present our final result for the exact partition function and compare it with the supersymmetric index. We find that the two quantities 
 match, up to a prefactor that defines a Casimir energy for a supersymmetric gauge theory on a curved background.

\subsection{Anomaly cancellations and the supersymmetric index}
\label{ssec:IndexAnom}

For the matrix model to be well-defined as an integral over the maximal torus $T^{r_G}$, we have pointed out that the sum of the 
anomalous parts must cancel
\be
\Psi^{(1)}_{\rm vec}(\scA_0)  + \sum_J \Psi^{(1)}_{{\rm chi,} (J)}(\scA_0) \ =\ 0 \ ,
\ee
where $\sum_J$ is a sum over the chiral multiplets of the theory. From~\eqref{Zvec},~\eqref{Zchi}, assuming arbitrary values of $b_1,b_2$, this gives rise to four constraints on the gauge group and matter content of the 
theory:\footnote{The translation into group theory language is the following: in a representation $\scR$ with weights $\{\rho^{j}\}$, the matrix representing $\scA_0$ in a weight basis is $\scA_0^{\scR} ={\rm diag}[\sum_k a_k \rho^{j}_k, \ 1 \le j \le |\scR|] = {\rm diag}[\, \rho^{j}_{\scA_0}, \ 1 \le j \le |\scR|\, ] $.  More generally $(\scA_0^{\scR})^n = {\rm diag}[\, (\rho^{j}_{\scA_0})^n, \ 1 \le j \le |\scR| \, ] $ and the trace in the representation $\scR$ is $\tr_{\scR} (\scA_0^n) = \tr ((\scA_0^{\scR})^n ) = \sum_{j=1}^{|\scR|} (\rho^{j}_{\scA_0})^n = \sum_{\rho \in \Delta_{\scR}} (\rho_{\scA_0})^n$.
}
\begin{align}
({\rm i}) & \qquad  \sum_{J} \tr_{\scR_J} \lp \scA_0^3 \rp \,=\,0 \ , \no\\
({\rm ii}) & \qquad \ \tr_{\rm Adj} \lp \scA_0^2 \rp \ + \ \sum_{J} (r_J -1) \, \tr_{\scR_J} \lp \scA_0^2 \rp \,=\, 0 \ , \no\\ 
({\rm iii}) & \qquad  \sum_{J} (r_J-1)^2 \, \tr_{\scR_J} \lp \scA_0 \rp \,=\, 0 \ , \no\\
({\rm iv}) & \qquad  \sum_{J} \tr_{\scR_J} \lp \scA_0 \rp \,=\, 0 \ ,
\label{cancel_conditions2}
\end{align}
where Adj denotes the adjoint representation of the gauge group $G$.
Using the Cartan decomposition $\scA_0 = \sum_{k=1}^{r_G} a_k H_k$, with $a_k \in \bR$, and requiring~\eqref{cancel_conditions2} for all $a_k$ leads to
\begin{align}
({\rm i})& \qquad  \sum_{J} \tr_{\scR_J} \lp H_{(k_1} H_{k_2} H_{k_3)} \rp \,=\, 0 \ ,   \no\\
({\rm ii})& \qquad \ \tr_{\rm Adj} \lp H_{(k_1} H_{k_2)} \rp \ + \ \sum_{J} (r_J -1) \, \tr_{\scR_J}\lp H_{(k_1} H_{k_2)} \rp \,=\, 0 \ ,  \no\\
({\rm iii}) & \qquad  \sum_{J} (r_J-1)^2 \, \tr_{\scR_J} \lp H_k \rp \,=\,0 \ ,  \no\\
 ({\rm iv})& \qquad  \sum_{J} \tr_{\scR_J} \lp H_k \rp \,=\,0 \ ,
\label{cancel_conditions3}
\end{align}
where $k = 1,\ldots, r_G$ for all $k$-indices.
These conditions can all be interpreted in terms of vanishing of triangle Feynman diagrams contributing to various anomalies.\footnote{See \cite{Spiridonov:2012ww} for a discussion of 
anomalies in relation to the supersymmetric index.} Condition (i) is implied by the requirement of the vanishing
of the non-Abelian gauge anomaly; condition (ii) is implied by the vanishing of the ABJ anomaly, responsible for non-conservation of the R-symmetry current in an instanton background; 
condition (iii) holds requiring the vanishing of the mixed gauge-R symmetry anomaly $G \times U(1)_R^2$; condition (iv) is equivalent to the vanishing of the mixed gauge-gravitational anomaly. 
All these anomalies arise from chiral fermions with  R-charge  $r_J-1$ in the $\scR_J$ representation. The contribution from the gauginos appears only in condition (ii), while  it drops out 
from the other ones, because the adjoint representation is real.

All the conditions are necessary for the preservation of the dynamical gauge symmetry at the quantum level, in a generic background. 
Notice that  the conditions (iii) and (iv) hold  automatically when the gauge group $G$ has no $U(1)$ factors. 
Moreover, the absence of the ABJ anomaly (condition (ii)), is equivalent to the vanishing of the NSVZ exact gauge beta functions of the theory \cite{NHVZ, Intriligator:2003jj}.
In particular, this is satisfied by all theories that flow to a SCFT in the infra-red (IR).  However, one can also consider theories exhibiting confinement in the IR, obtained for instance by suitable 
superpotential deformations \cite{Spiridonov:2010hh}. Pure ${\cal N}=1$  super Yang--Mills (SYM) is an example of a theory for which the partition function (and hence the supersymmetric index) is ill-defined.

Gathering the results of the vector and chiral multiplets (\ref{Zvec}), (\ref{Zchi}), the partition function on $S^1\times M_3$ is expressed by the exact formula
\begin{align}
Z [{\cal H}_{p,q} ] \ =\ \mathrm{e}^{- \scF(p,q)} \, \frac{(p;p)^{r_G} (q;q)^{r_G}}{|\scW|}\! \int\limits_{T^{r_G}}\!\! \frac{\diff z}{2\pi i z}\! 
 \prod_{\alpha \in \Delta_{+}} \! \theta \lp z^{\alpha}, p \rp  \theta \lp  z^{-\alpha}, q \rp \prod_{J}\! \prod_{\rho \in \Delta_J}\! \Gamma_e \big( z^{\rho}  (pq)^{\frac{r_J}{2}},p ,q \big)\, ,
 \label{PartFunction}
\end{align}
where $z^{\pm\alpha} = \mathrm{e}^{\pm 2\pi i \alpha_{\scA_0}}$, $z^{\rho} = \mathrm{e}^{2\pi i \rho_{\scA_0}}$, $J$ labels various chiral multiplets of R-charge $r_J$ transforming in representation $\scR_J$, $\Delta_J$ is the set of weights of $\scR_J$, and 
\bea
\scF(p,q) &=& \frac{\pi}{12} \lp |b_1| + |b_2| - \frac{|b_1| + |b_2|}{|b_1 b_2|} \rp \lp |G| + \sum_J (r_J-1)|\scR_J| \rp \no\\ [2mm]
&& \, + \,\frac{\pi}{24} \, \frac{(|b_1|+|b_2|)^3}{|b_1 b_2|} \sum_J \lp (r_J-1)^3 - (r_J-1) \rp |\scR_J|    \no\\ [2mm]
& =& \frac{4\pi}{3} \lp |b_1| + |b_2| - \frac{|b_1| + |b_2|}{|b_1 b_2|} \rp (\aaa - \ccc)  + \frac{4\pi}{27} \frac{(|b_1|+|b_2|)^3}{|b_1 b_2|}   (3\, \ccc - 2\, \aaa) \ ,\quad
 \label{CasimirE}
\eea
where  in the second line we have used the following definitions
\begin{align}
\aaa \ &=\ \frac{3}{32} \lp 3 \, \textrm{tr} R^3  - \textrm{tr} R \rp  \ =\ \frac{3}{32} \Big[   2 \, |G| + \sum_J  \Big( 3(r_J-1)^3 - (r_J-1) \Big) |\scR_J|  \, \Big] ~,\no\\
\ccc \ &=\ \frac{1}{32} \lp 9 \,  \textrm{tr} R^3  - 5 \, \textrm{tr} R \rp \ =\ \frac{1}{32}  \Big[   4 \, |G| + \sum_J  \Big( 9(r_J-1)^3 - 5 (r_J-1) \Big) |\scR_J| \, \Big]~,
\end{align}
with $R$ the R-symmetry charge and ``tr'' runs over the fermionic fields of the multiplets of the theory. When the theory flows to a fixed point, 
$\aaa$ and $\ccc$ are the central charges of the SCFT \cite{Anselmi:1997ys, Anselmi:1997am, Intriligator:2003jj}.

Comparing with the supersymmetric index  $\scI(p,q)$ with fugacities $p,q$ given for instance in \cite{Aharony:2013dha}, we obtain the relation advertised in the introduction
\be
Z  [{\cal H}_{p,q} ] \ =\  \mathrm{e}^{- \scF (p,q)} \ \scI(p,q)\ .
\label{ZandI}
\ee
The partition function depends on the geometry of $S^1 \times M_3$ only through the complex structure parameters $p = \mathrm{e}^{-2\pi |b_1|}$, $q = \mathrm{e}^{-2\pi |b_2|}$, as predicted by 
\cite{Closset:2013vra}.  More precisely, the authors of \cite{Closset:2013vra} have conjectured that the ratio $Z  [{\cal H}_{p,q} ]  /  \scI(p,q)  =  \mathrm{e}^{- \scF (p,q)} $ can be set to one by a choice of local counterterms.  However, by computing the partition function explicitly in a zeta function regularisation scheme, we have found that this ratio depends on the geometry 
only through the complex structure parameters, and thus generically it cannot be given in terms of integrals of densities local in the background fields. This is clear since generally such densities 
would depend on (functional) degrees of freedom  in the metric.

Notice that for supersymmetric field theories defined on Hopf surfaces the integrated Weyl anomaly vanishes \cite{Cassani:2013dba} and therefore the corresponding ``logarithmic'' 
term in the partition function, arising from conformal transformations of the functional measure \cite{Fujikawa:1980vr}, is absent. Thus (\ref{ZandI}) is the complete answer for the partition function.

In the reminder of this section we will discuss further the interpretation of $\scF (p,q)$. Firstly, we will show that this plays a role in the reduction of the partition function to
the partition function of a three-dimensional theory on $M_3$, upon taking the limit of small $S^1$. Following \cite{Aharony:2013dha}, the reduction along $S^1$ is performed by setting 
$b_1= \beta \, \check b_1$, $b_2 = \beta \, \check b_2$, $\scA_0 = \beta \, \sigma_0$ and taking the limit $\beta \rightarrow 0$ while keeping $\check b_1, \check b_2, \sigma_0$ fixed. 
In this limit the integration over $T^{r_G}$ for $\mathrm{e}^{2\pi i\scA_0}$ becomes an integration over the Cartan sub-algebra $\bR^{r_G}$ for $\sigma_0$. The limits of the various factors in the matrix model are discussed in \cite{Aharony:2013dha}, where it is shown that this reduces to the matrix model of the dimensionally reduced theory on $M_3$. However, 
 it was noticed that a divergent  overall factor appears in the reduction of the index $\scI(p,q)$, given by
\begin{align}
\exp \Bigg[ - \frac{\pi}{12 \, \beta} \, \frac{ |\check b_1|+  |\check b_2|}{|\check b_1| | \check b_2|} \lp |G| + \sum_J (r_J-1)|\scR_J|  \rp \Bigg]  \ = \ \exp \Big[ \, \frac{4 \pi}{ 3 \, \beta} \, \frac{ |\check b_1|+  |\check b_2|}{|\check b_1| | \check b_2|} \lp c- a \rp \Big]
\label{ExtraTerm}
\end{align}
and this was dropped to recover the exact three-dimensional partition function. 
Our results imply that to complete the reduction one should  take into account the contribution from the prefactor $\mathrm{e}^{-\scF(p,q)}$. 
The linear part in $\beta$ vanishes when $\beta \rightarrow 0$ (we discuss this part below), while the 
 part proportional to $\frac{1}{\beta}$  precisely cancels \eqref{ExtraTerm}. We conclude 
  that the full four-dimensional partition function reduces to the exact three-dimensional partition function, computed using the regularization in appendix \ref{app:1LoopDetReg}, reduces to 
the exact three-dimensional partition function of the dimensionally reduced theory.

\subsection{Supersymmetric Casimir energy and large $N$ limit}
\label{ssec:Casimir}

We now discuss how the term linear in $\beta$ appearing in $\scF$ may be interpreted as a Casimir energy, and then comment on the large $N$ limit.  
In general, the vacuum energy of a field theory defined on $S^1\times M_3$ may be defined from the path integral as 
\bea
E_\mathrm{Casimir} & = & - \lim_{\beta\to \infty} \frac{\diff }{\diff \beta} \log Z[\beta; M_3]~,
\eea
where one takes the limit of infinite radius of $S^1$, keeping all other parameters fixed. Using this definition, our partition function computed with supersymmetric boundary conditions 
for the fermions gives:
\bea
E_\mathrm{susy} ( \check b_1, \check b_2 ) & = &   \frac{4\pi}{3} \lp |\check b_1| + |\check b_2| \rp (\aaa - \ccc)  + \frac{4\pi}{27} \frac{(|\check b_1|+|\check b_2|)^3}{|\check b_1 || \check b_2|}   (3\, \ccc - 2\, \aaa) \ ,
\eea
that we refer to as  \emph{supersymmetric Casimir energy}.
This arises from the $\beta \to \infty$ limit of (\ref{CasimirE}), and we used the fact that $\lim_{\beta\to \infty} \tfrac{\diff}{\diff \beta}{\cal I} = 0\,$.  We see that $E_\mathrm{susy}$ depends on the complex structure parameters of the geometry, and on both the central charges $\aaa$
and $\ccc$, characterising the field theory. Since the parameter $\beta$ enters both in the $g_{\tau\tau}$ component of the metric and in $V_\tau$, one can see that $E_\mathrm{susy}$ receives contributions both from the energy-momentum tensor and from the currents in the R-multiplet.
When $p=q$, with $|b_1| = |b_2 |\equiv \frac{\beta}{2\pi}$, this reduces to 
\be
E_\mathrm{susy}  \ =\ \frac{4}{27}\lp \aaa + 3 \ccc \rp \ ,
\label{simplecasi}
\ee
which agrees\footnote{Up to a factor of $2/3$ noted in \cite{Cassani:2014zwa}. Note that (\ref{simplecasi})  holds for an \emph{arbitrary} metric on $M_3\simeq S^3$, as anticipated in 
\cite{Cassani:2014zwa}.} with the expression for the ``index Casimir energy'' given in
 appendix B of \cite{Kim:2012ava}. The latter was defined as tr$[(-1)^F H]$, where $H$ is the Hamiltonian commuting with the supercharges, and a particular supersymmetric regularisation was adopted. Extending to general $p,q$ a prescription given therein for $p=q$, we find that our $E_\mathrm{susy}$ can be expressed in terms of the letter indices \cite{Kinney:2005ej,Dolan:2008qi,Gadde:2010en}
\bea
f_{\rm chiral} (p,q)  =  \frac{(pq)^{\frac r 2} - (pq)^{\frac{2-r}{2}} }{ (1-p)(1-q) }~, \qquad \quad f_{\rm vector} (p,q) = \frac{2 pq - p - q }{ (1-p)(1-q) } \ ,
\eea
with $p = \mathrm{e}^{-2\pi \beta \check b_1}$, $q = \mathrm{e}^{-2\pi \beta \check b_2}$, as
\begin{align}
 E_\mathrm{susy}   ( \check b_1, \check b_2 )  &=  -  \half  \lim_{\beta \to 0} \frac{\diff}{\diff \beta} \sum_{\rm all ~fields} \Big(  f_{\rm chiral} (p,q) +  f_{\rm vector} (p,q)\Big) \ - \ \frac{4}{\pi\beta^2} \frac{|\check b_1|+|\check b_2|}{|\check b_1 || \check b_2|} (\aaa-\ccc)\,,
 \end{align}
where the finite part reproduces $E_\mathrm{susy}$ and the $O(\beta^{-2})$ term is proportional to $\aaa-\ccc$.

In order to compare our $E_\mathrm{susy} ( \check b_1, \check b_2 )$ with other  Casimir energies in the literature we should restrict to the sub-space 
$p=q$, and assume that the metric is the round one on $S^1\times S^3$. In this case, it was shown in \cite{Cappelli:1988vw,Herzog:2013ed}, 
that in a \emph{conformal field theory} (not necessarily supersymmetric) the
Casimir energy, defined as 
\bea
E_0 & = & \int_{S^3} \langle  T^{00} \rangle\, \mathrm{vol} (S^3)~,
\eea
is proportional to the trace anomaly coefficient $\aaa$,  namely
\bea
E_0 & = &  \frac{3}{4}\, \aaa \quad \qquad \mathrm{in~a~CFT}~. 
\eea
Note that this result is valid for an arbitrary CFT, where $\aaa$ and $\ccc$ are not necessarily related. 
For an ${\cal N}=1$  SCFT defined on the round $S^1\times S^3$,  when both can be computed, 
 $E_0$ and $E_\mathrm{susy}$ are two \emph{different} measures of the vacuum energy of a theory.

Notice that in the particular case of $\N=4$ SYM theory on $S^1 \times S^3$ with $G=SU(N)$, the Casimir energy, 
can be computed in the \emph{free field}  limit \cite{Balasubramanian:1999re} and agrees with  $E_0$, while it differs 
from  $E_\mathrm{susy}$ by a numerical factor, namely 
\bea
E_\mathrm{free} \ = \  \frac{3(N^2-1)}{16}   \  = \ E_0~, \quad \quad  E_\mathrm{susy}  \ =  \  \frac{4(N^2-1)}{27} \qquad \mathrm{for} \quad {\cal N}=4 ~~\mathrm{SYM}~.
\eea 
Although $E_\mathrm{susy}$ is valid for any value of the coupling constant (and for any $N$) and in particular at weak coupling in the ${\cal N}=4$ SYM theory, \emph{a priori} it does not have to coincide with $E_\mathrm{free}$  or $E_0$. It would be interesting to understand precisely the relationships between these Casimir energies.

Finally, let us discuss the implications of our results for field theories that admit a gravity dual. For concreteness, we will now assume that the gauge theory is a quiver, 
with gauge group $G = SU(N)^k$ and chiral fields transforming in bi-fundamental representations $(\mathbf{N},\mathbf{{\overline N}})$. We also assume that there is 
  a non-trivial superpotential, and that the theory flows to an interacting fixed point in the IR, with $\aaa=\ccc   +  \scO(1) = {\cal O}(N^2)$, in the limit $N\to \infty$.  These theories are expected to admit 
  a gravity dual solution in type IIB supergravity with geometry $M_5\times Y_5$, where $Y_5$ is a Sasaki--Einstein manifold \cite{Benvenuti:2004dw} and $M_5$ is a deformation of AdS$_5$, supported by $N$ units of five-form flux. Moreover, it should be possible to construct such solutions within the consistent truncation to minimal gauged supergravity and then uplift these to ten dimensions, as illustrated in \cite{Cassani:2014zwa}.  In these cases, at leading order in a large $N$ expansion, the prefactor (\ref{CasimirE}) in the partition function simplifies to 
\be
\scF (p,q)  \ = \  \frac{4\pi}{27} \frac{(|b_1|+|b_2|)^3}{|b_1 || b_2|} \,  \aaa   \ ,
\ee
and using the AdS/CFT relation $\mathrm{exp}(-S_\mathrm{gravity} [M_5] ) = Z_\mathrm{QFT} [\de M_5]$,
 we obtain the following prediction for the five-dimensional holographically renormalised on-shell action:
\bea
S_\mathrm{5d\, sugra} [M_5] & = &   \frac{\pi^2}{54 G_5} \frac{(|b_1|+|b_2|)^3}{|b_1 || b_2|}  \,  .
\label{5dprediction}
\eea
Here we used the relation $\aaa  = \ccc = \tfrac{\pi \ell^3}{8G_5}$ (at leading order in $N$), 
with $G_5$  denoting  the Newton constant of the five-dimensional  supergravity,  and we have set the AdS$_5$ radius $\ell = 1$.

In the  solution of~\cite{Cassani:2014zwa} this formula was found valid, up to some local counterterms. In particular, in that solution $p=q=\mathrm{e}^{-\beta}$, 
albeit the boundary metric comprises a biaxially squashed three-sphere (see appendix \ref{Berger:sec}) and hence it is not conformally flat. In the case of a solution of the 
form  AdS$_5\times Y_5$, the expression (\ref{5dprediction}) reduces to $S_\mathrm{5d\, sugra} [\mathrm{AdS}_5]  =    \tfrac{2\pi\beta}{27 G_5}$
 and again  this should be contrasted with the computation in \cite{Balasubramanian:1999re}, giving   $S_\mathrm{5d\, sugra} [\mathrm{AdS}_5]  =  \tfrac{3\pi\beta}{32G_5}$. 
 When $Y_5=S^5$  the latter agrees with the large $N$ limit of  $E_\mathrm{free} = E_\mathrm{CFT}$ above, while the former gives a different value. 
We expect that this difference can be traced to the use of different holographic regularisation procedures. 
However, this interesting problem deserves to be studied in a future occasion. 
 
Finally, it is tantalizing   to compare (\ref{5dprediction})  with analogous formulae for the on-shell actions in the case of four-dimensional and six-dimensional gauged supegravities,\footnote{The second formula was verified in several explicit examples in \cite{Alday:2014rxa}, and conjectured to hold for general solutions with the topology of the six-ball. In \cite{Alday:2014rxa} it is presented in terms of positive coefficients $b_1,b_2,b_3$, parameterising a contact structure on the 
 five-sphere.}
\bea
S_\mathrm{4d\, sugra} [M_4] \ = \   \frac{\pi}{8 G_4} \frac{(|b_1|+|b_2|)^2}{|b_1 || b_2|}  \,  , \qquad \quad S_\mathrm{6d\, sugra} [M_6] \ = \    \frac{\pi^2}{4G_6} \frac{(|b_1|+|b_2| + |b_3|)^3}{|b_1 || b_2| |b_3|}  \, ,
\label{comparisons}
\eea
put forward  in \cite{Farquet:2014kma}  and   \cite{Alday:2014rxa}, respectively. Here we simply note that these are expressions for the holographically renormalised on-shell action of supersymmetric solutions dual to field theories defined on backgrounds with topology of $S^3$ and $S^5$, respectively, referring to \cite{Farquet:2014kma} and \cite{Alday:2014rxa} for more details.

\section{Conclusions}
\label{conc:sec}

In this paper we have computed the partition function of ${\cal N}=1$ supersymmetric gauge theories --- comprising a vector multiplet for a general gauge group, chiral multiplets with generic R-charges and possibly a superpotential --- defined on a primary Hopf surface ${\cal H}_{p,q}$. We have found that this depends on the background only through the complex structure moduli $p,q$ of the Hopf surface, and is proportional to the supersymmetric index ${\cal I}(p,q)$ with fugacities $p,q$. 
We have carried out the computation reducing the path integral to a matrix integral over the holonomy of the gauge field around $S^1$, 
and evaluating explicitly the one-loop determinant using the method developed in \cite{Alday:2013lba}. 

Our result is essentially in agreement with the conjecture made in \cite{Closset:2013vra}, but we have also determined  the proportionality factor ${\mathrm{e}^{-\cal F}}$ by performing a careful regularisation of the infinite products, employing generalised zeta function techniques. 
This factor defines a \emph{supersymmetric Casimir energy}, depending on the anomaly coefficients $\aaa$, $\ccc$ and 
containing   the leading contribution of $\log Z$ in the large $N$ limit. 
We believe that this term cannot be expressed as a supersymmetric local counterterm and therefore it should be independent of the details of the regularisation scheme. We plan to investigate this further, for example by classifying the possible supersymmetric counterterms.

Perhaps a related question is that of clarifying the dependence of the partition function on the function $\kappa$, parametrising the freedom in choosing the background fields $A_\mu$, $V_\mu$ \cite{Closset:2013vra}. Throughout this paper we have worked with the specific choice of $\kappa$ in (\ref{kapparesult}), dictated by requiring that $A_\mu$ is \emph{real}.  
The general arguments presented in [18] imply that the partition function should not depend on $\kappa$, at least when the path integral is well defined. However, for a generic choice of $\kappa$ the Lagrangian (2.38) does not have positive-definite bosonic part, so that the localization arguments become more formal. It would be nice to analyse the dependence on $\kappa$ more explicitly.

There are several directions for future work. It would be interesting to apply our method  to compute other BPS observables, such as a supersymmetric Wilson loop. 
It should also be possible to prove factorisation of the index \cite{Yoshida:2014qwa,Peelaers:2014ima} using a generalisation of the arguments in section 5.2 of \cite{Alday:2013lba}. 
 As a simple generalisation of our analysis, it should be possible to consider non-direct-product metrics, thus allowing for general complex parameters $p,q$ (see appendix \ref{nondirect:sec}). 
A more challenging extension is that of performing a localization computation on Hermitian manifolds with different topologies, requiring only the existence of one supercharge.   

One of the motivations for this work was to clarify  the results of \cite{Cassani:2014zwa}, by obtaining a precise prediction for the holographically renormalised on-shell action in five-dimensional gauged supergravity, which we presented in (\ref{5dprediction}). It would be interesting to reproduce this formula directly from the dual gravitational perspective. We have noted that in dimensions four, five, and six, the relevant 
on-shell actions  appear to follow a precise pattern, and we expect that explaining this will improve our general understanding of the gauge/gravity duality.

\subsection*{Acknowledgments}
\noindent  
We thank J.~Sparks for discussions and communications on Ref.~\cite{Alday:2014rxa}. B.$\,$A. and D.$\,$M.  are  supported by the ERC Starting Grant N. 304806, ``The Gauge/Gravity Duality and Geometry in String Theory''.
D.$\,$M. also acknowledges partial support from the STFC grant ST/J002798/1. D.$\,$C. is supported by the STFC grant ST/J002798/1.

\appendix

\section{Conventions and identities}
\label{app:conventions}

In this appendix we spell out our conventions and give some identities, useful for the computations in the main text. 

Our spinor conventions are as in~\cite{Dumitrescu:2012ha}. A two-component notation is used:  left-handed spinors carry an undotted index, as $\zeta_\alpha$, $\alpha =1,2$, 
while right-handed spinors are denoted by a tilde and carry a dotted index, as $\widetilde \zeta^{\dot\alpha}$. 
These transform in the $({\bf 2}, {\bf 1})$ and $({\bf 1}, {\bf 2})$ representations of $Spin(4)= SU(2)_+ \times SU(2)_-$, respectively.
The Hermitian conjugate spinors have index structure
\be
(\zeta^{\dagger})^{\alpha}\; =\; (\zeta_{\alpha})^{\ast} \,,  \qquad  (\ti\zeta^{\dagger})_{\dot\alpha}\; =\; (\ti\zeta^{\dot\alpha})^{\ast} \ ,
\ee
and the spinor norms are given by $|\zeta|^2 \,=\, \zeta^{\dagger \, \alpha}\zeta_{\alpha}$ and $|\ti\zeta|^2 \,=\, \ti\zeta^{\,\dagger}_{\dot\alpha}\,\ti\zeta^{\dot\alpha}\,$.

The Clifford algebra is generated by $2\times 2$ sigma matrices
\be
\sigma^a_{\alpha \dot\alpha} \; =\; (\vec\sigma , -i\mathbbm{1}_2) \, ,\qquad \qquad \ti\sigma^{a \, \dot\alpha \alpha}\; =\; (-\vec\sigma , -i\mathbbm{1}_2)  \,,
\ee
where $a=1,\ldots,4$ is a frame index, and $\vec\sigma = (\sigma^1,\sigma^2,\sigma^3)$ are the Pauli matrices. The generators of $SU(2)_{+}$ and $SU(2)_{-}$ are given by
\begin{align}
\sigma_{a b} \,&=\, \frac{1}{4} \lp \sigma_{a} \ti\sigma_{b} - \sigma_{b} \ti\sigma_{a}  \rp \,, \qquad 
\ti\sigma_{a b} \,=\, \frac{1}{4} \lp \ti\sigma_{a} \sigma_{b} - \ti\sigma_{b} \sigma_{a}   \rp  \, ,
\end{align}
and satisfy 
\be
\half \epsilon_{abcd} \, \sigma^{cd} \,=\, \sigma_{ab}\,,\qquad\qquad \half \epsilon_{abcd}\, \ti\sigma^{cd} \,=\, -\ti\sigma_{ab}\, ,
\ee 
with $\epsilon_{1234} = 1$, namely they are self-dual and anti-self-dual, respectively. 
The sigma matrices have the following hermiticity properties
\be
(\sigma_{a})^{\dagger} \,=\, - \ti\sigma_{a} \, , \qquad (\sigma_{ab})^{\dagger} \,=\, - \sigma_{ab} \, , \qquad (\ti\sigma_{ab})^{\dagger} \,=\, - \ti\sigma_{ab}\;,
\ee
and satisfy the relations
\bea
&& \sigma_{a} \ti\sigma_{b} + \sigma_{b} \ti\sigma_{a} \;=\; -2 \delta_{ab} \, , \qquad\quad  \ti\sigma_{a} \sigma_{b} + \ti\sigma_{b} \sigma_{a} \;=\; -2 \delta_{ab}\,,\no \\ [1mm]
&& \sigma_{a} \ti\sigma_{b}\sigma_{c} \;=\; - \delta_{ab} \sigma_{c} 
+ \delta_{ac} \sigma_{b} - \delta_{bc} \sigma_{a} + \epsilon_{abcd} \sigma^d\,, \no \\ [1mm]
&& \ti\sigma_{a} \sigma_{b} \ti\sigma_{c} \;=\; - \delta_{ab} \ti\sigma_{c} 
+ \delta_{ac} \ti\sigma_{b} - \delta_{bc} \ti\sigma_{a} - \epsilon_{abcd} \ti\sigma^d \,,\no \\ [1mm]
&& \sigma_{ab}\sigma_{cd} \ =\ \tfrac{1}{4} \lp - \epsilon_{abcd} - 2 \delta_{ad}\sigma_{bc}+ 2 \delta_{ac}\sigma_{bd} - 2 \delta_{bc}\sigma_{ad} + 2 \delta_{bd}\sigma_{ac} - \delta_{ac}\delta_{bd} + \delta_{ad}\delta_{bc} \rp \,,\no \\ [1mm]
&& \ti\sigma_{ab} \ti\sigma_{cd} \ =\ \tfrac{1}{4} \lp + \epsilon_{abcd} - 2 \delta_{ad} \ti\sigma_{bc}+ 2 \delta_{ac} \ti\sigma_{bd} - 2 \delta_{bc} \ti\sigma_{ad} + 2 \delta_{bd} \ti\sigma_{ac} - \delta_{ac}\delta_{bd} + \delta_{ad}\delta_{bc} \rp.\qquad
\eea

Our supersymmetry parameters $\zeta$, $\ti\zeta$ are commuting spinors, with the supersymmetry variation $\delta_\zeta$, $\delta_{\ti\zeta}$ being Grassmann-odd operators; on the other hand, the dynamical spinor fields are assumed anti-commuting.
The spinor indices are raised or lowered acting from the left with the antisymmetric symbol
$\varepsilon^{\alpha \beta} = -\varepsilon_{\alpha \beta}=\varepsilon^{\dot \alpha\dot \beta}= - \varepsilon_{\dot \alpha\dot \beta}\,$, chosen such that $\varepsilon^{12} = +1$. When constructing a spinor bilinear, the indices are contracted as $\zeta \chi = \zeta^\alpha \chi_\alpha$ and $\ti\zeta \,\ti \chi = \ti\zeta_{\dot\alpha} \,\ti\chi^{\dot\alpha}$.
Then one has the following relations for commuting spinors
\bea
&& \zeta \chi \,=\, -\chi \zeta  \;, \qquad\qquad\qquad\quad \;\ti\zeta \,\ti\chi \,=\, - \ti\chi \,\ti\zeta \,,  \no\\ [2mm]
&& \zeta \sigma_{a} \ti\chi \,=\, \ti\chi \,\ti\sigma_{a} \zeta\;, \qquad\qquad\qquad \zeta \sigma_{ab}\chi \,=\,  \chi \sigma_{ab} \zeta \,,\no\\ [2mm]
&& (\sigma_{a} \ti\zeta\,)\, \chi \,=\, - \ti\zeta\, \ti\sigma_{a} \chi \;,\qquad\quad\quad (\sigma_{ab} \zeta)\, \chi \,=\,\, - \zeta \sigma_{ab} \chi\,, \no \\ [2mm]
&& (\zeta \chi)^{\dagger} \;=\; \chi^{\dagger} \zeta^{\dagger}  \; , \qquad\qquad\, \;\;\;\quad(\ti\zeta \,\ti\chi\,)^{\dagger} \;=\; \ti\chi^{\,\dagger} \,\ti\zeta^{\,\dagger}\,,  \no\\ [2mm]
&& (\zeta \sigma_{a} \ti\chi)^{\dagger} \,=\, - \ti\chi^{\dagger} \ti\sigma_a \zeta^{\dagger}\,,\qquad\quad  \;\;(\zeta \sigma_{ab} \chi)^{\dagger} \,=\, -  \chi^{\dagger} \sigma_{ab}\, \zeta^{\dagger}\,, 
\eea
as well as the Fierz identities
\bea
(\zeta\chi)(\ti\zeta \ti\chi) &=& - \tfrac 12 (\zeta \sigma_{a} \ti\chi)(\chi \sigma^{a} \ti\zeta)\,,\no \\ [2mm]
(\chi_1\chi_2)(\chi_3\chi_4) &=& -(\chi_1\chi_3)(\chi_4\chi_2) - (\chi_1\chi_4)(\chi_2\chi_3)\,.
\label{fierce}
\eea
When the spinors are anti-commuting one has to include an extra minus sign whenever the relation involves swapping two of them.

The spinor covariant derivative is given by
\be
\nabla_\mu \zeta \ = \  \partial_\mu \zeta - \frac{1}{2}\omega_{\mu ab}\sigma^{ab}\zeta  \,,\qquad\quad \nabla_\mu\ti\zeta \ = \ \partial_\mu \ti\zeta - \frac{1}{2}\omega_{\mu ab}\ti\sigma^{ab}\ti\zeta \,,
\ee
where $\omega_{\mu ab}$ is the spin connection, defined from the vielbein $e^a{}_\mu$ and its inverse $e^{\mu}{}_a$ as 
\be
\omega_\mu{}^{ab} \,=\,  2\,e^{\nu[a}\partial_{[\mu}e^{b]}{}_{\nu]} - e^{\nu[a} e^{b]\rho} e_{c\mu} \partial_\nu e^d{}_\rho\,.
\ee 
From the spin connection we can construct the Riemann tensor via\footnote{Our spin connection and Riemann tensor differ by a sign from those of~\cite{Dumitrescu:2012ha} (so our Ricci scalar is positive on a round sphere).}
\be
R_{\mu\nu ab} \ = \ \partial_\mu \omega_{\nu ab} -\partial_\nu \omega_{\mu ab} + \omega_{\mu a}{}^c \omega_{\nu cb} - \omega_{\nu a}{}^c \omega_{\mu cb} \,.
\ee

The integrability condition of the supersymmetry equation~\eqref{KeqnZeta} implies the following relations
\bea
\lp R + 6 V^{\mu}V_{\mu} \rp \zeta  &=& 4i \, \lp \p_{\mu}A_{\nu} - \p_{\nu}A_{\mu} \rp \sigma^{\mu\nu}\zeta \,,\no \\ [2mm]
R + 6 V^{\mu}V_{\mu} &=& 2 J^{\mu\nu} \lp \p_{\mu}A_{\nu} - \p_{\nu}A_{\mu}  \rp \, .\label{Prop1}
\eea 
The first is derived using $ [\nabla_{\mu},\nabla_{\nu}] \zeta = -\half R_{\mu\nu ab} \sigma^{ab} \zeta$, contracted with $\sigma^{\mu\nu}$, and implies the second.

\section{Weyl transformations}
\label{app:Weyl}

In this appendix we discuss how the supersymmetry transformations and Lagrangians are affected by a conformal rescaling of the geometry and of the dynamical fields, in the case when there exist two supercharges of opposite R-charge. This will explicitly show that the conformal factor $\Omega$ can be rescaled away from the localizing terms, and therefore does not affect the result of the computation of the one-loop determinants.

We consider a Weyl rescaling of the general metric~\eqref{seibmetric}, 
\be
g_{\mu\nu} \, = \, \OO^2\, \hat g_{\mu\nu}\;,
\ee
corresponding to redefining the conformal factor $\Omega$ as
\be
\Omega \, =\, \OO \,\hat\Omega\;,
\ee
where here and below a hat denotes the transformed quantities.
 We assume that $\OO$ is a real, positive function  depending on $z,\bar z$ only, so that 
 rescaled background still admits two supercharges of opposite R-charge.
If $\Lambda$ is chosen equal to $\Omega$, then the conformal factor of the new metric is simply $\hat \Omega = 1\,$.
The vielbein and the spin connection transform as 
\be
e^a{}_\mu \ =\ \OO \,\hat e^a{}_\mu\,,\qquad \omega_{\mu ab}\ = \ \hat \omega_{\mu ab}\, + \, \hat e^c{}_\mu \left( \delta_{ca} \hat e^\nu{}_b - \delta_{cb} \hat e^\nu{}_a \right) \partial_\nu \log\OO\,,
\ee
while the two-form $J_{\mu\nu}$ transforms in the same way as the metric,
$J_{\mu\nu} \, = \, \OO^2 \hat J_{\mu\nu}\,,$
and the complex structure $J^\mu{}_\nu$ remains invariant.
As a vector, $K$ is invariant, while as a one-form it transforms as
$K_\mu \, = \, \OO^2 \hat K_\mu\,.$
Starting from~\eqref{reexpressV},~\eqref{reexpressA}, we can now deduce how the background fields $A$ and $V$ transform. 
We will also assign a weight to $|s|$ and $\kappa$, 
\be
|s| \ = \  \OO \, |\hat s| \,,\qquad\qquad \kappa \ =\ \Lambda^{-2} \hat \kappa\,,
\ee
so that both the imaginary part of $A$ and the one of $V$ remain invariant.\footnote{The transformation of $|s|$ is necessary to make sure that the spinors transform correctly and that the imaginary 
part of $A$ does not transform. The transformation of $\kappa$ is imposed for simplicity: as explained in section \ref{sec:susy} any choice of $U_\mu = \kappa K_\mu$ drops from the supersymmetry variations and the localizing terms, as long as one defines $A_\mu^\ddagger=A_\mu^\dagger$.} Note that these conditions are consistent with those ensuring that $A$ is real, given in~\eqref{kapparesult}. 
Then from~\eqref{reexpressV} and~\eqref{reexpressA} we obtain
\be
V_\mu \ = \ \hat V_\mu + (\diff^c \log \OO)_\mu\,,\qquad \qquad A_\mu \ = \ \hat A_\mu + \frac 32 (\diff^c \log \OO)_\mu\,,
\ee
where $(\diff^c \log \OO)_\mu = J_{\mu}{}^\nu\partial_\nu \log\OO = -\widetilde J_{\mu}{}^\nu\partial_\nu \log\OO\,$.
Finally, from~\eqref{zetaTizetaSols} we see that the spinors transform as
\be
\zeta \ = \ \OO^{1/2}\, \hat \zeta\,,\qquad \quad \widetilde\zeta \ = \ \OO^{1/2} \,\hat{\widetilde\zeta}\,.
\ee

We now consider the variations of the fields in the supersymmetry multiplets, showing 
that these are covariant if the Weyl transformation is accompanied by suitable rescaling of the fields.
Let us start with the gauge multiplet, where we  assign the standard  conformal weights 
\be
\mathcal A_\mu \ = \ \hat{\mathcal A}_\mu\,,\qquad \lambda \ = \ \OO^{-3/2}\hat \lambda \,,\qquad \widetilde\lambda \ = \ \OO^{-3/2}\,\hat{\widetilde \lambda} \,,\qquad D\ = \ \OO^{-2} \hat D\,.
\ee
It is easy to see that the supersymmetry variations~\eqref{VecTransfo} transform covariantly as
\bea
\delta \mathcal A_\mu\ =\ \hat\delta \hat{\mathcal A_\mu} \,,\qquad
\delta\lambda \  = \ \OO^{-3/2}\,\hat \delta\hat \lambda\,,\qquad \delta\ti \lambda \ = \ \OO^{-3/2}\,\hat\delta\hat{\ti \lambda}\,,\qquad \delta D \ =\ \OO^{-2}\, \hat\delta\hat D \,,\quad\label{covarianceWeyl}
\eea
where the variation $\hat \delta$ uses $\hat \zeta$, $\hat{\widetilde \zeta}$, and is done on the transformed background defined by $\hat g_{\mu\nu}$, $\hat V$ and $\hat A$. The only non-trivial check is for the relation involving $D$: this follows using the fact that the 
$\Acks_\mu = A_\mu - \frac 32 V_\mu$ is invariant under the Weyl transformation, and the following identity
\be
\ti\zeta\, \ti\sigma^{\mu}\nabla_\mu \lambda\ \equiv \ \OO^{-1/2}\,\hat{\ti \zeta} \,\hat{\ti \sigma}{}^{\mu} \big(\hat \nabla_\mu - \hat \sigma_\mu{}^\nu \partial_\nu \log \OO \big) \big(\OO^{-3/2}\hat \lambda\big)  \ =\ \OO^{-2}\, \hat{\ti \zeta}\, \hat{\ti \sigma}{}^{\mu}\, \hat\nabla_\mu\hat \lambda\,,
\ee
where we used $\ti\sigma^a \sigma_{ab} = - \frac 32\, \ti\sigma_b\,$.

It is also easy to see that the localizing terms, as well as  the Lagrangian~\eqref{Lagrangians:vec} for the vector multiplet  scale
 as $\OO^{-4}$, so that the action is invariant, namely
\be\label{equalitydeltaVgauge}
\int \diff^4x \sqrt g\, \scL_{\rm vector} \ = \ \int \diff^4x  \sqrt{\hat g} \, \widehat{\scL}_{\rm vector} \,.
\ee

We then pass to the chiral multiplet, whose supersymmetry variations were given in~\eqref{ChiTransfo}.
For the scalar $\phi$ we take $\phi = \OO^{-w}\hat\phi$, and choose the conformal weight $w$ such as $w = 3r/2$. The conformal weight of $\psi$, $\widetilde \psi$ is $w + 1/2$, while the one of $F$, $\ti F$ is $w+1$. 
Again, one can show that the supersymmetry variations are covariant under the rescaling, namely
\be
\delta\phi \ = \ \OO^{-w}\,\hat\delta\hat\phi\, ,\qquad \delta \psi \ = \ \OO^{-w-1/2}\,\hat\delta\hat\psi \, ,\qquad \delta F \ = \ \OO^{-w-1}\,\hat\delta\hat F\,,
\ee
with exactly the same relations for $\ti\phi$, $\ti\psi$ and $\ti F$.
While this is straightforward for the variation of $\phi$, it is less obvious for the others. For instance, in the variation of $\ti\psi$ in~\eqref{ChiTransfo} 
 we have
\bea
\ti \sigma^\mu \zeta\, D_\mu \ti\phi & = & \OO^{-1/2}\,\hat{\ti \sigma}{}^\mu \hat\zeta \Big(\hat D_\mu + \frac 32 i\, r J_\mu{}^\nu\partial_\nu\log\OO\Big)\Big(\OO^{-w}\hat{\ti\phi}\Big)\no \\ [1mm]
&=& \OO^{-w-1/2}\,\hat{\ti \sigma}{}_\mu \hat\zeta\left[ \hat D^\mu \hat{\ti\phi} - \Big(w \,\delta^\mu{}_\nu  - \frac 32 r i J^{\mu}{}_\nu\Big)\partial^\nu\log\OO\,\hat{\ti\phi}\right]\,. \qquad
\eea
Since we set $w = \frac 32 r$, the second term vanishes because the vector
$
X^\mu =  \left(\delta^\mu{}_\nu  - i J^{\mu}{}_\nu \right)\partial^\nu\log\OO
$
is holomorphic, and therefore satisfies $X^\mu \ti\sigma_\mu \zeta = 0\,$. 
We can now discuss how the localizing term $\delta_\zeta (V_1 + V_2)$ for the chiral multiplet transforms. Given that this is constructed as a combination of supersymmetry variations, it is also covariant under the Weyl transformation. Specifically, it transforms as
\be
\delta_\zeta (V_1 + V_2) \ = \ \OO^{-2w-2}\,  \hat\delta_{\hat\zeta} (\widehat V_1 + \widehat V_2)\,. 
\ee

Now consider taking $\Lambda = \Omega$, so that $\hat\Omega =1$. If as a localizing term we consider the following \emph{modified} integral
 weighted by the suitable power of $\Omega$
\be
\int \diff^4x \sqrt g\, \Omega^{2w-2} \delta_\zeta (V_1 + V_2)\,,
\ee
then we see that this precisely equal to the original localizing term, in a background with $\Omega=1$, namely
\be\label{NewLocTermChiral}
\int \diff^4x \sqrt g\, \Omega^{2w-2}\, \delta_\zeta (V_1 + V_2) \ =\ \int \diff^4x \sqrt{\hat g}\, \,\hat\delta_{\hat\zeta} (\widehat V_1 + \widehat V_2)\,. 
\ee
In this way the background dependence on $\Omega$ in the localizing term can be reabsorbed by a redefinition of the dynamical fields. 

In conclusion, we have shown that the localizing terms on the left hand side of~\eqref{equalitydeltaVgauge} and~\eqref{NewLocTermChiral} are equivalent upon rescaling the dynamical fields to the same localizing terms defined on a background having $\Omega =1$. This is in agreement with the results of~\cite{Closset:2013vra}.


\section{$S^1 \times S^3_v$ with arbitrary $b_1,b_2$}
\label{Berger:sec}

In this appendix we apply the formulae of section~\ref{toric:sec} in a familiar example. We will consider a geometry comprising the Berger sphere $S^3_v$, namely the biaxially squashed 
three-sphere with $SU(2)\times U(1)$ isometry and squashing parameter $v$. For any value of $v$, this yields a family of four-dimensional supersymmetric backgrounds $S^1 \times S^3_v$, 
depending on the two parameters $b_1$ and $b_2$ which define the Killing vector~\eqref{torickv}. The results of the present paper show that the partition function depends on
 $b_1$, $b_2$, and not on $v$. A similar construction of three-dimensional backgrounds, obtained from a dual holographic perspective, has been presented in~\cite{Farquet:2014kma}.

We take a four-dimensional metric
\be
\diff s^2 \ = \ \Omega^2\, \diff \tau^2 + \diff s^2(S^3_v)\,,
\ee
where the metric on the Berger sphere in standard form is
\be
\diff s^2(S^3_v) \ = \ \diff \theta^2 + \sin^2\theta\, \diff \varphi^2 + v^2 ( \diff \anglepsi+ \cos\theta \,\diff \varphi)^2~,
\label{biax}
\ee
with $\theta \in [0,\pi]$, $\varphi \in [0,2\pi]$, $\anglepsi \in [0,4\pi]$, and $v > 0$ being the squashing parameter. 
This can be written in the toric form~\eqref{4dmetrictoric} by changing coordinates as
\bea
\varphi \ = \ \varphi_1 + \varphi_2 ~, \qquad \quad   \anglepsi \ = \ \varphi_1 - \varphi_2 ~.
\eea
Identifying $\theta = \pi\rho$, so that $f = \pi$, the matrix $m_{IJ}$ reads
\bea
m_{11} &=& 4 \cos^2\frac{\theta}{2} \left(\sin^2\frac{\theta}{2} + v^2  \cos^2\frac{\theta}{2}\right)\,,\qquad m_{12} \ =\ (1-v^2)\sin^2\theta\,, \no \\ [2mm]
m_{22} &=& 4 \sin^2\frac{\theta}{2} \left(v^2\sin^2\frac{\theta}{2} +  \cos^2\frac{\theta}{2}\right)\,.
\eea
Given the choice of Killing vector $K$ in~\eqref{torickv}, the supersymmetry condition $K_\mu K^\mu = 0$ yields
\be
\Omega^2 \ = \ b_+^2 \sin^2\theta + v^2 (b_-  + b_+ \cos\theta)^2\,,
\ee
where $b_\pm = b_1 \pm b_2\,$. 
The background fields $A$ and $V$ are obtained from eqs.~\eqref{VforRealAtoric}, \eqref{realAforS1M3} by first evaluating the functions $c$ and $a_\chi$ appearing in the form~\eqref{prodmetric} of the metric.
We find
\bea
c & = & \frac{4v|b_1b_2|}{\Omega^2}\sin\theta\;, \no \\ [2mm]
a_\chi & = & \frac{1}{\Omega^2} \left[ b_+b_- \sin^2\theta  + v^2 (b_-+ b_+ \cos\theta) (b_+ +  b_- \cos\theta)\right] \,,
\eea
with the map to the $\psi, \chi$ coordinates being
\bea
\varphi \ = \ b_+ \psi + b_-\chi ~, \qquad \quad   \anglepsi \ = \ b_- \psi + b_+ \chi ~.
\eea
One can also determine the complex coordinate $z = u(\theta) + \frac{i}{4b_1b_2}(b_+ \anglepsi - b_- \varphi)$ entering in~\eqref{prodmetric} by integrating \eqref{conditions_defz}, which takes the form 
\be
\frac{\diff \modz }{\diff \theta} \ =\ \frac{\Omega(\theta)}{4v|b_1b_2|\sin\theta} \;,
\ee
and can be solved in closed form. Then from~\eqref{realAforS1M3} we obtain
\bea
A  \!\!&=&\!\! \frac{v\,{\rm sgn}(b_1b_2)}{2\,\Omega^3}\left[ 2b_+^2 \cos\theta\sin^2\theta + v^2(b_- + b_+ \cos\theta)(b_- \cos\theta + b_+ \cos(2\theta))  \right] (b_+\diff \anglepsi -b_- \diff \varphi)\no \\ [1mm] 
\!\!&&\!\! +\, \frac{1}{2}\diff \betaph ,
\eea
with
\be
\betaph \ =\ \frac{1}{2}\left[{\rm sgn}(b_1)(\varphi+\anglepsi)+ {\rm sgn}(b_2)(\varphi- \anglepsi) \right]\,,
\ee
while~\eqref{VforRealAtoric} gives
\bea
V \!\!\!&=&\!\!\!  \frac{v}{48|b_1b_2|\Omega^3}\Big\{ -4v^4 (b_+ + b_- \cos\theta)(b_- + b_+ \cos\theta)^3 + v^2b_+ \! \big[ 8b_-^3 + 7 b_-b_+^2    \\[2mm] 
\!\!\!&+&\!\!\!  b_+ \left( (22b_-^2 + 4b_+^2 )\cos\theta + 16 b_-b_+ \cos(2\theta) + 2 (b_-^2 + 2b_+^2)\cos(3\theta) + b_-b_+ \cos(4\theta) \right) \big] \no \\ [2mm]
\!\!\!&+&\!\!\! 2\, b_+^2 \left[ 2( 3 b_+^2-b_-^2 ) \cos\theta +  b_- b_+ \left( 3 + \cos(2 \theta)\right) \right]\sin^2\theta\Big\}(b_+\diff \anglepsi -b_- \diff \varphi)\no \\[2mm]
\!\!\!&+&\!\!\! \frac{v\, {\rm sgn}(b_1b_2)}{3\,\Omega} \left[  b_+ \left( 2b_-\cos\theta + b_+ (1+\cos^2\theta) \right) - v^2 (b_- + b_+ \cos\theta)^2 \right]\!\left(\frac{b_+\diff\varphi  -b_- \diff \anglepsi}{4b_1b_2} + \frac{i}{2} \diff \tau \right).\no
\eea
These expressions simplify in the following two special cases.

\subsection*{Case $b_1 =-b_2$, with $v$ arbitrary}

If we choose $b_1=-b_2=b/2>0$, we obtain
\be
\Omega \,=\, b\,v\,, \qquad c \,=\, \frac{\sin\theta}{v}\,, \qquad a_\chi \,=\,\cos\theta\, .
\ee 
The complex coordinate $z$ is given by $z = \frac{1}{b}\left( \log\tan\frac{\theta}{2} + i \varphi\right)$. 
The background fields $A$ and $V$ reduce to the $SU(2)\times U(1)\times U(1)$ invariant expressions
\bea
A  &=&  \frac{1}{2} \left(\diff\anglepsi  + \cos\theta \, \diff\varphi \right)\,,\no \\[2mm]
V &=& \frac{v^2}{3} \left(\diff\anglepsi + \cos\theta\diff\varphi + \frac{i}{2}b\,\diff \tau \right),
\eea
with the conformally invariant combination being
\be
\Acks = A - \frac{3}{2}V \ = \ \frac{1}{2}(1-v^2) \left(\diff\anglepsi + \cos\theta\diff\varphi \right) - \frac{i}{4}b\,v^2\,\diff \tau\, .
\ee
The gravity dual of superconformal field theories on $S^1 \times S^3_v$ with this $SU(2)\times U(1)\times U(1)$ invariant choice of background one-forms has been studied in~\cite{Cassani:2014zwa}.

\subsection*{Case $v=1$, with $b_1$ and $b_2$ arbitrary}

Let us keep $b_1$ and $b_2$ arbitrary, and set $v=1$, so that the metric~\eqref{biax} becomes the one of the round three-sphere. Then $\Omega^2$, $c$ and $a_\chi$ simplify to
\bea
\Omega^2 &=& 4 \bigg(\,b_1^2 \cos^2\frac{\theta}{2} +  b_2^2 \sin^2\frac{\theta}{2}\,\bigg) ,\qquad c \ =\ \frac{4|b_1b_2|\sin\theta}{\Omega^2}\,,  \no \\ [2mm] 
a_\chi &=& \frac{4}{\Omega^2} \bigg(\,b_1^2 \cos^2\frac{\theta}{2} -  b_2^2 \sin^2\frac{\theta}{2}\,\bigg) \,,
\eea
and the background fields read
\be
A \ =\  \frac{{\rm sgn}(b_1b_2)}{4\,\Omega^3}\left[ 4\left(b_1^2 + b_2^2 \right) \cos\theta + \left(b_1^2 - b_2^2\right) \left(1+3\cos(2\theta)\right) \right](b_+\diff \anglepsi -b_- \diff \varphi) + \frac{1}{2}\diff\betaph\,, 
\ee
\be
\Acks \ =\ A - \frac{3}{2}V \ =\ - \frac{{\rm sgn}(b_1b_2)}{2\,\Omega}(b_+\diff\varphi  -b_- \diff \anglepsi) - i \frac{|b_1b_2|}{\Omega}\,\diff \tau  +\frac{1}{2}\diff \betaph \, .
\ee
 
\bigskip
 
As a final remark, we observe that the class of three-sphere metrics~\eqref{4dmetrictoric} also comprises the elliptically squashed three-sphere with $U(1)^2$ isometry.
This may be obtained redefining the coordinate $\rho$ into a coordinate $\vartheta \in [0,\pi/2]$ such that $f \diff\rho = [\gamma_1^{2}\sin^2\vartheta + \gamma_2^{2}\cos^2\vartheta]^{1/2}\, \diff\vartheta$, and taking $m_{11} = \gamma_1^{2}\cos^2\vartheta$, $m_{22} = \gamma_2^{2}\sin^2\vartheta$, $m_{12} = 0$; here, $\gamma_1$ and $\gamma_2$ are real parameters, with the squashing being controlled by $\gamma_2/\gamma_1$. The particular choice $\gamma_1 = 1/b_1$ and $\gamma_2 = 1/b_2$ leads to simpler expressions (for instance eq.~\eqref{OmegaToric} gives $\Omega =1$ and the background fields also simplify), however we stress that this choice is not necessary; again, the partition function depends on $b_1$, $b_2$ and not on $\gamma_1$, $\gamma_2$.


\section{Non-direct product metric}
\label{nondirect:sec}

In this paper we consider supersymmetric backgrounds having $S^1 \times S^3$ topology and admitting two supercharges of opposite R-charge. 
In the main text we focused on direct product metrics with $U(1)^3$ isometry, together with a complex Killing vector $K$ depending on two real parameters $b_1$, $b_2$, {\it cf.}\ 
eqs.~\eqref{4dmetrictoric} and~\eqref{torickv}, respectively. We discussed how these data are sufficient to characterize the supersymmetric background.
In this appendix, we relax the direct product condition and make a preliminary analysis of the more general case in which $S^1$ is fibered over $S^3$, still
preserving a $U(1)^3$ isometry.  As we show below, this generalization allows to consider complex values of the moduli $b_1$ and $b_2$ parametrising 
the complex structure on the Hopf surface and appearing in the supersymmetric partition function. 

The most general metric with $U(1)^3$ invariance on the topological product $S^1 \times S^3$ can be written as
\be
\diff  s^2 \ =\  \Omega^2 \left(\diff\tau + c_I \diff \varphi^I + \tilde c\, \diff \rho \right)^2 + f^2 \diff\rho^2 +  m_{IJ}   \left(\diff\varphi^I + n^I \diff \rho \right) \left(\diff\varphi^J + n^J \diff \rho \right)\,, 
\ee
where all the metric functions depend solely on the $\rho$ coordinate. An immediate semplification occurs by noting that one can set $n^I = \tilde c = 0$ by a suitable redefinition of the angular coordinates $\varphi^I$ and $\tau$; hence with no loss of generality we can restrict to the simpler metric
\be\label{4dmetrictoricFibered}
\diff  s^2 \ =\  \Omega^2 \left(\diff\tau + c_I \diff \varphi^I \right)^2 + f^2 \diff\rho^2 +  m_{IJ}   \diff\varphi_I \diff\varphi_J \, .
\ee
Further, the Killing vector $K$ in~\eqref{torickv} can be generalised by analytically continuing the parameters $b_1$ and $b_2$ to complex values
\be
K \ = \ \frac{1}{2} \left[\bbb_1 \frac{\partial}{\partial \varphi_1} + \bbb_2 \frac{\partial}{\partial \varphi_2} - i \frac{\partial}{\partial\tau}\right]\, ,
\label{torickvComplexb}
\ee
where $\bbb_I = b_I + i k_I$, with $b_I$ and $k_I$ real. Since $[K,\overline{K}\, ] = 0$ is still satisfied, for the background to be supersymmetric we just need to solve the condition $K_\mu K^\mu = 0\,$. This constrains the metric as
\be
\Omega^2 \left( 1 + i \,c_I \bbb^I \right)^2 \ =\ \bbb^I m_{IJ} \bbb^J\,.
\ee
Separating the real and imaginary parts, we obtain
\bea
 b^I c_I &=&  \Omega^{-1}\,{\rm Im} \sqrt{\bbb^I m_{IJ} \bbb^J}\,, \no \\ [2mm]
 k^I c_I &=&  1 - \Omega^{-1}\,{\rm Re} \sqrt{\bbb^I m_{IJ} \bbb^J}\,.\label{ConstrComplb}
\eea
In the generic case where the $2 \times 2$ matrix {\small{ $\left(\!\!\begin{array}{c}b_I\\ [-1mm] k_I\end{array}\!\!\right) = \left(\!\!\begin{array}{cc}b_1 & b_2\\ [-1mm] k_1 & k_2\end{array}\!\!\right)$}} is invertible, these equations can be solved for the $c_I$. 
In the main text we considered instead the non-generic case $k_I = 0$, with the second equation solved by $\Omega^2 = b^I m_{IJ} b^J$, and the first  satisfied by setting $c_I = 0$, namely assuming a direct product metric on $S^1 \times M_3$. 
Note that in the generic case one cannot set $c_I = 0\,$.
In both cases, the metric on $M_3$ remains arbitrary, in particular independent of the $\bbb_I\,$.

Let us discuss regularity of the metric in the generic case. In addition to the conditions stated in section~\ref{toric:sec}, ensuring regularity of the metric on $M_3$, we need that the one-form describing the $S^1$ fibration be well-defined on $M_3$. This amounts to requiring that $c_2 \to 0$ as $\rho \to 0$ (where the cycle dual to $\diff \varphi_2$ shrinks to zero size), and that $c_1 \to 0$ as $\rho \to 1$ (where the cycle dual to $\diff \varphi_1$ shrinks). Let us study the behavior at $\rho \to 0$, the case $\rho \to 1$ being completely analogous. Recalling the requirements \eqref{conditionsrhoto0}, from~\eqref{ConstrComplb} we see that as $\rho \to 0$,
\bea
c_1 &\to & \frac{\Omega(0)^{-1}\sqrt{m_{11}(0)}\,(b_1b_2 + k_1k_2) - b_2}{b_1k_2 - b_2k_1}\;, \no \\ [2mm]
c_2 &\to & \frac{-\Omega(0)^{-1}\sqrt{m_{11}(0)}\,|\bbb_1|^2 + b_1}{b_1k_2 - b_2k_1}\;.
\eea
The regularity condition $c_2(0) = 0$ fixes $\Omega(0) = \sqrt{m_{11}(0)} \,\frac{|\bbb_1|^2}{ b_1}$, which then gives $c_1 \to \frac{k_1}{|\bbb_1|^2}\,$. Apart for the behavior at the poles, in this generic case $\Omega(\rho)$ is arbitrary.

In order to complete the global analysis, and check regularity of the background fields $A$ and $V$ as well, we should proceed as done in the main text for the direct product case: define complex coordinates $w,z$ and then use the formulae in section~\ref{sec:2scharges}. 
Although straightforward, we will not pursue this in the present paper.


\section{Proof that $(z_1,z_2)\in \bC^2- (0,0)$}
\label{details:sec}

Below we complete the proof that the coordinates \eqref{z1z2glob}, namely
\bea
z_1  & =  & \mathrm{e}^{-|b_1| (iw+z)} \ = \ \mathrm{e}^{|b_1|\tau}  \mathrm{e}^{|b_1| (\qq-\modz)}    \mathrm{e}^{-i \, {\rm sgn}(b_1) \varphi_1}~,\nonumber\\
z_2 & =  & \mathrm{e}^{-|b_2| (iw-z)} \ = \  \mathrm{e}^{|b_2|\tau}  \mathrm{e}^{|b_2| (\qq+\modz)}  \mathrm{e}^{-i \, {\rm sgn}(b_2)\varphi_2}  ~,
\eea
where the function $\qq (\rho)$, $\modz (\rho)$ obey
\be
\qq' \ =\  \frac{f a_\chi}{\Omega\, c}\,,\qquad\qquad  \modz'\ = \   \frac{f}{\Omega\, c}\,,
\ee
span $\bC^2- (0,0)$. Recall that the functions appearing on the right hand side of these equations are given by 
\bea
c  & = &  \frac{ 2| b_1b_2|}{\Omega^2} \sqrt{\det( m_{IJ})} \,,\nonumber\\
a_\chi  & =&      \frac{1}{\Omega^2}\left(b_1^2 \, m_{11} - b_2^2\, m_{22} \right) \, ,\nonumber\\[2mm]
\Omega^2 & = & b^I m_{IJ} b^J \, ,
\eea
with $f$ arbitrary, and obey certain boundary conditions near to the end-points of the interval $[0,1]$.  
Fixing $|z_2|=\mathrm{e}^{|b_2| \delta_2}$ for finite $\delta_2 \in \bR$ and solving for $\tau = \delta_2 - \qq - u$, we obtain
\begin{align}
|z_1| = \mathrm{e}^{|b_1|\delta_2} \, \mathrm{e}^{-2|b_1|u} ~, \quad  \quad |z_2| = \mathrm{e}^{|b_2|\delta_2} \ ,
\end{align}
and similarly fixing $|z_1|=\mathrm{e}^{|b_1| \delta_1}$ for finite $\delta_1 \in \bR$ and solving for $\tau = \delta_1 - \qq + u$ we obtain
\begin{align}
|z_1| = \mathrm{e}^{|b_1|\delta_1} ~,  \qquad \quad  \quad  |z_2| = \mathrm{e}^{|b_2|\delta_1} \, \mathrm{e}^{2|b_2|u}  \ .
\end{align}
The expansion near to  $\rho \to 0$ and $\rho \to 1$ of the various metric functions imply
\bea
u'(\rho) & = & \frac{1}{2 |b_2| \rho} + \scO(\rho^0) \ , \no\\
u'(\rho) & = & \frac{1}{2 |b_1| (1-\rho)} + \scO((1-\rho)^0) \ ,
\eea
leading to
\bea
u(\rho) & = & \frac{1}{2 |b_2|} \log \rho + \scO(\rho^0)\ , \no\\
u(\rho) & = & -\frac{1}{2 |b_1|} \log (1-\rho) + \scO((1-\rho)^0) \ .
\eea
Using these, and noticing  that  $u(\rho)$ is a monotonically increasing function of $\rho$, since $u' = \frac{f}{\Omega c} \ge 0$, we see that 
$u (\rho)$ is a bijection $(0,1) \rightarrow (-\infty, + \infty)$.   
Therefore, at fixed non-zero $|z_2|$, the radial coordinate $|z_1|$ covers $\bR_{>0}$ (once) and at fixed non-zero $|z_1|$, the radial coordinate 
$|z_2|$ covers $\bR_{>0}$ (once).

So far we have seen that for $(\tau,\varphi_1,\varphi_2,\rho) \in \bR \times [0,2\pi) \times [0,2\pi) \times (0,1)$, the coordinates $(z_1,z_2)$ cover $\bC^2-\{(\bC,0)\}-\{(0,\bC)\}$. 
 The cases $u = \pm \infty$, corresponding to $\rho =0$ and $\rho=1$, must be considered separately, since we may not be able to solve for $\tau \in \bR$ in 
 those cases ($\tau = \pm \infty \notin \bR$ !).
Again solving for $\qq$  and $\modz$ near to $\rho \to 0$ and $\rho \to1$, we obtain
\bea
\qq - \modz \!&=&\! \scO(\rho^2) \,, \qquad\qquad\qquad\quad \qq - \modz \ =\  -\frac{1}{|b_1|} \log(1-\rho) + \scO((1-\rho)^0)\, , \no\\ [1mm]
\qq + \modz \!&=&\!  \frac{1}{|b_2|} \log \rho + \scO(\rho^0) \,, \qquad \qq + \modz \ = \ \scO((1-\rho)^2) \ .
\eea
In the limit $\rho =0$ we have
\begin{align}
|z_1| = \mathrm{e}^{|b_1|\tau} \, , \qquad |z_2| = 0 \ ,
\end{align}
while in the limit $\rho=1$ we have
\begin{align}
|z_1| = 0 \; , \qquad |z_2| = \mathrm{e}^{|b_2|\tau}  \ .
\end{align}
Then we observe that at $|z_1|=0$, $|z_2|$ covers $\bR_{>0}$ (once) and at $|z_2|=0$, $|z_1|$ covers $\bR_{>0}$ (once). This concludes the proof that $(z_1,z_2)$ covers $\bC^2-(0,0)$.


\section{Reduction of the 4d supersymmetry equations to 3d}
\label{app:4dto3dred}

In this appendix we revisit the 4d $\to$ 3d reduction of the supersymmetry equations~\eqref{KeqnZeta}, \eqref{KeqnTiZeta} discussed in~\cite[app.$\:$D]{Closset:2012ru} (see also~\cite{Klare:2012gn}), including a more general identification between the background fields as well as a non-trivial dilaton. Then we show that the 4d background described in section~\ref{toric:sec} reduces to the 3d background considered in~\cite{Alday:2013lba}.

\subsection*{General reduction}

Similarly to the four-dimensional case, in three dimensions the supersymmetry equation arising from the rigid limit of ``new minimal'' supergravity contains different signs depending on whether the spinor parameter has R-charge $+1$ or $-1$. In terms of a spinor $\epsilon$ with R-charge $+1$ and a spinor $\eta\,$ with R-charge $-1$, one has~\cite{Closset:2012ru}
\bea
&&\left(\check \nabla_i - i \check A_i  \right) \epsilon + \frac{i }{2}\check  h\, \gamma_i \epsilon + i\,\check V_i\, \epsilon  + \frac{1}{2}\epsilon_{ijk} \check V^{j} \gamma^k \epsilon \ =\ 0\,,\label{3deqepsilon} \\ [2mm]
&&\left(\check \nabla_i  + i \check A_i\right) \eta + \frac{i }{2}\check  h\,\gamma_i \eta  - i\, \check V_i \,\eta - \frac{1}{2}\epsilon_{ijk} \check V^{j} \gamma^k \eta \ =\ 0 \,,\label{3deqeta}
\eea
where $i, j, k$ are 3d curved indices, and we append a $\check\,$ on 3d quantities that may be confused with 4d ones. The 3d spinor covariant derivative is defined as 
\be
\check \nabla_i\epsilon \ =\ \Big(\partial_i + \frac{i}{4}\check \omega_{i\check a\check b}\epsilon^{\check a\check b\check c}\gamma_{\check c} \Big)\epsilon~,
\ee
(same for $\eta$), where $\check \omega_{i\check a\check b}$ is the 3d spin connection, and $\check a, \check b, \check c$ are 3d flat indices. Moreover, $\check A_i$ is the 3d background gauge field coupling to the R-current, while $\check V_i$ and $\check h$ are a background one-form and a background scalar, respectively.
Our 3d gamma matrices are defined as $(\gamma^{\check a})_\alpha{}^\beta = \sigma^{\check a}_{\rm Pauli}$. 
These are related to the 4d sigma matrices as
\be
\sigma^{\check a}_{\alpha\dot\alpha} \;=\; i\, (\gamma^{\check a})_\alpha{}^\beta \sigma^4_{\beta\dot\beta}\,, \qquad \ti\sigma^{\check a\,\dot\alpha\alpha} \;=\; - i\, \ti\sigma^{4\,\dot\alpha \beta} (\gamma^{\check a})_{\beta}{}^\alpha \,,
\ee
which imply
\bea
\sigma_{\check a4}\, =\, - \frac{i}{2} \gamma_{\check a}\,,&&\quad \sigma_{\check a \check b}\, =\, - \frac{i}{2} \epsilon_{\check a \check b \check c}\,\gamma^{\check c}\,, \,\no \\ [1mm]
\ti\sigma_{\check a4}\, =\, - \frac{i}{2} \ti\sigma_4 \gamma_{\check a} \sigma_4 \,,&&\quad \ti\sigma_{\check a \check b} \,=\, + \frac{i}{2} \epsilon_{\check a\check b\check c}  \ti\sigma_4 \gamma^{\check c} \sigma_4 \,.
\eea
In this way, a 4d left-handed spinor $\zeta_\alpha$ directly reduces to a 3d spinor, while a 4d right-handed spinor $\ti \zeta^{\dot\alpha}$ is mapped to a 3d spinor via $i\sigma^4_{\alpha\dot\alpha} \ti \zeta^{\dot\alpha}\,$.

Let us consider a 4d metric of the form
\be
\diff s^2 \ = \ \check g_{ij}(x)\diff x^i \diff x^j +  \mathrm{e}^{2\Phi(x)}\left( \diff\tau + c_i(x) \diff x^i \right)^2 \,,
\ee
where we are splitting the 4d coordinates as $x^\mu=(x^i,\tau)$, and $\check g_{ij}$, $c_i$, $\Phi$ are a 3d metric, a 3d one-form and a dilaton function, respectively, depending on the 3d coordinates only. 
The 4d vielbein and its inverse can be written as 
\be
e^a{}_\mu \; = \; \left( \begin{array}{cc} \check e^{\check a}{}_i & 0\\ \mathrm{e}^\Phi c_i & \mathrm{e}^\Phi\end{array}\right)\,,\qquad\qquad e^\mu{}_a \; = \; \left( \begin{array}{cc} \check e^i{}_{\check a} & 0\\ - c_j\check e^j{}_{\check a} & \mathrm{e}^{-\Phi}\end{array}\right)\,,
\ee
where $\check e^{\check a}{}_i$ is a vielbein for $\check g_{ij}$, with inverse $\check e^{i}{}_{\check a}$. The 4d spin connection $\omega_{cab}$ splits as
\bea
\omega_{\check c \check a\check b} &=& \check e^i{}_{\check c} \, \check \omega_{ i \check a \check b}\,,\qquad \qquad\quad\;\,\omega_{4\check a\check b} \ = \ -\mathrm{e}^\Phi \partial_{[i} c_{j]}\,\check e^i{}_{\check a}\, \check e^j{}_{\check b}\,, \no \\ [2mm]
\omega_{\check c 4\check b} &=& \mathrm{e}^\Phi\partial_{[i} c_{j]}\,\check e^i{}_{\check b}\, \check e^j{}_{\check c}\,,\qquad\quad \omega_{44\check b} \ = \ \check e^i{}_{\check b}\,\partial_{i} \Phi\,.
\eea

We now reduce the 4d equation for $\zeta$ given in~\eqref{KeqnZeta} along the Killing direction $\partial/\partial\tau$. Assuming that $\zeta$ is independent of $\tau$, we obtain the following 3d equations
\be
\left[\frac 14 \mathrm{e}^\Phi v^i \gamma_i  - \frac i2 \partial_i \Phi\, \gamma^i -i\, \mathrm{e}^{-\Phi} A_\tau +i\, \mathrm{e}^{-\Phi} V_\tau - \frac 12 (V_i -c_i V_\tau) \gamma^i \right]\zeta  \ = \ 0\,,
\ee
\be
\left[\check \nabla_i + \frac 14 \mathrm{e}^\Phi \epsilon_{i j k} v^{j} \gamma^k - i (A_i - c_i A_\tau) +i (V_i - c_i V_\tau) + \frac{1}{2} \mathrm{e}^{-\Phi} V_\tau \gamma_i + \frac{1}{2}\epsilon_i{}^{jk} (V_j -c_j V_\tau) \gamma_k \right]\! \zeta \, = \, 0\, ,\label{ReductionZetaEq2}
\ee
where we introduced
\be 
v^i = -i\,\epsilon^{ijk} \partial_j c_k\, .
\ee 
The first equation is solved by requiring that the 4d one-form
\be
\mathcal U_\mu \ =\ (\mathcal U_i \,,\, \mathcal U_\tau) \ = \ \Big( V_i- 3c_i V_\tau + 2c_i A_\tau - \frac 12 \mathrm{e}^\Phi v_i + i\, \partial_i \Phi  \;,\; 2A_\tau - 2V_\tau \Big)
\ee
satisfies
\be
\mathcal U_\mu \ti\sigma^\mu \zeta \,=\, 0 \qquad \Leftrightarrow \qquad (\mathcal U_i - c_i\, \mathcal U_\tau)\gamma^{i}\zeta + i\,\mathrm{e}^{-\Phi}\,\mathcal U_\tau \zeta \,=\,0 \,,
\ee 
which is equivalent to $J_\mu{}^\nu\,\mathcal U_\nu = i \,\mathcal U_\mu$, meaning that $\mathcal U_\mu$ is of type $(0,1)$ with respect to the complex structure $J$ defined by $\zeta$.
Then eq.~\eqref{ReductionZetaEq2} can be matched with either one of the 3d supersymmetry conditions~\eqref{3deqepsilon}, \eqref{3deqeta}. As we will need to precisely recover the solution studied in~\cite{Alday:2013lba}, 
we choose to match the equation~\eqref{3deqeta} for $\eta$, although this leads to a map between 4d and 3d background fields containing some awckward minus signs. Identifying the spinor parameters as $\eta = \zeta$, the 3d background fields are given by
\be\label{4d3drelation}
 \check A_i  =   -(A_i-c_iA_\tau) - \frac 12 \mathrm{e}^\Phi v_i \,,\qquad \check V_i = - (V_i - c_i V_\tau) -\frac 12\mathrm{e}^\Phi v_i \,, \qquad \check h = -i\,\mathrm{e}^{-\Phi} V_{\tau}\,.
\ee

The reduction of the equation~\eqref{KeqnTiZeta} for a spinor $\ti\zeta$ works similarly. In this case, we need to require that the one-form
\be
\mathcal{\ti U}_\mu \ = \ (\,\mathcal{\ti U}_i \,,\, \mathcal{\ti U}_\tau) \ = \ \Big( V_i - 3c_i V_\tau + 2c_i A_\tau - \frac 12 \mathrm{e}^\Phi v_i - i\, \partial_i \Phi  \;,\; 2A_\tau - 2V_\tau \Big)
\ee
(differing from $\mathcal U_\mu$ just by the sign of $\partial_i \Phi$) satisfies
\be
\ti{\mathcal U}_\mu \sigma^\mu \ti\zeta \,=\, 0 \qquad \Leftrightarrow \qquad \big(\ti{\mathcal U}_i - c_i \,\ti{\mathcal U}_\tau\big)\gamma^i \sigma_4\ti\zeta - i\, \mathrm{e}^{-\Phi}\,\ti{\mathcal U}_\tau \sigma_4\ti\zeta \,= \, 0 \,,
\ee
namely is of type $(0,1)$ with respect to the complex structure $\ti J$ defined by $\ti\zeta$.
Identifying the spinors as $\epsilon = i \sigma_4 \ti \zeta$, eq.~\eqref{3deqepsilon} is retrieved by taking exactly the same 3d background fields as in~\eqref{4d3drelation}.

From~\eqref{4d3drelation}, we see that if we want both the 4d $A$ and the 3d $\check A$ to be real, then the purely imaginary $v$ has to vanish. In this case, it is possible to set $c$ to zero by redefining the $\tau$ coordinate, so that the 4d metric takes a direct product form.

We observe that the 3d background fields are not uniquely determined though, as the 3d equations are invariant under certain shifts~\cite{Closset:2012ru}. This remains true even if the analogous shift freedom in 4d has been fixed.
For our purposes, it will be enough to discuss this for real 3d background fields $\check A^i,\check V^i, \check h$. In this case, given a solution $\epsilon$ to~\eqref{3deqepsilon}, one also has a solution to~\eqref{3deqeta} by taking the charge conjugate, $\eta = \epsilon^c$. This implies the existence of a real Killing vector $\check K^i = \epsilon^\dagger \gamma^i \epsilon$. Then the equations~\eqref{3deqepsilon}, \eqref{3deqeta} are invariant under shifting the background fields as 
\be\check A \to \check A + \frac{3}{2}\frac{\check \kappa\, \check K}{\sqrt{\check K^i\check K_i}}\,,\qquad  \check V \to \check V +  \frac{\check \kappa\,\check K}{\sqrt{\check K^i\check K_i}}\,, \qquad \check h \to \check h + \check \kappa\,,
\ee where $\check \kappa $ is a real function. 
The identifications~\eqref{4d3drelation} between 4d and 3d background fields for a general $\check \kappa$ become
\be\label{4d3drelationBIS}
 \check A + \frac{3}{2}\frac{\check \kappa\, \check K}{\sqrt{\check K^i\check K_i}}  =   -A_i \diff x^i \,,\qquad \check V +  \frac{\check \kappa\,\check K}{\sqrt{\check K^i\check K_i}} = - V_i\diff x^i  \,, \qquad \check h + \check \kappa = -i\,\mathrm{e}^{-\Phi} V_{\tau}\,,
\ee
where we have assumed that the 4d metric is in a direct product form, i.e. $c_i = v^i = 0$, as this is the case that will be relevant below.

\subsection*{Reduction of our background}

We now apply the formulae above and show that the $S^1\times M_3$ background given in section~\ref{toric:sec} reduces to the 3d background studied in~\cite{Alday:2013lba}. Here, neither the fact that the 3d metric admits $U(1)^2$ isometry, nor the global constraints discussed in section~\ref{toric:sec} will play any role. 
The solution in~\cite{Alday:2013lba} has real background fields and supercharges related by charge conjugation. The metric takes the general form
\be\label{metricAMRS}
\diff s^2_{\rm there} \ = \ \Omega^2\left[(\diff\psi + a)^2 + c^2 \diff z \diff\bar z  \right],
\ee
where $\Omega = \Omega(z,\bar z)$, $c=c(z,\bar z)$, $a = a_z(z,\bar z)\diff z + \bar a_{\bar z }(z,\bar z)\diff \bar z$, and for the spinors we take 
\be\label{spinorsAMRS}
\epsilon \,=\,\sqrt{s_{\rm there}}\left(\!\!\begin{array}{c} 1\\ 0 \end{array}\!\!\right),\qquad  \eta \,=\, -i\sigma_2 \epsilon^* \,=\, \sqrt{s_{\rm there}^*}\left(\!\!\begin{array}{c} 0\\ 1 \end{array}\!\!\right) \,,
\ee 
where $|s_{\rm there}| = \Omega$. Then the 3d Killing vector is $\check K = {\partial}/{\partial\psi}$, which as a one-forms reads $\check K = \Omega^2 (\diff \psi +a)$. Finally, the background fields given in~\cite{Alday:2013lba} read
\bea
A_{\rm there} &=& -{\rm Im}\!\left[\partial_z\log\left(\Omega^3c\right)\diff z\right]  + \frac{1}{2}\diff {\rm Arg} (s_{\rm there}) + \frac{*_2(\diff a)}{c^2}(\diff\psi + a)\,,\no\\ [2mm]
V_{\rm there} &=& -2\,{\rm Im}[\partial_z\log\Omega\, \diff z] + \frac{*_2(\diff a)}{c^2}(\diff\psi + a) \,,\no \\ [2mm]
h_{\rm there} &=& \frac{*_2(\diff a)}{2\Omega c^2}\,. \,\label{bckgndAMRS} 
\eea
These expressions are obtained expanding eqs.~(2.11)--(2.16) therein 
 and translating to our notation (in particular $c_{\rm there} = \Omega c$). 
 
Reducing our 4d metric~\eqref{prodmetric} along $\partial/\partial \tau$ clearly matches~\eqref{metricAMRS}. In order to match the spinors in~\eqref{spinorsAMRS} with our spinors~\eqref{zetaTizetaSolsRealA}, we need to identify 
\be
\zeta_\alpha \ = \ \frac{1}{\sqrt 2}\,\eta_\alpha\,,\qquad  i\sigma^4_{\alpha\dot\alpha}\widetilde\zeta^{\dot\alpha} \ = \ \frac{1}{\sqrt 2}\, \epsilon_\alpha \qquad {\rm \Rightarrow}\quad \betaph = -{\rm Arg}(s_{\rm there}) \,.
\ee
Using the formulae derived above, we can also check that the background fields reduce as needed. Since the 4d metric is a direct product, we set $c_i = v^i = 0$; in addition, we take \hbox{$\Phi = \log \Omega$}. Starting from our expressions~\eqref{finalA}, \eqref{Vinpsicoord} for the 4d fields $A$ and $V$, it is easy to check that the conditions on $\mathcal U$ and $\mathcal{\ti U}$ are indeed satisfied. 
Then~\eqref{4d3drelationBIS} gives
\bea
 \check A + \frac{3}{2}\check \kappa\,\Omega(\diff\psi +a) &=&  -A_i \diff x^i \ = \  -{\rm Im}\!\left[\partial_z\log\left(\Omega^3c\right)\diff z\right] \, - \frac{1}{2} \diff \betaph \,,\no\\
 \check V + \check \kappa\,\Omega (\diff\psi +a) &=& - V_i \diff x^i \ = \  -2\,{\rm Im}[\partial_z\log\Omega\, \diff z]  + \frac{1}{3 c^2} *_2\!(\diff a) (\diff\psi + a)  \,,\no \\
 \check h + \check \kappa &=& -\frac{i}{\Omega} V_\tau \,\ =\  -\frac{*_2(\diff a)}{6\Omega c^2} \,.\label{our4dto3dbackground}
\eea
Comparing~\eqref{bckgndAMRS} and~\eqref{our4dto3dbackground}, we see that $\check A$, $\check V$ and $\check h$ agree with $A_{\rm there}$, $V_{\rm there}$ and $h_{\rm there}$ if we pick $\check \kappa = -\frac{2}{3\Omega c^2}*_2\!(\diff a)\,$. However, in the main text it will be not necessary to fix $\check \kappa$, as it actually drops from the 3d supersymmetry transformations.

The condition on $\mathcal U$ translates into the relation
\be\label{3dprojection}
\left(\check V_i - i \,\partial_i \log \Omega \right) \gamma^i \eta + \frac{1}{\Omega} \check V_\psi\eta \ = \ 0\,. 
\ee
This is useful in the 4d $\to$ 3d reduction of the supersymmetry variations in section~\ref{subsec:1LoopDet}.

\section{Regularization of one-loop determinants}
\label{app:1LoopDetReg}

In this appendix we proceed with the regularization of one-loop determinants for the vector multiplet and for the chiral multiplet.

For the vector multiplet the one-loop determinant is given by the infinite product~\eqref{DetVec00}
\begin{align}
Z^{\rm vector}_{\rm 1{\textrm-}loop}
&= Z_{\rm Cartan} \ \Delta_1^{-1} \prod_{\alpha \in \textrm{roots}} \lp  \prod_{n_0 \in \bZ} \prod_{n_1,n_2 \ge 0} \frac{n_1 b_1 + n_2 b_2 + i (n_0 + \alpha_{\scA_0}) }{ -(n_1+1)b_1 -(n_2+1)b_2 + i (n_0 + \alpha_{\scA_0})}  \rp  \no\\
&= Z_{\rm Cartan} \ \Delta_1^{-1} \prod_{\alpha \in \textrm{roots}}  F[\alpha_{\scA_0},i b_1,i b_2] \ ,
\end{align}
with $b_1 >0, b_2 > 0$.

A natural regularization is to use the Barnes multiple zeta/gamma functions and we refer the reader to \cite{MR2078341,FelderVarchenko} for definitions, notations and useful formulae, in particular for the function $\Gamma_3$ and $\zeta_3$. The first step is to rewrite the infinite product above, labelled by a root $\alpha$, with triple gamma functions:
\begin{align}
F[w_{\alpha},\tau,\sigma] &\equiv \prod_{n_0 \in \bZ}  \prod_{n_1,n_2 \ge 0} \frac{ w_{\alpha} + n_0 - n_1 \tau - n_2 \sigma }{ w_{\alpha} + \tau + \sigma + n_0 + n_1 \tau + n_2 \sigma} \label{product1} \\
&=  \prod_{n_0,n_1,n_2 \ge 0} \frac{ 1 + w_{\alpha} + n_0 - n_1 \tau - n_2 \sigma }{  w_{\alpha} + \tau + \sigma + n_0 + n_1 \tau + n_2 \sigma}  
\prod_{n_0,n_1,n_2 \ge 0} \frac{ - w_{\alpha} + n_0 + n_1 \tau + n_2 \sigma }{ 1- w_{\alpha} - \tau - \sigma + n_0 - n_1 \tau - n_2 \sigma} \no\\
&= \frac{\Gamma_3(w_{\alpha} + \tau + \sigma \ | 1, \tau, \sigma) \ \Gamma_3(1- w_{\alpha} - \tau -\sigma \ | 1, -\tau, -\sigma) }{\Gamma_3( 1+w_{\alpha} \ | 1, -\tau, -\sigma) \  \Gamma_3( -w_{\alpha} \ | 1, \tau, \sigma)}  \no
\end{align}
where $w_{\alpha} = \alpha_{\scA_0}$, and we renamed the parameters $b_1$, $b_2$ into $\tau = i b_1$ and $\sigma = i b_2 $ for the ease of comparison with references~\cite{MR2078341,FelderVarchenko}. Then using formula (6.4) in \cite{MR2078341} we get:
\begin{align}
F[w_{\alpha},\tau,\sigma] &= \mathrm{e}^{i \pi \{ \zeta_3(0,-w_{\alpha}|1,\tau,\sigma) - \zeta_3(0,w_{\alpha}+\tau+\sigma|1,\tau,\sigma) \} } 
\ \prod_{n_1,n_2 \ge 0} \frac{1- \mathrm{e}^{2\pi i (-w_{\alpha} + n_1 \tau + n_2 \sigma)} }{1- \mathrm{e}^{2\pi i (w_{\alpha} + (n_1+1) \tau + (n_2+1) \sigma)}} \no\\
&= \mathrm{e}^{i \pi \{ \zeta_3(0,-w_{\alpha}|1,\tau,\sigma) - \zeta_3(0,w_{\alpha}+\tau+\sigma|1,\tau,\sigma) \} }  \frac{1}{\ti\Gamma_e(-w_{\alpha},\tau,\sigma)} \ ,
\end{align}
where $\ti\Gamma_e$ is the elliptic gamma function defined for $x, \tau,\sigma \in \bC$ and ${\rm Im}(\tau),\,{\rm Im}(\sigma) > 0$ by:
\begin{align}
\ti\Gamma_e(x,\tau,\sigma)= \prod_{n_1,n_2 \ge 0} \frac{1- \mathrm{e}^{2\pi i (-x + (n_1+1) \tau + (n_2+1) \sigma)} }{1- \mathrm{e}^{2\pi i (x + n_1 \tau + n_2 \sigma)}} \ .
\label{defGammae}
\end{align}
In the product over roots $\alpha$ we can combine the factors for the roots $\alpha$ and $-\alpha$ and use some formulae in \cite{FelderVarchenko}:
\begin{align}
F[w_{\alpha},\tau,\sigma] \, F[-w_{\alpha},\tau,\sigma] &= \frac{\mathrm{e}^{i \pi \Psi(w_{\alpha},\tau,\sigma) } }{\ti\Gamma_e(-w_{\alpha},\tau,\sigma)\ti\Gamma_e(w_{\alpha},\tau,\sigma)} 
\ =  \ \mathrm{e}^{i \pi \Psi(w_{\alpha},\tau,\sigma) }
\ \theta_0(w_{\alpha},\tau) \, \theta_0(-w_{\alpha},\sigma) \ ,
\end{align}
where 
\begin{align}
\Psi(w_{\alpha},\tau,\sigma) &=  \zeta_3(0,-w_{\alpha}|1,\tau,\sigma) - \zeta_3(0,w_{\alpha}+\tau+\sigma|1,\tau,\sigma) \no\\
& \quad +  \zeta_3(0,w_{\alpha}|1,\tau,\sigma) - \zeta_3(0,-w_{\alpha}+\tau+\sigma|1,\tau,\sigma)
\end{align}
and $\theta_0$ is the Jacobi theta function, defined for $w_{\alpha}, \tau \in \bC$, Im$(\tau)>0$ by
\begin{align}
\theta_0(w_{\alpha},\tau) &= \prod_{n \ge 0} \lp 1 - \mathrm{e}^{2\pi i (n\tau + w_{\alpha})} \rp \lp 1 - \mathrm{e}^{2\pi i ((n+1)\tau - w_{\alpha})} \rp  \ .
\label{defTheta0}
\end{align}
Formula (5.24) in \cite{MR2078341} gives:
\begin{align}
\Psi(w_{\alpha},\tau,\sigma) &= w_{\alpha}^2 \lp \frac{1}{\tau} + \frac{1}{\sigma} \rp + \frac{1}{6} \lp \tau + \sigma + \frac{1}{\tau} + \frac{1}{\sigma} \rp  \ .
 \label{prefactor1}
\end{align}
In total we have
\begin{align}
Z^{\rm vector}_{\rm 1{\textrm-}loop} &= Z_{\rm Cartan} \ \Delta_2^{-1} \  \prod_{\alpha \in \Delta_{+}} \mathrm{e}^{i\pi \Psi_{\alpha}}
\ \theta_0\lp \alpha_{\scA_0}, i b_1 \rp \, \theta_0 \lp -\alpha_{\scA_0}, i b_2 \rp  \  .
\end{align}
The contribution of a Cartan component corresponds to the contribution of a root $\alpha=0$. To evaluate it we can simply take the square root of the contribution of a positive root $\alpha$ and send $\alpha$ to zero,\footnote{We consider the square root because the $\alpha$-factor contains both the contribution of the roots $\alpha$ and $-\alpha$.} \begin{align}
Z_{\rm Cartan} &= \lp \lim_{\alpha \rightarrow 0} \sqrt{Z_{\alpha}} \rp^{r_G}  \ , \quad 
Z_{\alpha} = 
\frac{ \mathrm{e}^{i\pi \Psi_{\alpha}}  \   \theta_0 \lp \alpha_{\scA_0}, i b_1 \rp \,  \theta_0 \lp - \alpha_{\scA_0}, i b_2 \rp}{4 \sin(\pi \alpha_{\scA_0})^2 } \ ,
\end{align}
where $r_G$ is the rank of $G$ (\emph{i.e.} the number of Cartan generators).
This yields
\begin{align}
Z_{\rm Cartan} \ &=\ \mathrm{e}^{\frac{i \pi}{2} \Psi(0,\tau,\sigma) \, r_G } \ ( \mathrm{e}^{-2\pi b_1} ; \mathrm{e}^{-2\pi b_1} )^{r_G} \ ( \mathrm{e}^{-2\pi b_2} ; \mathrm{e}^{-2\pi b_2} ) ^{r_G} \ ,
\end{align}
with the Pochhammer symbol defined for $x,q \in \bC$, $|q|<1$, by $(x;q) = \prod_{n \ge 0} (1- x q^n)$.
With the change of notation $\theta_0(x,y) = \theta(\mathrm{e}^{2\pi i x},\mathrm{e}^{2\pi i y})$, we have
\begin{align}
Z^{\rm vector}_{\rm 1{\textrm-}loop}&=  \mathrm{e}^{i\pi \Psi^{(0)}_{\rm vec}} \, \mathrm{e}^{i\pi \Psi^{(1)}_{\rm vec}} \ (p;p)^{r_G} (q;q)^{r_G} \ \Delta_2^{-1} \ 
 \prod_{\alpha \in \Delta_{+}}  \theta \lp \mathrm{e}^{2\pi i \alpha_{\scA_0}}, p \rp \,  \theta \lp  \mathrm{e}^{-2\pi i \alpha_{\scA_0}}, q \rp ~,\no\\
\Psi^{(0)}_{\rm vec} &=  \frac{i}{12} \lp b_1 + b_2 - \frac{b_1+b_2}{b_1 b_2} \rp |G| ~,\no\\
\Psi^{(1)}_{\rm vec} &= -i \, \frac{b_1+b_2}{b_1 b_2} \sum_{\alpha \in \Delta_{+}} \alpha_{\scA_0}^2 \ ,
\label{VecDetApp}
\end{align}
with $p = \mathrm{e}^{-2\pi b_1}$, $q= \mathrm{e}^{-2\pi b_2}$ and $|G|$ is the dimension of $G$, and
we have split the prefactor into a part $\Psi^{(1)}_{\rm vec}$ depending on $\alpha_{\scA_0}$ and a part $\Psi^{(0)}_{\rm vec}$ independent of $\alpha_{\scA_0}$.

The regularization of the chiral multiplet one-loop determinant proceeds similarly
\begin{align}
Z^{\rm chiral}_{\rm 1{\textrm-}loop} &= \prod_{\rho \in \textrm{weights}} \  \prod_{n_0 \in \bZ}  \ \prod_{n_1,n_2 \ge 0} 
\frac{ \rho_{\scA_0} + i \frac{r-2}{2} (b_1 +b_2) +  n_0  - i n_1 b_1 - i n_2 b_2 }
{ -  \rho_{\scA_0} - \frac{i r}{2} (b_1 +b_2) -  n_0  - i n_1 b_1 - i n_2 b_2}  \no \\
& = \prod_{\rho \in \textrm{weights}} \ \frac{\Gamma_3(u_{\rho} \ | 1 , \tau, \sigma) \ \Gamma_3(1- u_{\rho} \ | 1 , -\tau, -\sigma)}{\Gamma_3(1 + u_{\rho} - \tau - \sigma \ | 1 , -\tau, -\sigma) \ \Gamma_3(- u_{\rho} + \tau + \sigma \ | 1 , \tau, \sigma)} \ , 
\end{align}
where we have regularized the infinite product using triple Gamma functions\footnote{The product $\prod_{n_0 \in \bZ}$ has been split into $\prod_{n_0 > 0} \times \prod_{n_0 \le 0}$ in the numerator and $\prod_{n_0 \ge 0} \times \prod_{n_0 < 0}$ in the denominator.}, and we have defined $\rho_{\scA_0} \equiv \rho(\scA_0)$, $u_{\rho} =  \rho_{\scA_0} + \frac{r}{2}(\tau + \sigma)$, and again $\tau = i b_1$, $\sigma = i b_2$. 

Using formula (6.4) of \cite{MR2078341} leads to
\bea
Z^{\rm chiral}_{\rm 1{\textrm-}loop} &=& \prod_{\rho \in \textrm{weights}}  \lp
 \mathrm{e}^{i \pi \Psi(u_{\rho},\tau,\sigma) }
\ \prod_{n_1,n_2 \ge 0} \frac{1- \mathrm{e}^{2\pi i (-u_{\rho} + (n_1+1) \tau + (n_2+1) \sigma)} }{1- \mathrm{e}^{2\pi i (u_{\rho} + n_1 \tau + n_2 \sigma)}} \rp \no \\ [2mm]
& =& \prod_{\rho \in \textrm{weights}} \mathrm{e}^{i \pi \Psi(u_{\rho},\tau,\sigma) } \, \ti\Gamma_e(u_{\rho},\tau,\sigma) \ ,
\eea
with
\begin{align}
\Psi(u_{\rho},\tau,\sigma) = \zeta_3(0,\tau + \sigma -u_{\rho}|1,\tau,\sigma) - \zeta_3(0,u_{\rho}|1,\tau,\sigma) 
 \ = \ \frac{(u'_{\rho})^{3}}{3 \tau\sigma} + \frac{2- \tau^2 - \sigma^2}{12 \tau\sigma} \, u'_{\rho} \, ,
 \label{prefactor2}
\end{align}
and $u'_{\rho} = u_{\rho} - \frac{\tau +\sigma}{2} = \rho_{\scA_0} + \frac{r-1}{2}(\tau + \sigma)$.
The full chiral multiplet one-loop determinant is 
\begin{align}
Z^{\rm chiral}_{\rm 1{\textrm-}loop} \ &=\  \mathrm{e}^{i\pi \Psi^{(0)}_{\rm chi}}\, \mathrm{e}^{i\pi \Psi^{(1)}_{\rm chi}} \ \prod_{\rho \in \Delta_{\cal R}}  \, \Gamma_e \lp \mathrm{e}^{2\pi i \rho_{\scA_0}} \, (pq)^{\frac{r}{2}},p ,q \rp ~,\no \\
\Psi^{(0)}_{\rm chi} \ &=\  \frac{i}{24}\frac{b_1 + b_2}{b_1 b_2} \left[ (r-1)^3 \, \lp b_1 + b_2 \rp^2 - (r-1) \,  \lp b_1^2 + b_2^2 + 2 \rp   \right] |{\cal R}| ~,\no\\
\Psi^{(1)}_{\rm chi} \ &=\ \sum_{\rho \in \Delta_{\cal R}} - \frac{\rho_{\scA_0}^3}{3 \, b_1 b_2}  -i (r-1)\frac{b_1 + b_2}{2\, b_1 b_2}  \rho_{\scA_0}^2   + [3(r-1)^2 (b_1 + b_2)^2 - 2 - b_1^2 - b_2^2] \frac{\rho_{\scA_0}}{12 \, b_1 b_2} \, ,
\label{ChiDetApp}
\end{align}
where $p = \mathrm{e}^{-2\pi b_1}$, $q= \mathrm{e}^{-2\pi b_2}$, $| {\cal R} |=\;$dim(${\cal R}$), $\Delta_{\cal R}$ is the set of weights of ${\cal R}$ and we have redefined the $\ti\Gamma_e$ function as $\ti\Gamma_e(x,\tau,\sigma) = \Gamma_e(\mathrm{e}^{2\pi i x}, \mathrm{e}^{2\pi i \tau}, \mathrm{e}^{2\pi i \sigma})$.

\bibliographystyle{JHEP}
\bibliography{Newbib}

\end{document}